\documentclass[11pt]{article}

\PassOptionsToPackage{nosort}{cite}

\usepackage{graphicx}
\usepackage{float}
\usepackage{mathtools}
\usepackage{scalerel}
\usepackage{amssymb}
\usepackage{amsfonts}
\usepackage{dsfont}
\usepackage{mathrsfs}
\usepackage{enumitem}
\usepackage{chet}
\usepackage{multirow}
\usepackage[font=small,format=hang,labelfont={sf,bf}]{caption}

\newcommand{\lsp}{\hspace{1pt}}
\newcommand{\llsp}{\hspace{0.5pt}}
\newcommand{\lnsp}{\hspace{-1pt}}

\newcommand{\veps}{\varepsilon}
\renewcommand{\ge}{\geqslant}
\renewcommand{\le}{\leqslant}

\newcommand{\oprp}{{\widetilde{\mathcal R}}}
\newcommand{\rp}{{\widetilde R}}

\allowdisplaybreaks

\definecolor{darkblue}{rgb}{0.1,0.1,0.7}
\hypersetup{colorlinks,
           linkcolor={darkblue},
           citecolor={darkblue},
           urlcolor={darkblue}
}

\usepackage{amsthm}

\theoremstyle{definition}
\newtheorem{invtool}{Inversion}

\newcommand{\hb}{{\bar{h}}}
\newcommand{\zb}{{\bar{z}}}
\newcommand{\1}{{\mathds{1}}}
\renewcommand{\O}{{\mathcal O}}
\newcommand{\G}{{\mathcal G}}

\renewcommand{\AA}{\mathbf A}
\renewcommand{\SS}{\mathbf S_{\boldsymbol 1}}

\DeclareMathOperator*{\dDisc}{dDisc}

\preprint{LA-UR-20-23147}

\date{April 2020}

\title{Analytic and Numerical Bootstrap of CFTs\\\vspace{4pt} with
$O(m)\times O(n)$ Global Symmetry in 3D}

\author{Johan Henriksson,$^{\!a,b}$ Stefanos R.\ Kousvos,$^{\!c,d}$ and Andreas Stergiou$^e$}

\affiliation{$^a$Mathematical Institute, University of Oxford, Andrew
Wiles Building, Radcliffe Observatory\\\vspace{-3pt}
Quarter, Woodstock Road, Oxford, OX2 6GG, UK\\
$^b$Lincoln College, University of Oxford, Turl Street, Oxford, OX1 3DR, UK\\
$^c$Department of Physics, University of Crete, Heraklion
GR-70013, Greece\\
$^d$Institute of Theoretical and Computational Physics (ITCP),
Department of Physics,\\\vspace{-3pt}
University of Crete, 70013 Heraklion, Greece\\
$^e$Theoretical Division, MS B285, Los Alamos National Laboratory, Los
Alamos, NM 87545, USA}

\abstract{Motivated by applications to critical phenomena and open
theoretical questions, we study conformal field theories with $O(m)\times
O(n)$ global symmetry in $d=3$ spacetime dimensions. We use both analytic
and numerical bootstrap techniques.  Using the analytic bootstrap, we
calculate anomalous dimensions and OPE coefficients as power series in
$\veps=4-d$ and in $1/n$, with a method that generalizes to arbitrary
global symmetry. Whenever comparison is possible, our results agree with
earlier results obtained with diagrammatic methods in the literature. Using
the numerical bootstrap, we obtain a wide variety of operator dimension
bounds, and we find several islands (isolated allowed regions) in parameter
space for $O(2)\times O(n)$ theories for various values of $n$. Some of
these islands can be attributed to fixed points predicted by perturbative
methods like the $\veps$ and large-$n$ expansions, while others
appear to arise due to fixed points that have been claimed to exist in
resummations of perturbative beta functions.}

\begin{document}

\maketitle

\toc
\newpage
\newsec{Introduction}
The Landau theory of phase transitions~\cite{Landau:1980mil} has provided a
powerful framework for the search of emergent critical behavior for
decades. It has served as a solid foundation for the development of
renormalization group (RG) methods like the $\veps$
expansion~\cite{Wilson:1971dc, Wilson:1973jj}, Monte Carlo~\cite{Swendsen}
and functional RG~\cite{Berges:2000ew, Delamotte:2007pf}. These methods
have been very successful in predicting critical behavior in a wide variety
of situations, but there is still a surprising number of discrepancies and
disagreements in the literature, pointing to potentially deep underlying
physical principles. More recently, following pioneering work of
\cite{Dolan:2000ut, Dolan:2003hv, Dolan:2011dv} and
especially~\cite{Rattazzi:2008pe}, old conformal bootstrap ideas have
morphed into an an entirely new and computationally rigorous approach for
the study of conformal field theories (CFTs)---for a review
see~\cite{Poland:2018epd} and for an introduction
\cite{Chester:2019wfx}.

In this work we use the conformal bootstrap method, both analytically and
numerically, to study CFTs with $O(m)\times O(n)$ symmetry. The importance
and relevance of this undertaking is highlighted both by the experimental
applications of such CFTs, as well as their inherent theoretical interest.
CFTs with symmetry $O(2)\times O(2)$ and $O(2)\times O(3)$ should describe
second order phase transitions in a wide variety of materials
\cite{KawamuraIII, Pelissetto:2000ek}, and indeed such transitions have
been claimed to be observed in various experiments.  Our work is also of
pure theoretical interest, since it attempts to shed light on the possible
existence of fixed points that arise due to resummations of perturbative
beta functions~\cite{Pelissetto:2000ne, DePrato:2003ra, Calabrese:2004nt}.
The existence of such fixed points has been questioned by some functional
RG studies~\cite{Tissier:2000tz, Tissier:2001uk, Delamotte:2003dw,
Delamotte:2016acs} and perturbative series analyses~\cite{Delamotte_2008,
Delamotte:2010ba, Delamotte_2010}. Thus, theoretically the state of affairs
regarding these fixed points has remained murky despite decades of effort,
with conflicting results obtained by different RG methods.

The most unequivocal results for $O(m)\times O(n)$ CFTs have been obtained
by taking $n$ large and $m$ finite. The existence of a well-defined
large-$n$ expansion was established in~\cite{KawamuraI}, and the strongest
results to date have appeared in~\cite{Pelissetto:2001fi, Gracey:2002pm,
Gracey:2002ze}.  The $\veps$ expansion has been widely used as well, see
e.g.~\cite{Paterson:1980fc, Pisarski:1980ix, KawamuraI, KawamuraII,
KawamuraIII, Pelissetto:2000ek, Osborn:2017ucf, Rychkov:2018vya}, with the
highest-loop study (six loops) performed recently
in~\cite{Kompaniets:2019xez}.\foot{Note that the $\varepsilon$ expansion
can be performed with $m,n$ generic, but the resulting expressions at
higher loops get very large. Below we will present $\varepsilon$ expansion
results expanded in $n$.} Here we will use the analytic bootstrap method of
large spin perturbation theory, introduced in
\cite{Alday:2016njk,Alday:2016jfr} and developed further in
\cite{Alday:2017zzv,Henriksson:2018myn,Alday:2019clp}, to confirm existing
results in the literature and obtain some new large-$n$ ones.  The analytic
bootstrap gives the same type of results as diagrammatic methods, but
simplifies the computation of certain quantities, such as scaling
dimensions of spinning operators, OPE coefficients and central charges.

\subsec{Analytic bootstrap}

The analytic bootstrap relies on the fact that conformal four-point
correlators can be computed from their double-discontinuities, up to
potential contributions from operators with spin zero or one. The
double-discontinuity measures the singularities that arise in the lightcone
limit, corresponding to operators approaching pairwise null separation in
Lorentzian signature, and is sensitive to operators of large spin in the
operator product expansion (OPE). Spinning conformal primary operators
group into twist families, where the scaling dimensions and OPE
coefficients, collectively the \emph{CFT-data}, are given by functions
analytic in spin extracted from the double-discontinuity. All operators in
a twist family have approximately equal value of the twist, defined as the
difference between scaling dimension and spin:
$\tau_\ell=\Delta_\ell-\ell$. In \cite{Fitzpatrick:2012yx,
Komargodski:2012ek} it was shown that such twist families must exist in any
CFT in dimension $d>2$, and that the CFT-data is sourced by operators
appearing in the OPE decomposition of the crossed channel. Specifically,
the identity operator $\1$ in the $\phi$ four-point function gives rise to
the leading order OPE coefficients for double-twist operators
$\phi\lsp\partial^\ell\phi$ with $\tau_\ell\to2\Delta_\phi$.  Other
crossed-channel operators induce corrections to the CFT-data in the twist
families.

Large spin perturbation theory \cite{Alday:2016njk,Alday:2016jfr} combined
with the Lorentzian inversion formula \cite{Caron-Huot:2017vep} constitutes
a systematic framework for analytic bootstrap for theories with a small expansion
parameter. It applies to both week coupling and strong coupling expansions, as well as to expansions in inverse number of degrees of freedom. At each order in
the expansion, the whole double-discontinuity can be generated from an
ansatz of contributions from a small set of crossed-channel operators, and
the undetermined constants of this ansatz can later be fixed by consistency
conditions, for instance conservation of symmetry currents. The method
applies to a wide range of theories, and in particular it has been used to study
the $\veps$ expansion for the Wilson--Fisher fixed point \cite{Alday:2017zzv}
and in the large-$N$ expansion for the $O(N)$ model
\cite{Alday:2019clp}.\footnote{The critical $O(N)$ model has also been studied from the bootstrap perspective by the methods of multiplet recombination \cite{Rychkov:2015naa,Liendo:2017wsn} and Mellin space bootstrap \cite{Dey:2016mcs}. We briefly revisit the Mellin space bootstrap in Appendix~\ref{sec:Mellinbootstrap}.} In this
paper we show how to generalize these implementations to critical $\phi^4$
theories with any global symmetry group.

Consider a field $\phi$ transforming in some (vector) representation $V$ of
the global symmetry. For the case of $O(m)\times O(n)$ we will take
$\phi=\phi_{ar}$ in the bifundamental representation, where $a=1,\ldots,m$
and $r=1,\ldots,n$. Operators and twist families in the OPE decomposition
of the $\phi$ four-point function will transform in all irreducible
representations $R$ in the tensor product $V\otimes V$. Looking first at
the $\veps$ expansion, the identity $\1$ and bilinear operators $\phi^2_R$
will source the complete CFT-data of double-twist families in all
representations to order $\veps^3$. Moreover, these bilinear operators are
the spin zero operators of the same twist families. Despite the fact that
spin zero is beyond the formal validity of the inversion formula, it was
shown in \cite{Alday:2017zzv} that it is possible to analytically continue
to the formula for the scaling dimensions in the twist families to include
the scalar operators. Encouraged by this observation, we conjecture that the
same is true for $\phi^4$ theories with any global symmetry, which leads to
a set of quadratic equations for $\Delta_{\phi^2_R}$. By solving these
equations we can determine all perturbative fixed points in the $\veps$
expansion consistent with the given global symmetry.\footnote{Apart from
$O(m)\times O(n)$ symmetry considered here, we have checked this reproduces
all known fixed points also in the case of $M\lnsp N$
\cite{Stergiou:2019dcv}, hypercubic and hypertetrahedral symmetry
\cite{Stergiou:2018gjj}.}

For the $\phi^4$ theories with $O(N)$ symmetry, it is well-known
 that the large-$N$ limit admits a description in terms of a
Hubbard--Stratonovich transform. In this
description, the bilinear singlet operator $\phi^2_S$ gets replaced by an
auxiliary field $\sigma$ of approximate dimension $2$. At criticality, the
large-$N$ expansion and the $\veps$ expansion are compatible---using the
perturbative results one can for instance confirm that $\Delta_{\phi^2_S}\to2+\text
O(1/N)$. For certain global symmetry groups, the Hubbard--Stratonovich
transformation can be generalized, where $N$ corresponds to a specific
group parameter.\footnote{For $O(N)$ symmetry, the analytic continuation in
$N$ was recently put on a more solid basis using Deligne categories
\cite{Binder:2019zqc}.} More precisely, if certain bilinear operators $\phi^2_R$
have dimensions approaching $2$, they should be promoted to auxiliary fields
$\mathcal R$. In large spin perturbation theory, these
auxiliary fields will, together with $\1$, source the generalized $1/N$
expansion, and play the role same role as $\sigma$ in the treatment of the
$O(N)$ model in \cite{Alday:2019clp}.

For the symmetry group of this paper, $O(m)\times O(n)$, we will
expand in $1/n$ for fixed $m$. From the results in the $\veps$
expansion, we note that two representations furnish scalars that can be
promoted to auxiliary fields: $\mathcal S$ in the singlet representation,
 and $\mathcal W_{ab}$ in the irrep that is a traceless symmetric
tensor in $O(m)$ and singlet in $O(n)$. Like in the $\veps$ expansion, we can derive
quadratic equations, and the solutions will determine all CFT-data
at order $1/n$. Apart from the free theory, we find that these equations
generate all three non-trivial fixed points from the literature: the
$O(mn)$ symmetric fixed point, where we have only $\mathcal S$; the
chiral fixed point, where we have both $\mathcal S$ and $\mathcal W$; and
the antichiral fixed point, where we have only $\mathcal W$. Amongst the
results that we derive are the order $1/n$ scaling dimensions of the
leading scalar operators in all (even) $O(m)\times O(n)$ irreps,
and the order $1/n$ corrections to the central charges.

\subsec{Numerical bootstrap}
The numerical implementation of our work involves the study of both single
and mixed correlator bootstrap systems. In our study we exclude values of
scaling dimensions of various operators that are not compatible with the
combined requirements of unitarity and crossing symmetry; the latter is
also known as associativity of the OPE.  First, we probe the correlator
$\langle \phi_{ar} \phi_{bs} \phi_{ct} \phi_{du} \rangle$ for self consistency; this is our single correlator system. Previous studies of
$O(2)\times O(n)$ single correlator systems have appeared in
\cite{Nakayama:2014lva} and \cite{Nakayama:2014sba}. We extend their
results and match with analytic predictions from the large-$n$ expansion.

For $m=2$ and sufficiently large $n$ (e.g.\ $n\gtrsim10$), we find
excellent agreement between the numerical bootstrap predictions and the
analytic ones. This can be clearly seen in Figs.~\ref{fig:Delta_X_O2n} and
\ref{fig:Delta_W_O2n}. The comparison is performed by comparing the
position of the kinks in the exclusion bounds in
Figs.~\ref{fig:Delta_X_O2n} and \ref{fig:Delta_W_O2n} with the values of
the analytically predicted scaling dimensions of the corresponding
operators. This reinforces the empirical notion that kinks in bootstrap
bounds correspond to the position in parameter space of actual CFTs. We
find that the antichiral fixed points appear as kinks in our $W$
sector, which is a representation furnished by operators that transform in
the two-index traceless symmetric representation of $O(2)$ and the singlet
representation of $O(n)$.  The chiral fixed points coincide with kinks in
our $X$ sector; operators in this representation transform as singlets of
$O(2)$ and two-index traceless symmetric tensors of $O(n)$.  As expected,
for smaller values of $n$ the agreement between large-$n$ and numerical
predictions becomes progressively worse. For $n=2$ we find a pronounced
kink that appears to correspond to a known fixed point of the
$\veps$ expansion as discussed in~\cite{Stergiou:2019dcv}. In the
$n=3$ case, there exist mild kinks that we study extensively with a mixed
correlator system.

A mixed-correlator bootstrap for $O(2)\times O(n)$ CFTs is studied for the
first time in this work.\footnote{The manuscript \cite{Dowens:2020cua},
which studies the $O(3)\times O(15)$ case with the same mixed-correlator
bootstrap as us, appeared on the arXiv the same day as our original
submission.} It consists of probing self consistency for four-point
functions involving both $\phi$ and $S$, where $S$ denotes the smallest
dimension scalar operator (above $\1$) in the singlet representation. Our
goal is to obtain closed isolated regions (islands) in parameter space,
which may correspond to physical CFTs. This method has so far produced
extremely accurate calculations of critical exponents in the Ising and $
O(2)$ critical theories \cite{Kos:2016ysd, Chester:2019ifh}. Islands have
also been discovered in other theories with relevance to three dimensional
statistical field theory~\cite{Kos:2015mba, Kousvos:2018rhl,
Atanasov:2018kqw, Rong:2018okz}---for a more comprehensive list of
references we refer the reader to \cite{Poland:2018epd}.

We find two sets of islands, which we identify with two qualitatively
different types of fixed points. The first set corresponds to the theories
predicted by the large-$n$ and $\veps$ expansions; these are found by
saturating bounds in the $W$ and $X$ sectors as discussed in the previous
paragraph. We have found these islands for $n$ as low as $6$, which appears
to be in agreement with the predictions of
\cite{Kompaniets:2019xez}---below $n=6$ these fixed points are expected to
be nonunitary. The second set of islands appears to correspond to fixed
points that have been claimed to arise after resummations of perturbative
beta functions \cite{Pelissetto:2000ne, DePrato:2003ra, Calabrese:2004nt}.
These are not the same fixed points that are found in the standard large
$n$ and $\varepsilon$ expansions, and their existence is not widely
accepted~\cite{Tissier:2000tz, Tissier:2001uk, Delamotte:2003dw,
Delamotte_2008, Delamotte:2010ba, Delamotte_2010, Delamotte:2016acs}. In
our bootstrap studies these islands are found by saturating bounds in the
$W$ and $Z$ sectors, where operators in the $Z$ sector transform in the
antisymmetric representation of both the $O(2)$ and the $O(n)$. We find
such islands for $n=3$, which is an experimentally relevant value of $n$.
The corresponding fixed points are called chiral and collinear.

For the $O(2)\times O(3)$ chiral fixed point, we find an island by
saturating a bound in the $W$ sector. The associated critical exponents are
\eqn{\beta = 0.344(5) \,,\qquad \nu = 0.639(7)\,.
}[exponents]
This result is of particular interest, since the experimentally observed
critical point of certain frustrated Heisenberg antiferromagnets is
conjectured to be the $O(2)\times O(3)$ chiral fixed point---see
\cite{Calabrese:2004at, KawamuraI}. Our exponent $\nu$ in \exponents agrees
very well with experimental determinations~\cite[Table
37]{Pelissetto:2000ek}, while $\beta$ does not. Despite our results, which
appear to support the existence of the $O(2)\times O(3)$ chiral and
collinear fixed points, we believe that further research is required to
conclusively settle outstanding issues related to criticisms of some
authors in the functional RG community~\cite{Tissier:2000tz,
Tissier:2001uk, Delamotte:2003dw, Delamotte:2016acs}.  Another issue that
remains unresolved is related to the assertion of some authors that the
$O(2)\times O(3)$ chiral and collinear fixed points are of the focus type
and thus nonunitary~\cite{Pelissetto:2000ne, DePrato:2003ra,
Calabrese:2004nt, Nagano_2019}.

We note that in the $O(2)\times O(3)$ case, the two islands we find are
consistent with a second scalar singlet that has a scaling dimension above
three, i.e.\ the corresponding fixed points are both stable in the context
of the RG. This is something that cannot hold for fixed points found in the
$\varepsilon$ expansion~\cite{Michel:1983in, Rychkov:2018vya,
zinn2007phase}. We also note that all previously mentioned islands are
obtained by making assumptions on the second $B$ sector operator, which
contains odd-spin operators among which the first spin-one operator is the
conserved vector of the $O(n)$ contained in $O(2) \times O(n)$. The fact
that the allowed region presents a sensitivity to assumptions specifically
on the $B$ sector was observed empirically. The dependence of bootstrap
bounds on assumptions in sectors that contain conserved operators have been
studied in other cases in e.g.~\cite{Li:2017kck, Reehorst:2019pzi,
Kousvos:2019hgc, delaFuente:2019hbl}.

The structure of this paper is as follows. In section \ref{revPert} we
review known perturbative results obtained in the $ \varepsilon$ and large
$n$ expansions. In section \ref{sec:analyticgeneral} we lay out the general
formalism of the analytical bootstrap, applicable to any $\phi^4$ theory.
In section \ref{sec:analyticOmOn} we apply the formalism of section
\ref{sec:analyticgeneral} specifically to $O(m)\times O(n)$ theories and
present explicitly numerous results. In section \ref{numerics} we study
$O(2)\times O(n)$ theories with the numerical bootstrap and compare to all
previous results. We conclude in section \ref{conc}.

\newsec{Review of perturbative results}[revPert]
\subsec{\texorpdfstring{$\veps$}{epsilon} expansion}
Here we review results regarding CFTs with global symmetry
$O_{m,n}=O(m)\times O(n)$. In the $\veps=4-d$ expansion, such CFTs are
reached as endpoints of the RG flow of the two-coupling
Lagrangian~\cite{Paterson:1980fc, Pisarski:1980ix, KawamuraI, KawamuraII,
KawamuraIII, Pelissetto:2000ek, Osborn:2017ucf}
\eqn{\mathscr{L}=\tfrac12\lsp\partial_\mu\phi_{ar}\lsp\partial^\mu\phi_{ar}
+\tfrac18\lambda\lsp(\phi_{ar}\phi_{ar})^2
+\tfrac{1}{24}\lsp g\lsp\phi_{ar}\phi_{br}\phi_{as}\phi_{bs}\,.}[Lag]
The $mn$ scalar fields are arranged into a matrix, $\phi_{ar}$, with row
indices running from 1 to $m$ and column indices from 1 to $n$.  The
standard summation convention for repeated indices is used in \Lag. The
$O(m)$ part of the symmetry group acts on the row indices and the $O(n)$
part on the column indices. Since both $O(m)$ and $O(n)$ contain the same
$\mathbb{Z}_2$ symmetry generated by $\phi\to-\phi$, the correct global
symmetry group is obtained by modding $O(m)\times O(n)$ out by a
$\mathbb{Z}_2$. In this work we will use $O_{m,n}$, $O(m)\times O(n)$ and
$O(m)\times O(n)/\mathbb{Z}_2$ interchangeably.

An equivalent Lagrangian, introduced in~\cite{KawamuraII} and commonly used
in subsequent literature, takes the form
\eqn{\mathscr{L}=\tfrac12\sum_a\partial_\mu\vec{\phi}_{a}\cdot
\partial^\mu\vec{\phi}_{a}
+\tfrac{1}{24}u\Big(\sum_a\vec{\phi}_a^{\,2}\Big)^2
+\tfrac{1}{24}v\sum_{a,b}\lsp
\big((\vec{\phi}_a\cdot\vec{\phi}_b)^2-\vec{\phi}_a^{\,2}
\vec{\phi}_b^{\,2}\big)\,,}[LagII]
where $\vec{\phi}_a$ are $m$ vectors of size $n$ each. The couplings
$u,v$ of \LagII are related to the couplings $\lambda,g$ of \Lag by
\eqn{u=3\lsp\lambda+g\,,\qquad v=g\,.}[]
Fixed points with $v<0$ are called collinear or sinusoidal and fixed points
with $v>0$ chiral, helical or noncollinear.  With $m\le n$ stability of the
scalar potential requires $u>0$ if $v\le0$ and $u>(1-1/m)v$ if $v>0$.

\begin{figure}[ht]
  \centering
  \includegraphics{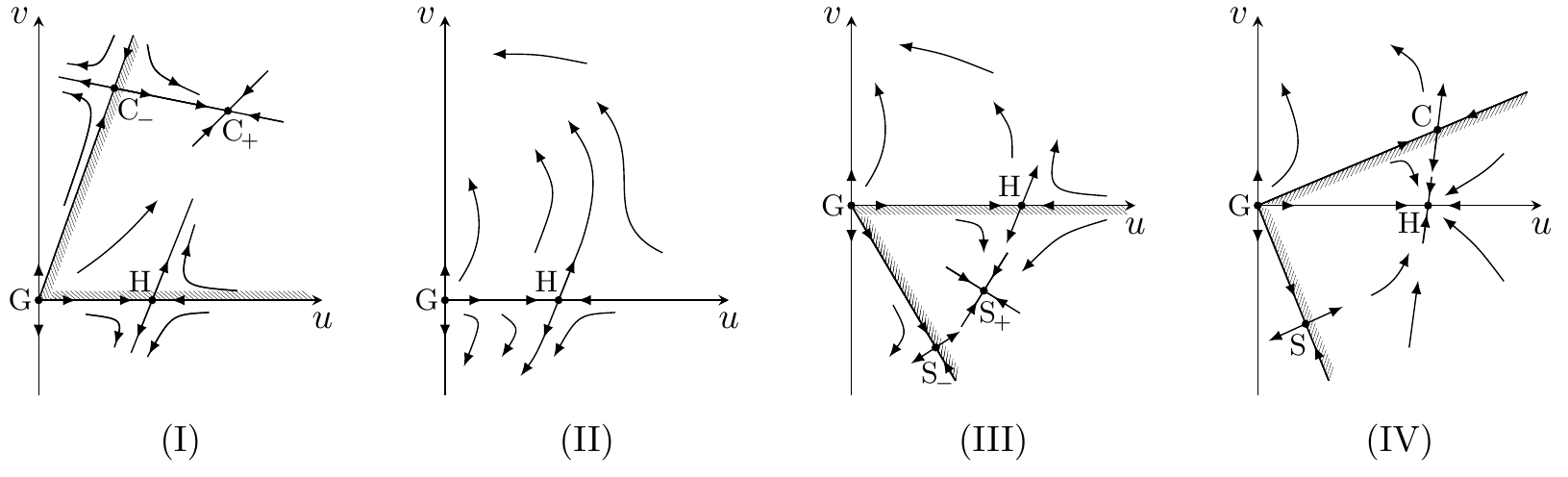}
  \caption{Flow diagrams corresponding to the various regimes mentioned in
    the text. The hatched regions represent the basins of attraction of the
    stable fixed point.  This figure is a reproduction of
    \cite[Fig.~7]{KawamuraIII}.}
  \label{fig:Omn_fixed_points}
\end{figure}

The number of fixed points of the Lagrangian \LagII depends on the values
of $m$ and $n$. There are four regimes:
\begin{enumerate}[label=(\Roman*)]
  \item For $n>n^+(m)$ there are four fixed points (Gaussian, $O(mn)$,
    chiral, antichiral). Stable\foot{A fixed point with only one relevant
    scalar singlet operator, namely the mass operator $\phi_{ar}\phi_{ar}$, is
    called stable.} fixed point: chiral.
  \item For $n^-(m)<n<n^+(m)$ there are two fixed points (Gaussian and
    $O(mn)$). They are both unstable.
  \item For $n_H(m)<n<n^-(m)$ there are four fixed points (Gaussian,
    $O(mn)$, sinusoidal, antisinusoidal). Stable fixed point: sinusoidal.
  \item For $n<n_H(m)$ there are four fixed points (Gaussian, $O(mn)$,
    chiral, sinusoidal). Stable fixed point: $O(mn)$.
\end{enumerate}
The Gaussian fixed point has $u=v=0$, while the $O(mn)$ fixed point has
$u>0,v=0$. The fully-interacting fixed points (i.e.\ the ones besides
Gaussian and $O(mn)$) both have $uv\ne0$ and $O_{m,n}$ global symmetry.
These fixed points move around in the $\lambda$-$g$ coupling plane as $m,n$
change. For every $m$ there is a value of $n$, indicated by $n^+(m)$ above,
for which the chiral and antichiral fixed points collide in the real
$u$-$v$ plane and subsequently move to the complex $u$-$v$ plane. For
$n>n^+(m)$ the chiral fixed point is stable, but for $n<n^+(m)$ there is no
stable fixed point. However, for some $n^-(m)<n^+(m)$ two fixed points
reappear in the $u$-$v$ plane---this time they are called sinusoidal and
antisinusoidal because they have $v<0$, and the sinusoidal fixed point is
stable.  Furthermore, there is a value $n^H(m)<n^-(m)$ below which the
$O(mn)$ fixed point is stable, for one of the fully interacting fixed
points of the $n^H(m)<n<n^-(m)$ regime crosses the $v=0$ line and acquires
$v>0$ (chiral), while the other remains with $v<0$ (sinusoidal).  The
situation is summarized in Fig.~\ref{fig:Omn_fixed_points}.

The values of $n^\pm(m)$ and $n^H(m)$ can be
estimated in the $\veps$ expansion\cite{KawamuraI, KawamuraII,
Pelissetto:2001fi, Osborn:2017ucf}:
\eqna{n^\pm(m)&=5m+2\pm 2\sqrt{6(m-1)(m+2)}
-\bigg(5m+2\pm\frac{25m^2+22m-32}{2\sqrt{6(m-1)(m+2)}}\bigg)\lsp\veps
+\text{O}(\veps^2)\,,\\
n_H(m)&=\frac{2}{m}\big(2-\veps+\text{O}(\veps^2)\big)\,.}[]
In a recent paper, these results have been extended to six loops, or order
$\veps^5$~\cite{Kompaniets:2019xez}. After resummation techniques are
employed, the authors of~\cite{Kompaniets:2019xez} give, for $m=2$, the
estimates
\eqn{n^+(2)=5.96(19)\,,\quad n^-(2)=1.970(3)\,,\quad n^H(2)=1.462(13)\,.
}[npmHmTwo]
Numerical estimates for other values of $m$ can be found in~\cite[Table
8]{Kompaniets:2019xez}. In this work we will attempt to use our bootstrap
bounds to independently estimate these quantities, particularly $n^+(2)$,
and compare with \npmHmTwo. For $m=3$ a similar study was performed
in~\cite{Nakayama:2014lva}, while the existence of $O(2)\times O(n)$
theories in $d=5$ was examined with perturbative methods
in~\cite{Gracey:2017oqf}.

As mentioned in the introduction, it has been suggested that fixed points
beyond the ones we just reviewed exist in $O(2)\times O(2)$ and $O(2)\times
O(3)$ theories. Confusingly, the terminology ``chiral'', ``collinear'',
etc., is still used for those fixed points, depending on their sign of the
coupling $v$.

For scalar theories in the $\veps$ expansion below $d=4$, it is a theorem
that a stable fixed point, if it exists, is unique~\cite{Michel:1983in,
Rychkov:2018vya, zinn2007phase}. For the numerical studies in this work, we
will fix $m$ to a small value, specifically $m=2$, and obtain bounds for
increasing $n$.  Thus, we expect that if kinks appear at large $n$, they
will be due to fixed points of regime (I). In that case, we expect from the
$\veps$ expansion that since the chiral fixed point is stable, the
antichiral is unstable.  This prediction is also expected to hold in the
large-$n$ limit in $d=3$, to which we now turn.

\subsec{Large \texorpdfstring{$n$}{n}}\label{sec:largenlitt}
As mentioned in the introduction, it was realized a long time ago that
$O_{m,n}$ theories admit a large-$n$ expansion~\cite{KawamuraI}. For the
chiral fixed point in $d=3$, large-$n$ computations
give~\cite{Pelissetto:2001fi, Gracey:2002pm, Gracey:2002ze}\foot{These
computations are done in the more general setting of arbitrary $d$ at large
$n$, but here we present the $d=3$ results only. In \cite{Gracey:2002pm},
the operator $C$ corresponds to our $Z$, and $\eta_c$ computed there is
given by $\eta_c=\Delta_Z-1$. In \cite{Gracey:2002ze}, the operator $T$
corresponds to our $W$, and $\chi_T$ computed there is given by
$\chi_T=3-2\Delta_\phi-\Delta_W$.}
\newcommand{\LargenChiral}{\eqref{LargenChiral}\xspace}
\begin{align}
\Delta_{\phi+}&=\frac12+\frac{2(m+1)}{3\pi^2}\frac{1}{n}
+\frac{8(m^2-7m-26)}{27\pi^4}\frac{1}{n^2}+\text{O}\Big(\frac{1}{n^3}
\Big)\,,\nonumber\\
\Delta_{S+}&=2-\frac{16(m+1)}{3\pi^2}\frac{1}{n}
-\frac{64(7m^2+5m-20)+108(m^2+3m+4)\pi^2}{27m^2\pi^4}\frac{1}{n^2}
+\text{O}\Big(\frac{1}{n^3}\Big)\,,\nonumber\\
\Delta_{S'+}&=4-\frac{32(m+1)}{3\pi^2}\frac{1}{n}
+\text{O}\Big(\frac{1}{n^2}\Big)\,,\nonumber\\
\Delta_{S''+}&=4-\frac{8(m+4)}{3\pi^2}\frac{1}{n}
+\text{O}\Big(\frac{1}{n^2}\Big)\,,\nonumber\\
\Delta_{W+}&=2-\frac{4(m+4)}{3\pi^2}\frac{1}{n}
+\text{O}\Big(\frac{1}{n^2}\Big)\,,\nonumber\\
\Delta_{Z+}&=1+\frac{4(m-2)}{3\pi^2}\frac{1}{n}
+\frac{2\big(8(m+1)(m^2-7m-26)-27(5m+11)\pi^2\big)}
{27(m+1)\pi^4}\frac{1}{n^2}+\text{O}\Big(\frac{1}{n^3}\Big)\,,
\label{LargenChiral}
\end{align}
while for the antichiral fixed point in $d=3$ at large $n$ the results
are~\cite{Pelissetto:2001fi, Gracey:2002ze}
\newcommand{\LargenAntichiral}{\eqref{LargenAntichiral}\xspace}
\begin{align}
\Delta_{\phi-}&=\frac12+\frac{2(m-1)(m+2)}{3m\pi^2}\frac{1}{n}
+\frac{8(m-1)(m+2)(m^2-8m-2)}{27m^2\pi^4}\frac{1}{n^2}
+\text{O}\Big(\frac{1}{n^3}\Big)\,,\nonumber\\
\Delta_{S-}&=1+\frac{16(m-1)(m+2)}{3m\pi^2}\frac{1}{n}
\nonumber\\&\hspace{48pt}+\frac{4(m-1)(m+2)
\big(16(7m^2-2m+40)+27(m-2)(m+4)\pi^2\big)}
{27m^2\pi^4}\frac{1}{n^2}+\text{O}\Big(\frac{1}{n^3}\Big)\,,\nonumber\\
\Delta_{S'''-}&=4-\frac{8(m^2+4m-8)}{3m\pi^2}\frac{1}{n}
+\text{O}\Big(\frac{1}{n^2}\Big)\,,\nonumber\\
\Delta_{W-}&=2-\frac{4(m^2+4m-8)}{3m\pi^2}\frac{1}{n}
+\text{O}\Big(\frac{1}{n^2}\Big)\,.
\label{LargenAntichiral}
\end{align}
Here we denote singlet operators in the $\phi\times\phi$ OPE with the
letter $S$, operators that transform as two-index traceless symmetric
tensors under $O(m)$ and singlets under $O(n)$ by the letter $W$, and
operators that transform as two-index antisymmetric tensors under both
$O(m)$ and $O(n)$ with the letter $Z$.\foot{Note that there are in total
nine irreps containing bilinears of $\phi$, and five of them contain
scalars. We will determine dimensions of these operators in the $\veps$ and
large-$n$ expansions in section~\ref{sec:analyticOmOn} using analytic
bootstrap techniques.}  As usual, primes denote the order in scaling
dimension of these operators, i.e.\ $S$ is the leading singlet, $S'$ the
next-to-leading singlet and so on. We have not found large-$n$ results for
$\Delta_{S'-}$ in the literature, but it is widely believed that
$\Delta_{S'-}<3$, i.e.\ the antichiral fixed point is unstable. By
explicitly constructing singlet operators we find the ones in
Table~\ref{table:lowestscalars}, where we tabulate the spectrum of the
lowest dimension scalar singlet operators in the three fixed points at
large $n$.

The results in \LargenChiral and \LargenAntichiral were obtained with the
use of a Hubbard--Stratonovich transformation, extending a procedure used
first in the $O(N)$ models by~\cite{Vasiliev:1981yc, Vasiliev:1981dg}.  Two
Hubbard--Stratonovich auxiliary fields, $\mathcal{S}$ and
$\mathcal{W}_{ab}$, are introduced in this case. $\mathcal{S}$ is a
singlet, while $\mathcal{W}_{ab}$ transforms as a traceless symmetric
tensor under $O(m)$ and a singlet under $O(n)$. The Lagrangian
is~\cite{Pelissetto:2001fi, Gracey:2002ze}
\begin{equation}
\mathscr{L}=\frac{1}{2}\partial_\mu
\vec{\phi}_a\cdot\partial^\mu\vec{\phi}_a+\frac{1}{2}\mathcal{S}
\vec{\phi}_a\cdot\vec{\phi}_a+\frac{1}{2}\mathcal{W}_{ab}\lsp\vec{\phi}_a\cdot
\vec{\phi}_b-\frac{3\lsp\mathcal{S}^2}{2w}-\frac{3}{2v}\mathcal{W}_{ab}
\mathcal{W}_{ab}\,,
\end{equation}
where repeated indices are summed over and $w=u+(1-1/m)v$ with $u,v$ the
couplings in \eqref{LagII}. The equations of motion for $\mathcal{S}$ and
$\mathcal{W}_{ab}$ can be used to go back to \eqref{LagII}.  Below we will
reproduce \LargenChiral and \LargenAntichiral and obtain more results at
order $1/n$ following the analytic bootstrap logic of~\cite{Alday:2019clp}.

\begin{table}[H]
\centering
\caption{The lowest scalar singlet operators at the three non-trivial fixed points and
their scaling dimensions. In the chiral fixed point, the operators $S'$ and $S''$ arise from resolving a mixing, where $\O$ is either $\mathcal S$ or $\mathcal W$.}\label{table:lowestscalars}
\begin{tabular}{|lc|lc|lc|}\hline
\multicolumn{2}{|c|}{$\boldsymbol{O(mn)}$}&
\multicolumn{2}{c|}{\textbf{Chiral}}&\multicolumn{2}{c|}{\textbf{Antichiral}}
\\
\hline
$S=\sigma$ & $2+\text{O}(\tfrac1{mn})$
&
$S=\mathcal S $ & $2+\text{O}(\tfrac1n)$
&
$S=\phi^2_S$ & $2\Delta_\phi+\text{O}(\tfrac1n)$
\\
$S'=\sigma^2$ & $4+\text{O}(\tfrac{1}{mn})$
&
$S'=\langle[\O,\O]_{S,0,0}\rangle_1$ & $4+\text{O}(\tfrac1n)$
&
$S'=\phi^4_S$ & $4\Delta_\phi+\text{O}(\tfrac1n)$
\\
$S''=[\sigma,\sigma]_{1,0}$ & $6+\text{O}(\tfrac{1}{mn})$
&
$S''=\langle[\O,\O]_{S,0,0}\rangle_2$ & $4+\text{O}(\tfrac1n)$
&
$S''=\phi^6_S$ & $6\Delta_\phi+\text{O}(\tfrac1n)$
\\
$S'''=\sigma^3$ & $6+\text{O}(\tfrac1{mn})$
&&&
$S'''=[\mathcal W,\mathcal W]_{S,0,0}$ & $4+\text{O}(\tfrac1{n})$\\
\hline
\end{tabular}
\end{table}

\newsec{Analytic bootstrap for any global symmetry}
\label{sec:analyticgeneral}

In this section we outline an implementation of the analytic bootstrap that
can be applied to $\phi^4$ theories with any global symmetry. We begin with a brief review of large spin perturbation theory. This is followed by a summary of relevant results from the literature, in terms of a toolbox containing the explicit solution to the inversion problems we will encounter. We then give a general recipe for applying these tools, first to the $\veps$ expansion and then to the large-$N$ expansion, where the latter is applicable to $\phi^4$ theories which admit a Hubbard--Stratonovich description.

For the analytic bootstrap, we will consider the four-point function of $\phi^i\in V$, written in the form
\begin{equation}
\langle
\phi^i(x_1)\phi^j(x_2)\phi^k(x_3)\phi^l(x_4)
\rangle
=\frac{1}{(x_{12}^{2}x_{34}^{2})^{\Delta_\phi}}\sum_R \mathbf
T_R^{ijkl}\G_R(u,v)\,.
\label{eq:correlatordef}
\end{equation}
In this expression, $\mathbf T_R^{ijkl}$ are tensor structures for the
irreducible representations $R\in V\otimes V$,\foot{More precisely, the
$\mathbf T_R^{ijkl}$ are the projectors that can be used to decompose the
four-point function into invariant subspaces labeled by $R$.} and $u,v$
are the usual cross-ratios defined by
\eqn{u=z \zb=\frac{x_{12}^2x_{34}^2}{x_{13}^2x_{24}^2}\,,\qquad
v=(1-z)(1-\zb)=\frac{x_{14}^2x_{23}^2}{x_{13}^2x_{24}^2}\,.}[crossratios]
The crossing equation follows from exchanging operators at $x_2$ and $x_4$,
and can we written as
\begin{equation}\label{eq:crossingGen}
  \G_R(u,v)=M_{R\rp }\left(\frac uv\right)^{\Delta_\phi}\G_{\rp }(v,u)\,,
\end{equation}
where the explicit form of the matrix $M$ can be worked out from the tensor
structures for the symmetry group under consideration. Here we choose
normalizations in agreement with \cite{Alday:2019clp}, so that the matrix in the $O(N)$ case takes the form
\begin{equation}
M^{O(N)}=
\begin{pmatrix}
1/N\ &\ (N+2)(N-1)/(2N^2) \ &\ (1-N)/(2N)\\
1&(N-2)/(2N)&1/2\\
-1&(N+2)/(2N)&1/2
\end{pmatrix}\,,
\label{eq:crossingOn}
\end{equation}
where the representations are the singlet $S$, rank-two traceless symmetric
$T$ and antisymmetric $A$ representations of $O(N)$. For any symmetry group, we reserve the letter $S$ for the singlet representation.

Each of the functions $\G_R(u,v)$ admits a decomposition in conformal blocks,
\begin{equation}\label{eq:CBecomp}
\G_R(u,v)=\sum_{\O\in R} c_{\phi\phi\O}^2\lsp G_{\Delta_\O,\ell_\O}(u,v)
\,,\end{equation}
where the sum runs over conformal primary operators $\O$ appearing with OPE coefficient $c_{\phi\phi\O}$ in the OPE $\phi^i\times\phi^j|_R$, and the conformal blocks $G_{\Delta,\ell}(u,v)$ are functions which sum up the contributions of descendants to that primary.

The OPE expansion \eqref{eq:CBecomp} is regular in the limit $z,\zb\to0$.
However, we will expand in the \emph{lightcone limit}, defined by
$z\ll1-\zb\ll1$. Taking $z\to0$, the conformal blocks as functions of
$z,\zb$ simplify as
\begin{equation}\label{eq:collinearblocks}
  G_{\Delta,\ell}(z,\zb)=z^{\frac{\Delta-\ell}2}k_{\frac{\Delta-\ell}2}(\zb)+\text{O}\left(z^{\frac{\Delta+\ell}2+1}\right)\,,
\end{equation}
where $k_{\beta}(\zb)=\zb^\beta{_2F_1}(\beta,\beta;2\beta;\zb)$, and
$_2F_1$ is Gauss's hypergeometric function. The lightcone limit therefore emphasizes the contribution from the operators of smallest value of the twist, defined as $\tau=\Delta-\ell$, which shows that it is useful to organize the OPE in terms of \emph{twist families} of operators of approximately equal twist. In addition, the specific form of the hypergeometric function contains a single logarithmic divergence at $\zb\to1$, but no power or $\log^2$ divergence.

A generic CFT contains families of double-twist operators, written as
$[\O_1,\O_2]_{R,n,\ell}$, where $n=0,1,2,\ldots,$ where $\O_1\in R_1$ and
$\O_2\in R_2$ are operators in the theory and $R\in R_1\otimes R_2$. For
Lagrangian theories these operators have the schematic form
$\O_1\lsp\partial^{\mu_1}\cdots \partial^{\mu_\ell}\square^n\O_2$, up to
contributions from descendants. In the theories we consider, $\phi$ is near
the unitarity bound, $\Delta_\phi=\frac{d-2}2+\gamma_\phi$, which means
that the leading double-twist operators are weakly broken currents
$\mathcal J_{R,\ell}=[\phi,\phi]_{R,0,\ell}$. Our main objective is to
determine the CFT-data of these operators, which consist of their scaling
dimensions $\Delta_{R,\ell}$ and their OPE coefficients
$a_{R,\ell}=c_{\phi\phi\mathcal J_{R,\ell}}^2 $. Of particular interest are
the conserved currents: the stress-energy tensor $T^{\mu\nu}=\mathcal
J_{S,2}$, and in the case of continuous global symmetry, Noether currents
$J^{\mu}_R=\mathcal J_{R,1}$ in one or several representations $R$. They
have conserved dimensions, $\Delta_{S,2}=d$, $\Delta_{R,1}=d-1$ and their
OPE coefficients are related to central charges $C_T$ and $C_{J_R}$ by the
relations\footnote{We use the conventions of \cite{Alday:2019clp}, where the normalization of the conformal blocks differ with a factor $(-2)^\ell$ from e.g.\ \cite{Dolan:2000ut}.}
\begin{equation}\label{eq:centralchargesconventions}
  a_{S,2}=\frac{d^2\Delta_{\phi}^2}{4(d-1)^2C_T}\,, \qquad
  a_{R,1}=-\frac1{C_{J_R}}\,,
\end{equation}
following from conformal Ward identities \cite{Petkou:1994ad, Dolan:2000ut}.

Our main tool is the Lorentzian inversion formula, derived in
\cite{Caron-Huot:2017vep}:
\begin{equation}\label{eq:fullinversionformula}
C_R(\ell,\Delta)=\left(
1\pm(-1)^\ell
\right)\frac{\kappa_{\Delta+\ell}}4 \int\limits_{[0,1]^2}\mathrm d z\llsp\mathrm
d\zb\, \mu(z,\zb)G_{d-1+\ell,1-d+\Delta}(z,\zb)\dDisc[\mathcal
G_R(z,\zb)]\,,
\end{equation}
where the double-discontinuity $\dDisc$ is defined as the difference
between the correlator and its analytic continuations around $\zb=1$,
$\dDisc[f(\zb)]=f(\zb) -\frac{1}{2}
f^\circlearrowleft(\zb)-\frac{1}{2}f^\circlearrowright(\zb)$. In particular, a single conformal block has vanishing double-discontinuity. The sign in \eqref{eq:fullinversionformula} depends on the symmetry of $\G_R(z,\zb)$ under exchanging $x_1$ and $x_2$, and the normalization constants are given by $\kappa_\beta=\frac{\Gamma(\beta/2)^4}{2\pi^2\Gamma(\beta)\Gamma(\beta-1)}$ and $\mu(z,\zb)=|z-\zb|^{d-2}(z\zb)^{-d}$. The poles in $\Delta$ of the function $C_R(\ell,\Delta)$ are located at the scaling dimensions of the operators $\O_R$, with OPE coefficients given by the residue; more precisely
\begin{equation}\label{eq:aellextraction}
a_{R,\ell}=-\int \mathrm d \ell\oint \frac{\mathrm d \Delta}{2\pi i}C_R(\ell',\Delta)\delta(\ell-\ell')\,.
\end{equation}
The inversion formula is valid for $\ell>1$.
If a power $z^{\tau/2}$ appears in $\dDisc[\G(z,\zb)]$, it signals the
existence of a family of operators of twist near that value. This follows
from the scaling $\mu(z,\zb)G_{d-1+\ell,1-d+\Delta}(z,\zb)\sim
z^{\frac{\ell-\Delta-2}2}$, which induces poles in $C_R(\ell,\Delta)$ from the $z\to0$ limit of the $z$ integral:
\begin{equation}\label{eq:deltasinglepole}
  C_R(\ell,\Delta)\sim -\frac{a_{R,\ell}}{\Delta-(\tau+\ell)}\,.
\end{equation}

We now focus on the leading twist family in each representation, and assume
that $\tau_{R,\ell}=2\Delta_\phi+\gamma_{R,\ell}$ for small anomalous
dimensions $\gamma_{R,\ell}$. In that case, following the manipulations of
\cite[Sec.~4]{Caron-Huot:2017vep}, the $z$ integral can be computed and \eqref{eq:fullinversionformula} reduces to the one-dimensional inversion problem given in \cite{Alday:2017vkk}: the CFT-data $a_{R,\ell}$ and $\gamma_{R,\ell}$ are given by
\begin{equation}\label{eq:extractCFTdata}
\hat a_{R,\hb}(\gamma_{R,\ell})^p=U_{R,\hb}^{(p)}+\frac12 \partial_\hb
U_{R,\hb}^{(p+1)}+\frac18 \partial^2_\hb U_{R,\hb}^{(p+2)}+\ldots, \quad
a_{R,\ell}=\frac{\Gamma\big(\frac{\Delta_{R,\ell}+\ell}2\big)^2}{\Gamma(\Delta_{R,\ell}+\ell)}\hat a_{R,\hb}
,\quad \hb=\Delta_\phi+\ell
\,,
\end{equation}
where
\begin{equation}\label{eq:onediminversion}
U^{(p)}_{R,\hb}=\frac{2^pp!\Gamma(\hb)}{\pi^2\Gamma(2\hb-1)}\int\limits_0^1\frac{d\zb}{\zb^2}\lsp k_\hb(\zb)\dDisc\left[\G_R(z,\zb)\big|_{z^{\Delta_\phi}\log^pz}\right]\,.
\end{equation}
These expressions were derived in \cite{Alday:2017vkk}, by assuming that
$\gamma_{R,\ell}\sim g\ll1$ and expanding all quantities in $g$. This expansion generates terms proportional to $z^{\Delta_\phi}\log^pz$ in the double-discontinuity, which under the $z$ integral are converted to higher order poles in $\Delta$ in \eqref{eq:deltasinglepole}. These poles are responsible for the derivatives $\partial_\hb$ appearing in \eqref{eq:extractCFTdata}, following from changing variables from $(\Delta,\ell)$ to $(\tau,\hb)$ in \eqref{eq:aellextraction}. An alternative heuristic derivation of \eqref{eq:onediminversion} starting from the collinear conformal blocks is given in \cite{Alday:2017zzv}.

The success of large spin perturbation theory stems from the fact that
$\dDisc[\G_R(z,\zb)]$ can be computed through the crossing equation
\eqref{eq:crossingGen}. At each order in the expansion parameter, the whole
double-discontinuity is computed from the conformal blocks of a small set
of crossed-channel operators. In particular, the double-twist operators
$[\phi,\phi]_{R,n,\ell}$ themselves do not contribute at leading order. This is
because the power $(1-\zb)^{-\Delta_\phi}$ from crossing is cancelled by
$(1-\zb)^{\frac{\tau_{n,\ell}}2}$ from the conformal blocks, as seen from
the expansion \eqref{eq:collinearblocks} evaluated in the crossed channel,
$z\to1-\zb$. Expanding $\tau_{R,n,\ell}=2\Delta_\phi+n+\gamma_{R,n,\ell}$, we note that
the first non-zero double-discontinuity arises from the term
$\frac{\gamma_{R,n,\ell}^2}8\log^2(1-\zb)$. In the two cases relevant in
this paper, the leading double-discontinuities will be generated by scalar
operators, and the weakly broken currents $\mathcal J_{R,\ell}=[\phi,\phi]_{R,0,\ell}$ will only contribute at subleading order.

\subsection{Inversion toolbox}
In this section we collect all inversion formulas from the literature that are required to find the leading order CFT-data in the $\veps$ expansion and in the large-$N$ expansion. The entries take the form
\begin{equation}
\{\O\}\longrightarrow U_{\hb}^{(0)}+\frac12 U^{(1)}_{\hb}\log z+\ldots\,,
\end{equation}
where $\{\O\}$ is a crossed channel operator or a family of such. The results have
been computed from the inversion integral \eqref{eq:onediminversion} where
$\G_R(z,\zb)$ is replaced by the crossing factor $\left(\frac uv\right)^{\Delta_\phi}$ multiplied by a (sum of) conformal blocks in crossed variables.
Up to an overall prefactor, the resulting functions $U^{(p)}_\hb$ expand in inverse integer powers of the conformal spin
\begin{equation}
  J^2=\hb(\hb-1)\,,
\end{equation}
a statement referred to in the literature as reciprocity and proved in the context of CFT in \cite{Alday:2015eya}.

\begin{invtool}\label{inv:identity}
The identity operator $\1$ contributes with
\begin{equation}
\left\{\1\right\}\longrightarrow
\AA[\Delta_\phi](\hb)=\frac{2(2\hb-1)\Gamma(\hb+\Delta_\phi-1)}{\Gamma(\Delta_\phi)^2\Gamma(\hb-\Delta_\phi+1)}\,,
\end{equation}
which contributes to $U^{(0)}_\hb$
only.
This holds for generic $\Delta_\phi$, and applies in both the $\veps$
expansion and in the large-$N$ expansion. This result can be directly computed from the integral
\eqref{eq:onediminversion} using an integral
representation for $k_\hb(\zb)$ (see e.g.\ eq.\ (4.7) of \cite{Caron-Huot:2017vep}).
\end{invtool}
\begin{invtool}\label{inv:scalarEps}
In the $\veps$ expansion, a bilinear scalar
$\Delta_{\phi^2}=2\Delta_\phi+\gamma$ with OPE coefficient
$c^2_{\phi\phi\phi^2}$, assuming $\gamma=g\lsp\veps\lsp(1+g^{(2)}\veps+\ldots)$, expanded to order $\veps^3$, contributes with
\begin{align}\nonumber
\{\phi^2\}\longrightarrow & \, \frac{c^2_{\phi\phi\phi^2}}2 \gamma^2\frac{2\hb-1}{J^2}\left(-1-\gamma+\veps+\gamma S_1(\hb-1)\right)\log z\\&+\frac{c^2_{\phi\phi\phi^2}}2 \gamma^2\frac{2\hb-1}{J^4}\left(
-1+(J^2\zeta_2+1)\veps+(S_1(\hb-1)-J^2\zeta_2-1)\gamma
\right)\,.
\end{align}
This was derived in eq.\ (2.34) of \cite{Alday:2017zzv} using the explicit form of the scalar conformal block as an infinite sum \cite{Dolan:2000ut} and taking the small $z$ limit. Here $S_1(\hb-1)$ denotes the harmonic numbers.
\end{invtool}
\begin{invtool}\label{inv:scalarN}
The leading contribution from a scalar $\O$ with $\Delta_\O=2$, where $\Delta_\phi=\mu-1$ in generic spacetime dimension $d=2\mu$ is
\begin{equation}
\left\{\O\right\}\longrightarrow (\mu-2)^2c^2_{\phi\phi\O}\frac{\AA[\mu-1](\hb)}{J^2}\left(
-\log z+\SS[\mu-1](\hb)-\frac1{J^2}
\right)\,,
\end{equation}
where
\begin{equation}
  \SS[\alpha](\hb)=2S_1(\hb-1)-S_1(\hb-2+\alpha)-S_1(\hb-\alpha)\,.
\end{equation}
This was derived in eqs.\ (2.25) and (2.29) of \cite{Alday:2019clp}.
\end{invtool}
\begin{invtool}\label{inv:fromcurrents}
The leading contribution from an infinite sum over $\ell\in \mathbb I_\pm$ of
broken currents $\mathcal J_\ell$ with anomalous dimensions
$\gamma_\ell=\frac \kappa{J^2}$ and OPE coefficients $\alpha\lsp a^{\mathrm{GFF}}_{0,\ell}|_{\mu-1}$ (where $a^{\mathrm{GFF}}_{n,\ell}|_{\Delta_\phi}$ are the generalized free field OPE coefficients given in \eqref{eq:GFFOPE}) is
\begin{equation}
\{\mathcal J_\ell\}_{\ell\in\mathbb I_\pm,\, \gamma=\frac \kappa{J^2}}\longrightarrow
-\alpha
\kappa^2\frac{2\hb-1}{2(\mu-2)^2J^2}\left[\pm1+(\mu-2)\pi\csc(\pi\mu)\right]\ln
z+E_\pm\,,\quad\mathbb I_\pm=\begin{cases}
\{0,2,4,\ldots\}
\\
\{1,3,5,\ldots\}
\end{cases}\!\!\!\!\!\!.
\end{equation}
Here $E_\pm$ are lengthy expressions given explicitly in \eqref{eq:Eplusminus}. This formula was derived in \cite{Alday:2019clp} by summing over blocks on the unitarity bound and subsequently inverting the sum.
\end{invtool}

\subsection{General solution in the
\texorpdfstring{$\veps=4-d$}{epsilon=4-d} expansion}
\label{sec:generalsoleps}
Consider first the contribution from the identity operator, appearing in
the singlet ($S$) representation. This will give rise to the leading
contribution to $U^{(0)}_{R,\hb}$ in all representations. Since this is the
only operator contributing until order $\veps^2$, we get, using
Inversion~\ref{inv:identity},
\begin{equation}
  U^{(0)}_{R,\hb}=M_{RS}\lsp\AA[\Delta_\phi](\hb)+\text{O}(\veps^2)\,,
\end{equation}
where $\Delta_\phi=1-\veps/2+\gamma_\phi$ with $\gamma_\phi=\text{O}(\veps^2)$.
From this expression, the leading order OPE coefficients can be extracted:
\begin{equation}\label{eq:treeOPE}
c^2_{\phi\phi\mathcal
J_{R,\ell}}=\frac{2\lsp\Gamma(\ell+1)^2}{\Gamma(2\ell+1)}M_{RS}+\text{O}(\veps)\,.
\end{equation}
Here $\ell$ takes even (odd) values for $R$ being an even (odd) representation. The scalar bilinears $\phi^2_R$ in the even representations have OPE coefficients
\begin{equation}
  c^2_{\phi\phi\phi^2_{R}}=2M_{RS}+\text{O}(\veps)\,.
\end{equation}
These scalars are the next operators to contribute to the
double-discontinuity. Assume that they have dimension
$\Delta_{\phi^2_{R}}=2\Delta_\phi+g_{R}\lsp\veps+\text{O}(\veps^2)$. Then,
using Inversion~\ref{inv:scalarEps} we get the order $\veps^2$ corrections
\begin{align}
  U^{(1)}_{R,\hb}&=-M_{RS}\lsp\Gamma^{\{2\}}_R\frac{2(2\hb-1)}{J^2}\veps^2+\text{O}(\veps^3)\,,
\\
U^{(0)}_{R,\hb}&=M_{RS}\lsp\AA[\Delta_\phi](\hb)-M_{RS}\lsp\Gamma^{\{2\}}_R\frac{2\hb-1}{J^4}\veps^2+\text{O}(\veps^3)\,,
\end{align}
where
\begin{equation}
  \Gamma_R^{\{2\}}=\frac{1}{M_{RS}}\sum_{\rp \text{ even}}M_{R\rp }\lsp
  g_{\rp }^2\lsp M_{\rp S}\,.
\end{equation}
Using \eqref{eq:extractCFTdata} we can thus write down the leading correction to the anomalous dimension,
\begin{equation}
\Delta_{R,\ell}=2\Delta_\phi+\ell+\gamma_{R}(\hb), \qquad
\gamma_{R}(\hb)=-\frac{\Gamma^{\{2\}}_R\veps^2}{J^2}\,,
\end{equation}
where $\hb=\Delta_\phi+\ell$ and $J^2=\hb(\hb-1)$.

Next, as observed in \cite{Alday:2017zzv}, we assume that it is possible to
analytically continue the result $\gamma_R(\hb)$ to spin zero, by making
the change of variables $\hb\to \hb_{\mathrm f}=\frac{\Delta+\ell}2$, i.e.\
we replace the \emph{bare} with the \emph{full} conformal spin (eigenvalue
of the quadratic Casimir). For spin zero we should evaluate this at
$\hb_\mathrm{f}=\Delta_{\phi^2_{R}}/2=1-\veps/2+g_{R}\lsp\veps/2+\text{O}(\veps^2)$. This leads to a system of equations
\begin{equation}\label{eq:matchingEps}
g_{R} \overset!=\left.\gamma_{R}(\hb)\right|_{\hb=\Delta_\phi+\frac{g_{R}}2
  }, \qquad \text{$R$ even\,,}
\end{equation}
at order $\veps$, where now one power of $\veps$ in the $\gamma_{R}(\hb)$ cancels against the factor $\hb_{\mathrm f}-1=(g_{R}-1)\veps/2$ in the denominator. This simplifies to
\begin{equation}\label{eq:eqsystScalars}
M_{RS}\, g_{R}(g_{R}-1)+2\sum_{\rp \text{ even}}M_{R\rp }M_{\rp S}\,
g_{\rp }^2=0\,, \qquad \text{$R$ even\,,}
\end{equation}
which is a system of $k$ quadratic equations for the $k$ constants $g_R$, where $k$ is the number of even representations, or equivalently the number of scalar bilinears. Solving \eqref{eq:eqsystScalars} gives all possible fixed points in the $\veps$ expansion with the given symmetry.

As an example, consider the $O(N)$ model with crossing matrix \eqref{eq:crossingOn}. The even representations are $S$ and $T$, and the bilinear scalars are $\phi^2_S=\phi^i\phi^i$ and $\phi^2_T=\phi^{\{i}\phi^{j\}}$. There are two solutions to \eqref{eq:eqsystScalars}, $g_S=g_T=0$, which is the Gaussian theory, and
\begin{equation}\label{eq:ONscalardims}
  g_S=\frac{N+2}{N+8}\,, \qquad g_T=\frac{2}{N+8}\,,
\end{equation}
which is exactly the critical $O(N)$ model \cite{Henriksson:2018myn}. With these values we have $\Gamma^{\{2\}}_S=\frac{3(N+2)}{(N+8)^2}$, $\Gamma^{\{2\}}_T=\frac{N+6}{(N+8)^2}$ and $\Gamma^{\{2\}}_A=\frac{N+2}{(N+8)^2}$.

The singlet spin-two current in any global symmetry group is the
stress-energy tensor with dimension $\Delta_{S,2}=d=4-\veps$. This gives the constraint
\begin{equation}
  \gamma^{(2)}_\phi=\tfrac{1}{12}\Gamma^{\{2\}}_S\,,
\end{equation}
where $\Delta_\phi=1-\veps/2+\gamma^{(2)}_\phi\veps^2+\text{O}(\veps^3)$. Using this we write down the full dimension of the broken currents to order $\veps^2$:
\begin{equation}
  \Delta_{R,\ell}=2-\veps+\ell+2\gamma_\phi^{(2)}\veps^2-\frac{\Gamma^{\{2\}}_R\veps^2}{\ell(\ell+1)}+\text{O}(\veps^3)\,.
\end{equation}
The OPE coefficients are extracted using \eqref{eq:extractCFTdata},
\begin{equation}\label{eq:OPEcoefseps2}
a_{R,\ell}=M_{RS}a^{\mathrm{GFF}}_{0,\ell}+M_{RS}\frac{\Gamma_R^{\{2\}}\veps^2}{\ell(\ell+1)}\left(
S_1(2\ell)-S_1(\ell)+\frac{1}{\ell+1}
\right)\frac{2\Gamma(\ell+1)^2}{\Gamma(2\ell+1)}+\text{O}(\veps^3)\,,
\end{equation}
where $a^{\mathrm{GFF}}_{0,\ell}$ is the generalized free field OPE
coefficients for $\Delta_\phi=1-\veps/2+\gamma^{(2)}_\phi\veps^2$ expanded
to order $\veps^2$, which we give in \eqref{eq:aGFF2} in
Appendix~\ref{app:explicitresults}.

From the spin two singlet OPE coefficient we can extract the central charge correction given by \eqref{eq:centralchargesconventions}
\begin{equation}
\frac{C_T}{C_{T,\mathrm{free}}}=1-\frac{5\gamma_\phi^{(2)}}{3}\veps^2+\text{O}(\veps^3)=1-\frac{5\lsp\Gamma_S^{\{2\}}}{36}\veps^2+\text{O}(\veps^3)\,,
\end{equation}
which is consistent with~\cite[Eq.~(E.1)]{Osborn:2017ucf}.  We emphasize
that the considerations here are valid with any global symmetry group. The
input needed to specialize to a given symmetry group is the crossing matrix
$M_{R\rp }$ and the division of the representations into even and odd spin.
By solving the system of equations \eqref{eq:eqsystScalars} one finds all
fixed points in the $\veps$ expansion compatible with that symmetry group
and derives the leading (order $\veps$) anomalous dimensions of the
bilinear scalars.  Conservation of the stress-energy tensor allows one to
compute the leading (order $\veps^2$) anomalous dimension of $\phi$.

In one or several of the odd representations $R$, the current at spin $\ell=1$ may be conserved, being a generator of global symmetry. This therefore gives further constraints $\Delta_{R,\ell}=d-1$, which must be explicitly checked. By \eqref{eq:centralchargesconventions} the corresponding OPE coefficient is related to the $C_J$ of that symmetry current:
\begin{equation}
  \frac{C_{J_R}}{C_{J_R,\mathrm{free}}}=1-3\gamma_\phi^{(2)}\veps^2+\text{O}(\veps^3)=1-\frac{3\lsp\Gamma_R^{\{2\}}}4\veps^2+\text{O}(\veps^3)\,.
\end{equation}

Let us discuss the extension to higher orders in the $\veps$ expansion. To
order $\veps^3$, the operators contributing with a nonzero
double-discontinuity are the same as at the previous order, namely the
bilinear scalars $\phi^2_R$. At higher orders, infinite families of
operators contribute. In the $O(N)$ model, the only such families at order $\veps^4$ are operators of approximate twist $2$ and $4$, and subsequently the problem was solved there.
We expect that this generalises to any global symmetry. However, to compute the contribution from approximate twist $4$ requires detailed knowledge of the operator content of the theory in question. This was done in the case of the $O(N)$ model in \cite{Henriksson:2018myn}.
To order $\veps^5$ the same operators will contribute but now with subleading corrections. To work this out, even in the $N=1$ (Ising) case, is still an open problem.

We have seen that at order $\veps^2$ all constants that enter the problem
can be fixed using continuation to spin zero and conservation of the
stress-energy tensor. This is no longer true at higher orders. At order
$\veps^3$ a total of $2k+1$ new constants appear: $\gamma_\phi^{(3)}$, the
second order correction to $\gamma_{\phi^2_R}=g_R\lsp\veps(1+g^{(2)}_R\veps)+\ldots,$ and the corrections $\alpha_R$ to the OPE coefficients defined by
\begin{equation}\label{eq:c2phi2def}
  c^2_{\phi\phi\phi^2_R}=2M_{RS}(1+\alpha_R\lsp\veps)+\text{O}(\veps^2)\,.
\end{equation}
Based on experience from the $O(N)$ model \cite{Henriksson:2018myn}, the
order $\veps^2$ continuation to spin zero requires order $\veps^4$ results
for the currents, so the only new equations at order $\veps^3$ are the
conservation of the symmetry currents (including the stress-energy tensor).
In general this will not provide enough equations to fix all constants, but
in many cases we can still make progress.  Firstly, we will use that the
correction to the OPE coefficient takes the form $\alpha_R=-g_R$. This
holds for any global symmetry, and follows from analytic bootstrap in
Mellin space, as we show in Appendix~\ref{sec:Mellinbootstrap}.  Secondly,
the second order corrections $g_R^{(2)}$ to the bilinear scalar dimensions
are in many cases known from the literature, and one can proceed using
these as input.

Using Inversion~\ref{inv:scalarEps}, it is straightforward to derive expressions for $U^{(p)}_{R,\hb}$ at order $\veps^3$. The anomalous dimensions extracted from these expressions take the form
\begin{equation}
  \gamma_R(\hb)=-\frac{\Gamma_R^{\{2\}}}{J^2}\veps^2+\frac{\Gamma_R^{\{2\}}-2\lsp\Gamma_R^{\{2,1\}}+\big(\Gamma_R^{\{3\}}-\Gamma_R^{\{2\}}\big)S_1(\hb-1)}{J^2}\veps^3+\text{O}(\veps^4)\,,
\end{equation}
where $\hb=1-\frac\veps2+\ell+O(\veps^2)$ and
\begin{equation}\label{eq:defG3G21}
\Gamma_R^{\{3\}}=\frac{1}{M_{RS}}\sum_{\rp \text{ even}}M_{R\rp }\lsp
g_{\rp }^3M_{\rp S}\,, \qquad
\Gamma_R^{\{2,1\}}=\frac{1}{M_{RS}}\sum_{\rp \text{ even}}M_{R\rp }\lsp g_{\rp }^2g^{(2)}_{\rp }M_{\rp S}\,.
\end{equation}

From the corresponding expression for the OPE coefficients using Inversion~\ref{inv:scalarEps}, we can extract the central charge correction:
\begin{equation}
  \frac{C_T}{C_{T,\mathrm{free}}}=1-\frac53\left(\gamma_\phi^{(2)}\veps^2+\gamma_\phi^{(3)}\veps^3\right)-\frac{29}{18}\gamma_\phi^{(2)}\veps^3+\frac5{48}\Gamma^{\{3\}}_S\veps^3+\text{O}(\veps^4)\,.
\end{equation}
Here we used that the stress-energy tensor conservation eliminates the
dependence on $g_R^{(2)}$ in favour of $\gamma_\phi^{(3)}$. For
two-coupling theories as considered in \cite{Osborn:2017ucf} we may find
\eqn{\Gamma_S^{\{3\}}=\tfrac{1}{90}(N+2)\big(58(a\lsp g_*^2+6\lsp\lambda_*^2)
-258(N+8)\lambda_*^3-129\lsp a\lsp(b\lsp g_*+6\lsp\lambda_*)g_*^2\big)\,,}[]
where $\lambda_*,g_*$ are the coefficients of the order-$\veps$ values of
the two couplings at the fixed point, i.e.\
$\lambda=\lambda_*\lsp\veps+\text{O}(\veps^2)$ and
$g=g_*\lsp\veps+\text{O}(\veps^2)$, and $a,b$ are defined in
\cite[Eq.~(5.2)]{Osborn:2017ucf}.  For the $O(N)$ model, where
$\lambda_*=1/(N+8)$ and $g_*=0$, this gives
$\Gamma_S^{\{3\}}=(N+2)/(N+8)^2$, in complete agreement with \eqref{eq:defG3G21}.

Similarly, for the current central charges we derive the expression
\begin{equation}
  \frac{C_{J_R}}{C_{J_R,\mathrm{free}}}=1-3\left(\gamma_\phi^{(2)}\veps^2+\gamma_\phi^{(3)}\veps^3\right)-\frac94\gamma_\phi^{(2)}\veps^3+\frac{1}{4}\Gamma^{\{3\}}_R\veps^3+\text{O}(\veps^4)\,.
\end{equation}

\subsection{General solution in the large \texorpdfstring{$N$}{N} expansion}\label{sec:GeneralN}
Let us now describe the computation of CFT-data in the large-$N$ expansion for a
generic symmetry group, parametrised by some number $N$. Compared to the
$\veps$ expansion the situation is a bit more complicated, since the
parameter $N$ enters in the crossing matrix $M_{R\rp }$ itself. In a given
even representation $R$, we assume that there are two possibilities for the smallest dimension
scalar. It is either a scalar bilinear $\phi^2_R$ with dimension
$2\Delta_\phi+\text{O}(N^{-1})$, or a Hubbard--Stratonovich field $\mathcal
R$ with dimension $2+\text{O}(N^{-1})$. If one has access to results in the $\veps$ expansion, one
can assess the situation by taking the large $N$ limit of the order $\veps$
scalar dimensions. For instance, in the $O(N)$ model we get, using
\eqref{eq:ONscalardims},
\begin{equation}
  \Delta_{\phi^2_S}=2-\veps+g_S\lsp\veps\to2+\text{O}(N^{-1})\,, \qquad
\Delta_{\phi^2_T}=2-\veps+g_T\lsp\veps\to d-2+\text{O}(N^{-1})\,,
\end{equation}
so we see that the singlet representation admits a Hubbard--Stratonovich
field $\mathcal S$ (in the literature denoted by $\sigma$), but not the traceless symmetric representation (whence we keep the notation $\phi^2_T$).
We assume that the Hubbard--Stratonovich fields $\mathcal R$ have dimension $\Delta_{\mathcal R}=2+\text{O}(N^{-1})$ and OPE coefficient $c^2_{\phi\phi\mathcal R}={a_{\mathcal R}}/N+\text{O}(N^{-2})$.

In order to provide some structure of the subsequent computations we define the following subsets of the representations in $V\otimes V=\mathrm I\cup\mathrm{II}$:

\begin{itemize}
\item Group $\mathrm I$: Representations whose only corrections at order $1/N$ come from crossed channel Hubbard--Stratonovich fields.
\item Group $\mathrm{II}$: Representations where the corrections at order $1/N$  come from Hubbard--Stratonovich fields as well as from broken currents in Group $\mathrm I$ representations in the crossed channel.
\item Group $\mathrm{III}$: Representations that admit a Hubbard--Stratonovich field. Typically $\mathrm{III}\subset\mathrm{II}$.
\end{itemize}
As an example, in the $O(N)$ model we have $S\in\mathrm{II}\cap\mathrm{III}$ and $T,A\in\mathrm I$. Our strategy will then be the following. First, as in the $\veps$ expansion, the identity operator creates the leading contribution to $U^{(0)}_{R,\hb}$ for all representations. Next we turn to the representations in Group $\mathrm I$. The contributions from Hubbard--Stratonovich fields will give the order $1/N$ anomalous dimensions in these representations. Using Inversion~\ref{inv:scalarN} we see that these corrections will be proportional to $1/J^2$. Finally we turn to representations in the Group $\mathrm{II}$. Here we get contributions from both the Hubbard--Stratonovich fields, using Inversion~\ref{inv:scalarN}, and from the currents in Group $\mathrm{I}$. Due to the particular form of the anomalous dimensions of these currents, we can use Inversion~\ref{inv:fromcurrents} to find the complete order $1/N$ CFT-data.

The expressions will depend on $|\mathrm{III}|+1$ free parameters: the OPE
coefficients $a_{\mathcal R}=c^2_{\phi\phi\mathcal R}$ for $R\in
\mathrm{III}$, and the leading order anomalous dimension of $\phi$.
The only consistency conditions available to fix these constants are the
conservation of the symmetry currents (including the stress-energy tensor), and
depending on the number of conserved currents this may or may not be
enough. As in the order $\veps^3$ results above, literature values can be
used to fix the remaining constants if the conservation equations are not
sufficient. Finally, the leading anomalous dimensions of the
Hubbard--Stratonovich fields may be extracted by imposing a shadow relation
similar to the one observed in the $O(N)$ model \cite{Alday:2019clp}:
\begin{equation}
  \Delta_{\mathcal R}+\Delta_{R,0}\overset !=d\,.
\end{equation}

Let us now execute the strategy in full generality.
The contribution from the identity operator gives
\begin{equation}
  U^{(0)}_{R,\hb}=M_{RS}\lsp\AA[\Delta_\phi](\hb)\,,
\end{equation}
where now $\Delta_\phi=\mu-1+\gamma^{(1)}_\phi/N+\text{O}(N^{-2})$ with
$d=2\mu$. For the representations in Group $\mathrm I$ we get the contributions from Hubbard--Stratonovich fields in Group $\mathrm{III}$. Using Inversion~\ref{inv:scalarN} we get
\begin{equation}
U^{(1)}_{R,\hb}=-\sum_{ \rp \in\mathrm{III}}M_{R \rp
}\lsp 2(\mu-2)^2\frac{a_{\oprp}}N\frac{\AA[\mu-1](\hb)}{J^2}\,, \quad R\in \mathrm
I\,,
\end{equation}
and a corresponding expression for $U^{(0)}_{R,\hb}$. From this we extract the order $N^{-1}$ anomalous dimensions of currents in Group $\mathrm I$ representations:
\begin{equation}
  \gamma_{R,\hb}=-\frac{2(\mu-2)^2K_R}{J^2N}+\text{O}(N^{-2})\,, \quad
  K_R=\frac{1}{M_{RS}}\sum_{\rp \in\mathrm{III}}M_{R\rp }\, a_{\oprp}\,,
  \quad R\in \mathrm I\,,
\label{eq:gammaGpI}
\end{equation}
where the scaling dimensions are given by $\Delta_{R,\ell}=2\Delta_\phi+\ell+\gamma_{R,\ell}$
In step 3 we consider the second group of operators, $\mathrm{II}$. They get contributions both from $\mathcal R$ for $R\in \mathrm{III}$ and from $\mathcal J_{R,\ell}$ for $R\in\mathrm I$. We get
\begin{align}\nonumber
U^{(1)}_{R,\hb}&=-\sum_{ \rp \in\mathrm{III}}2M_{R \rp }(\mu-2)^2\frac{a_{\oprp}}N\frac{\AA[\mu-1](\hb)}{J^2}
\\
&\quad-\sum_{\rp _\pm\in\mathrm{I}}4M_{R\rp }K_{\rp }^2M_{\rp
S}\frac{(\mu-2)^2(2\hb-1)}{J^2N^2}\left(\pm1+(\mu-2)\pi\csc(\pi\mu)\right),\quad
R\in \mathrm{II}\,,
\end{align}
where the $+$ ($-$) sign is used if the operators in the $\rp $ representations have even (odd) spin. This means that the anomalous dimensions of the group $\mathrm{II}$ double-twist operators are
\begin{equation}
\gamma_{R,\ell}=-\frac{2(\mu-2)^2K_R}{J^2N}-\frac{\widehat
K_{R}}{J^2N^2}\frac{(\mu-2)^2\Gamma(\mu-1)^2\Gamma(\ell+1)}{\Gamma(2\mu+\ell-3)}\,,\quad
R\in\mathrm{II}\,.
\label{eq:gammaGpII}
\end{equation}
In the above expressions $J^2=(\mu-1+\ell)(\mu-2+\ell)$ and
\begin{equation}
\widehat K_R=\frac{1}{M_{RS}}\sum_{\rp _\pm\in\mathrm I}2M_{R\rp }K_{\rp
}^2M_{\rp S}\left(\pm1+(\mu-2)\pi\csc(\pi\mu)\right), \quad R\in
\mathrm{II}\,.
\end{equation}

As an example, let us explicitly evaluate $K_R$ and $\widehat K_R$ in the $O(N)$ model. We get
\begin{equation}
  K_{R}=a_{\mathcal S}\,, \ R=S,T,A\,, \qquad \widehat K_{S}=2N
(\mu-2)\pi\csc(\pi\mu) a_{\mathcal S}^2+\text O(N^0)\,.
\end{equation}
We have two conservation equations, a linear relation due to the global symmetry current $\Delta_{A,1}=d-1$, and a quadratic relation due to the stress tensor $\Delta_{S,2}=d$. There are two solutions, free theory $a_{\mathcal S}=\gamma_\phi^{(1)}=0$ and the known $O(N)$ model result \cite{Alday:2016jfr,Alday:2019clp}
\begin{equation}
\gamma_\phi^{(1)}=\frac{(\mu-2)^2}{\mu(\mu-1)}a_{\mathcal
S}=\eta_1^{O(N)}:=\frac{(\mu-2)\Gamma(2\mu-1)}{\Gamma(\mu)\Gamma(\mu+1)\pi
\csc(\pi\mu)}\,.
\label{eq:eta1def}
\end{equation}
The extension to subleading orders in $1/N$ is a complicated task, which was achieved in \cite{Alday:2019clp} for the $T$ and $A$ representations. This involved computing the contributions from operators $[\sigma,\sigma]_{n,\ell}$,which was found in the form of a Mellin space amplitude, using OPE coefficients derived from the mixed correlator $\langle\phi\phi\sigma\sigma\rangle$. We do not attempt to generalize it for generic global symmetry group.

\newsec{Analytic bootstrap of \texorpdfstring{$\boldsymbol{O(m)\times
O(n)}$}{O(m)xO(n)} CFTs}\label{sec:analyticOmOn}
In this section we will apply the methods of section~\ref{sec:analyticgeneral} to the global symmetry group $O(m)\times O(n)$.

The crossing matrix $M_{R\rp }$ in the basis $\{S,W,X,Y,Z,A,B,C,D\}$, where
in terms of products of representations of each orthogonal group we have
\eqn{\begin{gathered}
S=(S,S)\,,\quad W=(T,S)\,,\quad X=(S,T)\,,\quad Y=(T,T)\,,\quad Z=(A,A)\,,\\
A=(A,S)\,,\quad B=(S,A)\,,\quad C=(A,T)\,,\quad D=(T,A)\,,
\end{gathered}}[Representations]
where by $S=(S,S)$ we mean that we take the singlet of each of $O(m)$ and
$O(n)$ to form the representation $S$ of $O_{m,n}=O(m)\times O(n)$ and
similarly for the rest, takes the form
\begin{equation}
\begin{pmatrix}
 \frac{1}{m n} & \frac{m^2+m-2}{m^2 n} & \frac{n^2+n-2}{m n^2} &\hspace{-2mm} \frac{(m^2+m-2) (n^2+n-2)}{m^2 n^2} & \hspace{-2mm}\frac{(m-1) (n-1)}{m n} \hspace{-2mm}&  \frac{m-1}{m n} & \frac{n-1}{m n} &\hspace{-3mm} \frac{(m-1) (n^2+n-2)}{m n^2} \hspace{-3mm}& \frac{(m^2+m-2) (n-1)}{m^2 n} \\
 \frac{1}{2 n} & \frac{m-2}{2 m n} & \frac{n^2+n-2}{2 n^2} & \frac{(m-2) (n^2+n-2)}{2 m n^2} & -\frac{n-1}{2 n} & -\frac{1}{2 n} & \frac{n-1}{2 n} & -\frac{n^2+n-2}{2 n^2} & \frac{(m-2)
   (n-1)}{2 m n} \\
 \frac{1}{2 m} & \frac{m^2+m-2}{2 m^2} & \frac{n-2}{2 m n} &
 \frac{(m^2+m-2) (n-2)}{2 m^2 n} & -\frac{m-1}{2 m} & \frac{m-1}{2 m} &
 -\frac{1}{2 m} & \frac{(m-1) (n-2)}{2 m n} & -\frac{m^2+m-2}{2 m^2} \\
 \frac{1}{4} & \frac{m-2}{4 m} & \frac{n-2}{4 n} & \frac{(m-2) (n-2)}{4 m n} & \frac{1}{4} & -\frac{1}{4} & -\frac{1}{4} & -\frac{n-2}{4 n} & -\frac{m-2}{4 m} \\
 \frac{1}{4} & -\frac{m+2}{4 m} & -\frac{n+2}{4 n} & \frac{(m+2) (n+2)}{4 m n} & \frac{1}{4} & \frac{1}{4} & \frac{1}{4} & -\frac{n+2}{4 n} & -\frac{m+2}{4 m} \\
 \frac{1}{2 n} & -\frac{m+2}{2 m n} & \frac{n^2+n-2}{2 n^2} & -\frac{(m+2) (n^2+n-2)}{2 m n^2} & \frac{n-1}{2 n} & \frac{1}{2 n} & \frac{n-1}{2 n} & \frac{n^2+n-2}{2 n^2} & -\frac{(m+2)
   (n-1)}{2 m n} \\
 \frac{1}{2 m} & \frac{m^2+m-2}{2 m^2} & -\frac{n+2}{2 m n} &
 -\frac{(m^2+m-2) (n+2)}{2 m^2 n} & \frac{m-1}{2 m} & \frac{m-1}{2 m} &
 \frac{1}{2 m} &\hspace{-2mm} -\frac{(m-1) (n+2)}{2 m n} \hspace{-2mm}&
 \frac{m^2+m-2}{2 m^2} \\
 \frac{1}{4} & -\frac{m+2}{4 m} & \frac{n-2}{4 n} & -\frac{(m+2) (n-2)}{4 m n} & -\frac{1}{4} & \frac{1}{4} & -\frac{1}{4} & \frac{n-2}{4 n} & \frac{m+2}{4 m} \\
 \frac{1}{4} & \frac{m-2}{4 m} & -\frac{n+2}{4 n} & -\frac{(m-2) (n+2)}{4 m n} & -\frac{1}{4} & -\frac{1}{4} & \frac{1}{4} & \frac{n+2}{4 n} & \frac{m-2}{4 m}
\end{pmatrix}.
\label{eq:crossingOmOn}
\end{equation}
This matrix is simply the tensor product $M^{O(m)}\otimes M^{O(n)}$ for $M^{O(N)}$ given in \eqref{eq:crossingOn}.
The representations are either even ($S$, $W$, $X$, $Y$ and $Z$) or odd
($A$, $B$, $C$ and $D$) under $x_1\leftrightarrow x_2$. The
even (odd) representations have intermediate operators of even (odd)
spins.

\subsec{Results in the \texorpdfstring{$\veps$}{epsilon} expansion} In the
$\veps$ expansion, the operators $\phi^2_R$ in the five even
representations $R$ introduce corrections to weakly broken currents in all
nine representations.  Solving equations~\eqref{eq:eqsystScalars} for the
constants $g_R$ we find four sets of solutions, corresponding to the free
theory and to the $O(mn)$ (Heisenberg), chiral and antichiral fixed points.
For the latter two fixed points, of interest to this paper, the explicit
expressions for the $g_R$ are rather complicated, containing square roots.
We give complete results in an ancillary data file, which we describe in
Appendix~\ref{sec:ancillarydata}. For presentation purposes the expressions
in the $\veps$ expansion in this section are expanded for at large $n$, but
at each order in $\veps$ presented here the complete function of $m$ and
$n$ has been determined.

The constants $g_R$ correspond to the scaling dimensions of the scalar operators, which take the form
\begin{align}
\Delta_{\phi^2_S+}&=2-3 (m+1)
\frac{\veps}{n}
-3 \left(m^2-3 m-14\right)
\frac{\veps}{n^2}
+\text{O}\left(\veps^2,n^{-3}\right),
\nonumber\\
\Delta_{\phi^2_W+}&=2-(m+3)
\frac{\veps}{n}
+\left(m^2+7 m+42\right)
\frac{\veps}{n^2}
+\text{O}\left(\veps^2,n^{-3}\right),
\nonumber\\
\Delta_{\phi^2_X+}&=2-\veps+(m+1)
\frac{\veps}{n}
-\left(m^2+5 m+10\right)
\frac{\veps}{n^2}
+\text{O}\left(\veps^2,n^{-3}\right),
\nonumber\\
\Delta_{\phi^2_Y+}&=2-\veps+
\frac{\veps}{n}
-(m+10)
\frac{\veps}{n^2}
+\text{O}\left(\veps^2,n^{-3}\right),
\nonumber\\
\Delta_{\phi^2_Z+}&=2-\veps-
\frac{\veps}{n}
+(m-2)
\frac{\veps}{n^2}
+\text{O}\left(\veps^2,n^{-3}\right),
\label{eq:ansChiralEps}
\end{align}
for the chiral fixed point, and
\begin{align}
\Delta_{\phi^2_S-}&=2-\veps+\frac{3 (m-1) (m+2)}{m}
\frac{\veps}{n}
+\frac{3 (m-1) (m+2) \left(m^2-4 m+16\right)}{m^2}
\frac{\veps}{n^2}
+\text{O}\left(\veps^2,n^{-3}\right),
\nonumber\\
\Delta_{\phi^2_W-}&=2-\frac{(m-2) (m+5)}{m}
\frac{\veps}{n}
+\frac{(m-4) \left(m^3+11 m^2+14 m-40\right)}{m^2}
\frac{\veps}{n^2}
+\text{O}\left(\veps^2,n^{-3}\right),
\nonumber\\
\Delta_{\phi^2_X-}&=2-\veps+\frac{(m-1) (m+2)}{m}
\frac{\veps}{n}
-\frac{(m-1) (m+2) \left(m^2+4 m-16\right)}{m^2}
\frac{\veps}{n^2}
+\text{O}\left(\veps^2,n^{-3}\right),
\nonumber\\
\Delta_{\phi^2_Y-}&=2-\veps+\frac{m-2}{m}
\frac{\veps}{n}
-\frac{m^3-2 m^2-24 m+32}{m^2}
\frac{\veps}{n^2}
+\text{O}\left(\veps^2,n^{-3}\right),
\nonumber\\
\Delta_{\phi^2_Z-}&=2-\veps-\frac{m+2}{m}
\frac{\veps}{n}
+\frac{(m+2) \left(m^2+8 m-16\right)}{m^2}
\frac{\veps}{n^2}
+\text{O}\left(\veps^2,n^{-3}\right),
\label{eq:ansAntiEps}
\end{align}
 for the antichiral fixed point.
These results agree with \cite[Eqs.\ (5.92) and
(5.93)]{Osborn:2017ucf}.\foot{In the notation of \cite{Osborn:2017ucf},
$\rho_1, \rho_2, \rho_3, \rho_4$ correspond to $\phi_X^2, \phi_W^2,
\phi_Y^2, \phi_Z^2$, respectively. There is a typo in
$\gamma_{\rho_1\llsp\pm}$ in \cite[Eq. (5.93)]{Osborn:2017ucf}: the sign
before the $1/n$ term there should be positive.} The operators $\phi_S^2$,
$\phi_W^2$ and $\phi_Z^2$ correspond to $S$, $W$ and $Z$, respectively, in
\LargenChiral and \LargenAntichiral.

Having identified the fixed points we move on to a determination of the CFT-data to order $\veps^3$. As described in the previous section, we need to take as input the second order corrections $g^{(2)}_R$ to the anomalous dimensions $\gamma_{\phi^2_R}$ of bilinear scalars, given in \cite{Osborn:2017ucf}.

We present only a subset of the data computed at order $\veps^3$. The
dimension of $\phi$,
\begin{align}
\Delta_{\phi+}&=1-\frac\veps2+\frac{m+1}{8 n}\veps^2-\frac{2 m^2+9 m+17}{8 n^2}\veps^2-\frac{m+1}{32 n}\veps^3+\frac{14 m^2+57 m+101}{32
   n^2}\veps^3+\text{O}\left(\veps^4,n^{-3}\right),
\nonumber\\
\Delta_{\phi-}&=1-\frac\veps2+\frac{(m-1) (m+2)}{8 m n}\left(1-\frac{2 m^2+7 m-22}{m n}
\right)\veps^2\nonumber\\&\qquad\qquad\qquad\qquad-\frac{(m-1) (m+2)}{32 m n}\left(1-\frac{14 m^2+43 m-158}{m n}\right)\veps^3
+\text{O}\left(\veps^4,n^{-3}\right),
\end{align}
agrees with the literature values \cite{Pelissetto:2001fi}, whereas the results of the spinning operators are new, as far as we are aware. These include
the dimensions of the non-conserved spin-one operators
\begin{align}
\Delta_{C,1+}&=3-\veps
+\frac{m+2 }{4 n}\veps ^2-\frac{m^2+6 m+8}{2 n^2} \veps ^2-\frac{m+2 }{16 n}\veps ^3+\frac{7 m^2+36 m+44}{8 n^2}\veps ^3
+\text{O}\left(\veps^4,n^{-3}\right),
\nonumber\\
\Delta_{D,1+}&=3-\veps+\frac{m}{4 n}\veps ^2-\frac{m (m+3)}{2 n^2}\veps ^2-\frac{m}{16 n}\veps ^3+\frac{7 m (m+3)}{8
   n^2}\veps ^3
+\text{O}\left(\veps^4,n^{-3}\right),
\nonumber\\
\Delta_{C,1-}&=3-\veps+\frac{m+2}{4 n}\left(1-\tfrac{2 (m^2+4 m-12)}{m n}\right)\veps^2-\frac{m+2}{16 n}\left(1-\tfrac{2 (7 m^2+22 m-80)}{m n}\right)\veps^3
+\text{O}\left(\veps^4,n^{-3}\right),
\nonumber\\
\Delta_{D,1-}&=3-\veps+\frac{m}{4 n}\veps^2-\frac{m^2+3
   m-12}{2 n^2}\veps^2-\frac{m}{16 n}\veps^3+\frac{7 m^2+21
   m-80}{8 n^2}\veps^3
+\text{O}\left(\veps^4,n^{-3}\right),
\end{align}
and the central charges
\begin{align}
\frac{C_{T+}}{C_{T,\mathrm{free}}}&=1-\frac{5 (m+1)}{24 n}\veps^2+\frac{5 \left(2 m^2+9 m+17\right)}{24 n^2}\veps^2\nonumber\\&\qquad\qquad\qquad\qquad-\frac{7 (m+1)}{72 n}\veps^3-\frac{31
   m^2+117 m+196}{72 n^2}\veps^3
+\text{O}\left(\veps^4,n^{-3}\right),
\nonumber\\
\frac{C_{T-}}{C_{T,\mathrm{free}}}&=1-
\frac{5 (m+2) (m-1)}{24 m n}\left(1-\frac{2 m^2+7 m-22}{m n}\right)\veps^2\nonumber\\&\qquad\qquad\qquad\qquad-\frac{7  (m+2)(m-1)}{72 m n}\left(1+\frac{31 m^2+86 m-356}{7 m n}\right)\veps^3
+\text{O}\left(\veps^4,n^{-3}\right).
\end{align}
More results can be found in the ancillary data file, as described in
Appendix~\ref{sec:ancillarydata}.

\subsec{Results at large \texorpdfstring{$n$}{n}}

As mentioned in the introduction, we can give a description at large $n$ by introducing Hubbard--Stratonovich operators in the $S$ and $W$ representations. This is in agreement with the results \eqref{eq:ansChiralEps} and \eqref{eq:ansAntiEps} in the $\veps$ expansion above. As a starting point for the analytic bootstrap at large $n$ we will therefore assume that these two representations contain scalar operators $\mathcal S$ and $\mathcal W$, with dimensions and OPE coefficients given by
\begin{equation}
\Delta_{\mathcal R}=2+\text O\left(\frac1n\right), \qquad
c_{\phi\phi\mathcal R}^2=\frac{a_{\mathcal R}}n+\text
O\left(\frac1{n^2}\right),\qquad \mathcal R=\mathcal S,\mathcal W\,.
\end{equation}
We will take these representations to consitute group $\mathrm{III}$ in our implementation of the recipe of section~\ref{sec:GeneralN}. The next task is to determine what operators contribute at order $1/n$ to the broken currents in all of the nine reprensentations in \Representations. This is done by expanding the crossing matrix \eqref{eq:crossingOmOn} at large $n$ and studying the relative scaling of elements in the first column, where $\1$ contributes, and the other columns. We identify that for representations in group $\mathrm I=\{X,Y,Z,B,C,D\}$, only the auxiliary fields generate order $1/n$ corrections, whereas for group $\mathrm{II}=\{S,W,A\}$ also the currents in group $\mathrm I$ need to be taken into account.

Having identified the groups $\mathrm I$, $\mathrm{II}$ and $\mathrm{III}$, we follow the implementation of section~\ref{sec:GeneralN} and generate CFT-data at order $1/n$ in all representations. In particular, the scaling dimensions are given by
\begin{equation}\label{eq:scalingdimensionslargeN}
\Delta_{R,\ell}=2(\mu-1)+\ell+\frac{2\gamma_\phi^{(1)}}n+\gamma_{R,\ell}+\text O\left(\frac1{n^2}\right),
\end{equation}
for $\gamma_{R,\ell}$ given by \eqref{eq:gammaGpI} and \eqref{eq:gammaGpII}
and $\mu=d/2$. These expressions depend on three undetermined constants,
$a_{\mathrm S}$, $a_{\mathrm W}$ and $\gamma_\phi^{(1)}$. Fortunately, the
stress-energy tensor and the two global symmetry currents $J_A^\mu$ and
$J_B^\mu$ provide three consistency equations for these unknowns, namely
\begin{equation}
  \Delta_{S,2}=d\,,\qquad \Delta_{A,1}=d-1\,, \qquad \Delta_{B,1}=d-1\,.
\end{equation}
There are four solutions to these equations, which we can identify with the
four fixed points of the $\veps$ expansion,
\begin{align}
&\text{Free:} & a_{\mathcal S}&=0\,, & a_{\mathcal W}&=0\,,  & \gamma^{(1)}_\phi&=0\,,
\nonumber\\
&\text{$O(mn)$:} & a_{\mathcal S}&=\frac{\mu(\mu-1)}{(2-\mu)^2}\frac{\eta_1^{O(N)}}m\,,  & a_{\mathcal W}&=0\,,  & \gamma^{(1)}_\phi&=\frac{\eta_1^{O(N)}}{m}\,,
\nonumber\\
&\text{Chiral:} & a_{\mathcal S}&= \frac{\mu(\mu-1)\eta_1^{O(N)}}{m(2-\mu)^2}\,,  & a_{\mathcal W}&= \frac{\mu(\mu-1)\eta_1^{O(N)}}{2(2-\mu)^2} \,, & \gamma^{(1)}_\phi&=\frac{(m+1)\eta_1^{O(N)}}{2}\,,
\nonumber\\
&\text{Antichiral:} & a_{\mathcal S}&=0 \,, & a_{\mathcal W}&= \frac{\mu(\mu-1)\eta_1^{O(N)}}{2(2-\mu)^2} \,, & \gamma^{(1)}_\phi&=\frac{(m+2)(m-1)\eta_1^{O(N)}}{2m}\,,
\end{align}
where $\eta_1^{O(n)}$ is the anomalous dimension of $\phi$ in the $O(N)$ model, given in \eqref{eq:eta1def}. The values for $\gamma^{(1)}_\phi$ are consistent with the literature results quoted in \LargenChiral and \LargenAntichiral.

In Table~\ref{table:twistfamilies} we summarize the twist families of the $O(m)\times O(n)$ symmetric theory at large $n$ in the chiral and antichiral fixed points. We give the leading twist family in each representation, and we also display a couple of subleading families in the singlet representation. The existence of each of these subleading families follows from the initial analytic bootstrap considerations of \cite{Fitzpatrick:2012yx,
Komargodski:2012ek}, since they are the double-twist operators in a suitable four-point function. Importantly, these families contain more than one operator at each spin and therefore participate in mixing. In the cases where there is an operator at spin zero, we match it with the scalar singlets presented in Table~\ref{table:lowestscalars}.

In Table~\ref{table:twistfamilies} we also explain how the scaling dimension of each twist family relates to the corresponding scalar. In similarity with the $O(N)$ model, we assume that the expressions~\eqref{eq:scalingdimensionslargeN} can be analytically continued to spin zero, giving the dimension or the shadow dimension of the corresponding scalar. Including also $\Delta_\phi$, this gives for the chiral fixed point
\begin{align}\label{largenNewChiral}
\Delta_{\phi+}&=\mu-1+\frac{m+1}{2}\frac{\eta_1^{O(N)}}n+\ldots
&&\overset{\mathrm{3d}}=\frac12+\frac{2(m+1)}{3\pi^2n}+\ldots\,,
\nonumber\\
\Delta_{S+}&=2-\frac{2 (\mu -1) (2 \mu -1) (m+1)}{2-\mu}     \frac{\eta_1^{O(N)}}{n}+\ldots
&&\overset{\mathrm{3d}}=2-\frac{16 (m+1)}{3 \pi ^2 n}+\ldots\,,
\nonumber\\
\Delta_{W+}&=2+\left(\frac{2 (m+3)}{\mu -2}+2 \mu  (m+2)+2\right)    \frac{\eta_1^{O(N)}}{n}+\ldots
&&\overset{\mathrm{3d}}=2-\frac{4 (m+4)}{3 \pi ^2 n}+\ldots\,,
\nonumber\\
\Delta_{X+}&=2\Delta_{\phi+}+\frac{\mu  (m+1)}{2-\mu}      \frac{\eta_1^{O(N)}}{n}+\ldots
&&\overset{\mathrm{3d}}=1+\frac{16 (m+1)}{3 \pi ^2  n}+\ldots\,,
\nonumber\\
\Delta_{Y+}&=2\Delta_{\phi+}+\frac{\mu }{2-\mu }      \frac{\eta_1^{O(N)}}{n}+\ldots
&&\overset{\mathrm{3d}}=1+\frac{4 (m+4)}{3 \pi ^2 n}+\ldots\,,
\nonumber\\
\Delta_{Z+}&=2\Delta_{\phi+}-\frac{\mu }{2-\mu }      \frac{\eta_1^{O(N)}}{n}+\ldots
&&\overset{\mathrm{3d}}=1+\frac{4 (m-2)}{3 \pi ^2 n}+\ldots\,,
\end{align}
and for the antichiral fixed point
\begin{align}\label{largenNewAntichiral}
\Delta_{\phi-}&=\mu-1+\frac{(m+2)(m-1)}{2m}\frac{\eta_1^{O(N)}}n+\ldots
&&\overset{\mathrm{3d}}=\frac12+\frac{2(m+2)(m-1)}{3\pi^2mn}+\ldots\,,
\nonumber\\
\Delta_{S-}&=2\Delta_{\phi-}+\frac{\mu  (4 \mu -5) (m-1) (m+2)}{(2-\mu) m}\frac{\eta_1^{O(N)}}n+\ldots
&&\overset{\mathrm{3d}}=1+\frac{16 (m-1) (m+2)}{3 \pi ^2 m n}+\ldots\,,
\nonumber\\
\Delta_{W-}&=2+\!\left((1+2\mu)(m-4)+\tfrac{m^2+3 m-10}{\mu -2}+m^2\mu\right)\!\frac{2\eta_1^{O(N)}}{mn}\!+\ldots
&&\overset{\mathrm{3d}}=2-\frac{4 \left(m^2+4 m-8\right)}{3 \pi ^2 m
   n}+\ldots\,,
\nonumber\\
\Delta_{X-}&=2\Delta_{\phi-}+\frac{\mu  (m-1)
   (m+2)}{(2-\mu) m}\frac{\eta_1^{O(N)}}n+\ldots
&&\overset{\mathrm{3d}}=1+\frac{16 (m-1) (m+2)}{3 \pi ^2 m n}+\ldots\,,
\nonumber\\
\Delta_{Y-}&=2\Delta_{\phi-}+\frac{\mu  (m-2)}{(2-\mu) m}\frac{\eta_1^{O(N)}}n+\ldots
&&\overset{\mathrm{3d}}=1+\frac{4 \left(m^2+4 m-8\right)}{3 \pi ^2 m
   n}+\ldots\,,
\nonumber\\
\Delta_{Z-}&=2\Delta_{\phi-}-\frac{\mu  (m+2)}{(2-\mu) m}   \frac{\eta_1^{O(N)}}n+\ldots
&&\overset{\mathrm{3d}}=1+\frac{4 (m-4) (m+2)}{3 \pi ^2 m n}+\ldots\,.
\end{align}
The values for $\phi$, $S$, $W$ and $Z$ agree with those quoted in section~\ref{sec:largenlitt}, whereas we are not aware of any previous results for the remaining operators. We also give results for the non-conserved spin one operators
\begin{align}
\Delta_{C,1+}&=2\Delta_{\phi+}+1+\frac{\eta_1^{O(N)}}{n}+\ldots&&\overset{\mathrm{3d}}=2+\frac{4(m+2)}{3\pi^2n}+\ldots\,,
\nonumber\\
\Delta_{D,1+}&=2\Delta_{\phi+}+1-\frac{\eta_1^{O(N)}}{n}+\ldots&&\overset{\mathrm{3d}}=2+\frac{4m}{3\pi^2n}+\ldots\,,
\nonumber\\
\Delta_{C,1-}&=2\Delta_{\phi-}+1+\frac{(m+2)\eta_1^{O(N)}}{mn}+\ldots&&\overset{\mathrm{3d}}=2+\frac{4(m+2)}{3\pi^2n}+\ldots\,,
\nonumber\\
\Delta_{D,1-}&=2\Delta_{\phi-}+1-\frac{(m-2)\eta_1^{O(N)}}{mn}+\ldots&&\overset{\mathrm{3d}}=2+\frac{4m}{3\pi^2n}+\ldots\,.
\end{align}

The computation of the OPE coefficients provides results for the central charges, by \eqref{eq:centralchargesconventions}.
For the chiral fixed point we get
\begin{align}
\frac{C_{T+}}{C_{T,\mathrm{free}}}&=1-\frac{(m+1)c_1}n+\ldots
&&\overset{\mathrm{3d}}=1-\frac{20(m+1)}{9\pi^2n}+\ldots\,,
\nonumber\\
\frac{C_{J_A+}}{C_{J_A,\mathrm{free}}}&=1+\frac{c_2}{n}-\frac{(m+2)c_3}{2n}+\ldots
&&\overset{\mathrm{3d}}=1-\frac{44+38m}{9\pi^2n}+\ldots\,,
\nonumber\\
\frac{C_{J_B+}}{C_{J_B,\mathrm{free}}}&=1-\frac{(m+1)c_2}{n}+\ldots
&&\overset{\mathrm{3d}}=1-\frac{32(m+1)}{9\pi^2n}+\ldots\,,
\label{eq:chargesChiral}
\end{align}
and for the antichiral fixed point
\begin{align}
\frac{C_{T-}}{C_{T,\mathrm{free}}}&=1-\frac{(m+2)(m-1)c_1}{mn}+\ldots
&&\overset{\mathrm{3d}}=1-\frac{20(m+2)(m-1)}{9\pi^2mn}+\ldots\,,
\nonumber\\
\frac{C_{J_A-}}{C_{J_A,\mathrm{free}}}&=1+\frac{(m+2)c_2}{mn}-\frac{(m+2)c_3}{2n}+\ldots
&&\overset{\mathrm{3d}}=1-\frac{2(m+2)(19m-16)}{9\pi^2 mn}+\ldots\,,
\nonumber\\
\frac{C_{J_B-}}{C_{J_B,\mathrm{free}}}&=1-\frac{(m+2)(m-1)c_2}{mn}+\ldots
&&\overset{\mathrm{3d}}=1-\frac{32(m+2)(m-1)}{9\pi^2mn}+\ldots\,,
\label{eq:chargesAnti}
\end{align}
where the precise form of the constants $c_i$ is given in \eqref{eq:cconstants} in Appendix~\ref{app:explicitresults}.

\begin{table}[H]
\centering
\caption{Twist families in the large-$n$ expansion. We give a couple of subleading twist families in the singlet case, and the leading family in the other irreps. We denote degenerate operators by $\langle\ \cdot \ \rangle$.}\label{table:twistfamilies}
\begin{tabular}{|c|cccl|cccl|}\hline

&\multicolumn{4}{c|}{\textbf{Chiral}}&\multicolumn{4}{c|}{\textbf{Antichiral}}
\\
$R$& Gp & $\O_\ell$ & $\tau_\infty$ & constraints& Gp & $\O_\ell$ & $\tau_\infty$ & constraints
\\\hline
\multirow{3}{*}{$S$}
&$\mathrm{III}$& $\mathcal J_{S,\ell}$ & $2\Delta_\phi$ & $\hspace{-9pt}\left\{\hspace{-4.8pt}\begin{array}{l}\Delta_0=d-\Delta_S\\\Delta_2=d\end{array}\right.$ & $\mathrm{II}$ & $\mathcal J_{S,\ell}$ & $2\Delta_\phi$ & $\hspace{-9pt}\left\{\hspace{-4.8pt}\begin{array}{l}\Delta_0=\Delta_S\\\Delta_2=d\end{array}\right.$
\\
&--- & $\left\langle[\phi,\phi]_{S,1,\ell}\right\rangle$ & $2\Delta_\phi+2\!\!$ & $\hspace{9.2pt} \ell\geqslant 2$
& --- & $\left\langle\partial^\ell\phi^4_S\right\rangle$ & $4\Delta_\phi$ & $\Delta_0=\Delta_{S'}$
\\
&--- & $\left\langle
\begin{smallmatrix}[\mathcal S,\mathcal S]_{S,0,\ell}\\ [\mathcal W,\mathcal W]_{S,0,\ell}\end{smallmatrix}\right\rangle$ & 4 &
$\Delta_0=\langle\Delta_{S'},\Delta_{S''}\rangle$
& --- & $\left\langle[\phi,\phi]_{S,1,\ell}\right\rangle$ & $2\Delta_\phi+2\!\!$ & $\hspace{9.2pt} \ell\geqslant 2$
\\\hline
$W$& $\mathrm{III}$ & $\mathcal J_{W,\ell}$& $2\Delta_\phi$ &$\Delta_0=d-\Delta_W$&
$\mathrm{III}$ & $\mathcal J_{W,\ell}$& $2\Delta_\phi$ & $\Delta_0=d-\Delta_W$
\\\hline
$X$& $\mathrm{I}$ & $\mathcal J_{X,\ell}$& $2\Delta_\phi$ &$\Delta_0=\Delta_X$&
$\mathrm{I}$ & $\mathcal J_{Z,\ell}$& $2\Delta_\phi$ & $\Delta_0=\Delta_X$
\\\hline
$Y$& $\mathrm{I}$ & $\mathcal J_{Y,\ell}$& $2\Delta_\phi$ &$\Delta_0=\Delta_Y$&
$\mathrm{I}$ & $\mathcal J_{Y,\ell}$& $2\Delta_\phi$ & $\Delta_0=\Delta_Y$
\\\hline
$Z$& $\mathrm{I}$ & $\mathcal J_{Z,\ell}$& $2\Delta_\phi$ &$\Delta_0=\Delta_Z$&
$\mathrm{I}$ & $\mathcal J_{Z,\ell}$& $2\Delta_\phi$ & $\Delta_0=\Delta_Z$
\\\hline
$A$& $\mathrm{II}$ & $\mathcal J_{A,\ell}$& $2\Delta_\phi$ &$\Delta_1=d-1$&
$\mathrm{II}$ & $\mathcal J_{A,\ell}$& $2\Delta_\phi$ & $\Delta_1=d-1$
\\\hline
$B$& $\mathrm{I}$ & $\mathcal J_{B,\ell}$& $2\Delta_\phi$ &$\Delta_1=d-1$&
$\mathrm{I}$ & $\mathcal J_{B,\ell}$& $2\Delta_\phi$ & $\Delta_1=d-1$
\\\hline
$C$& $\mathrm{I}$ & $\mathcal J_{C,\ell}$& $2\Delta_\phi$ &&
$\mathrm{I}$ & $\mathcal J_{C,\ell}$& $2\Delta_\phi$ &
\\\hline
$D$& $\mathrm{I}$ & $\mathcal J_{D,\ell}$& $2\Delta_\phi$ &&
$\mathrm{I}$ & $\mathcal J_{D,\ell}$& $2\Delta_\phi$ &
\\\hline
\end{tabular}
\end{table}

\begin{figure}[H]
  \centering
  \includegraphics[scale=0.85]{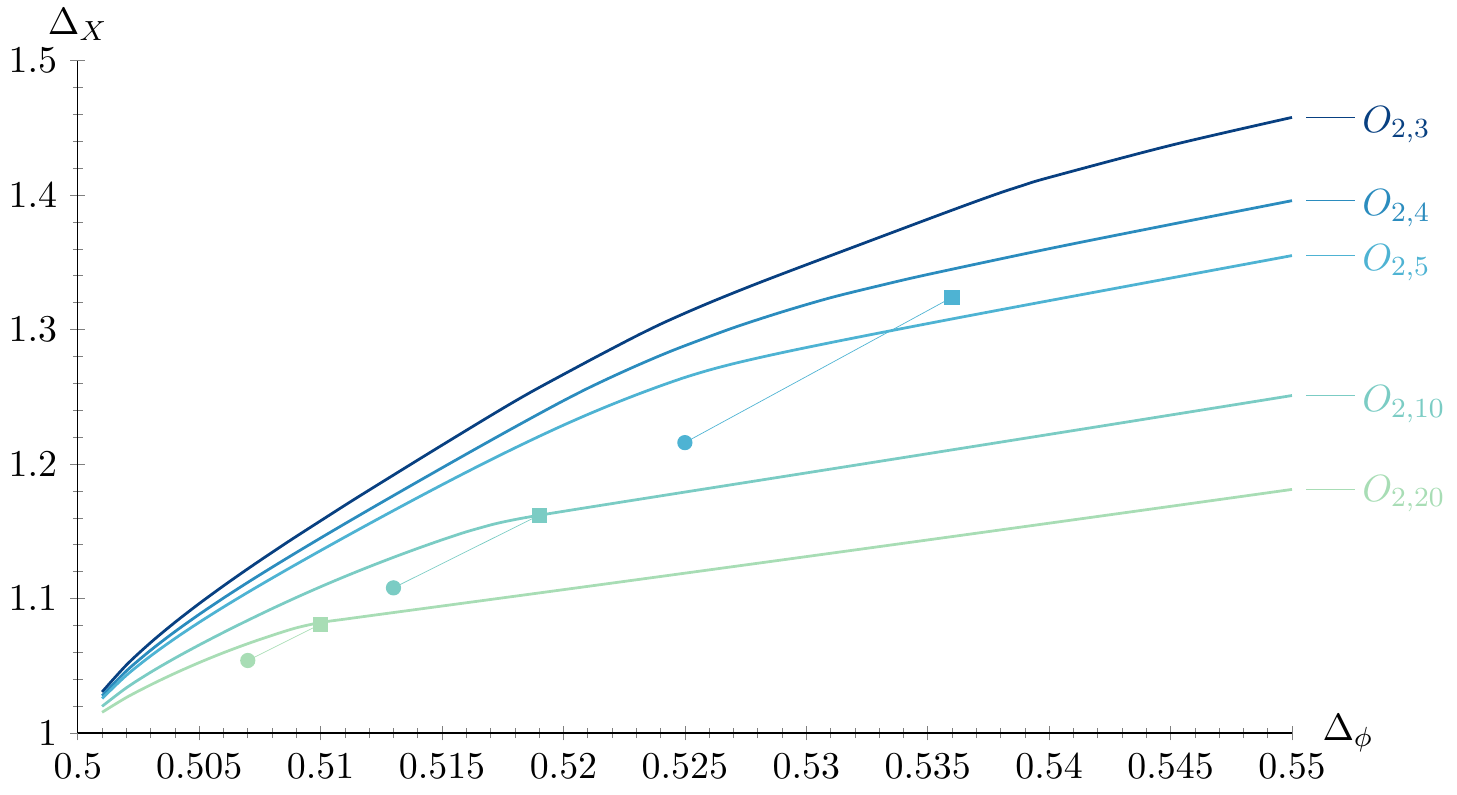}
  \caption{Upper bound on the dimension of the first scalar $X$ operator in
  the $\phi_{ar}\times\phi_{bs}$ OPE as a function of the dimension of
  $\phi$.  Areas above the curves are excluded in the corresponding
  theories.  The positions of the fixed points as predicted by the
  large-$n$ results \LargenChiral, \LargenAntichiral for $\Delta_\phi$ and
  \eqref{largenNewChiral}, \eqref{largenNewAntichiral} for $\Delta_X$ for
  $n=5,10,20$ are also given as squares and circles for the chiral and
  antichiral fixed points, respectively. (The lines between squares and
  circles are added to illustrate fixed points with the same symmetry.)}
  \label{fig:Delta_X_O2n}
\end{figure}

\begin{figure}[H]
  \centering
  \includegraphics[scale=0.85]{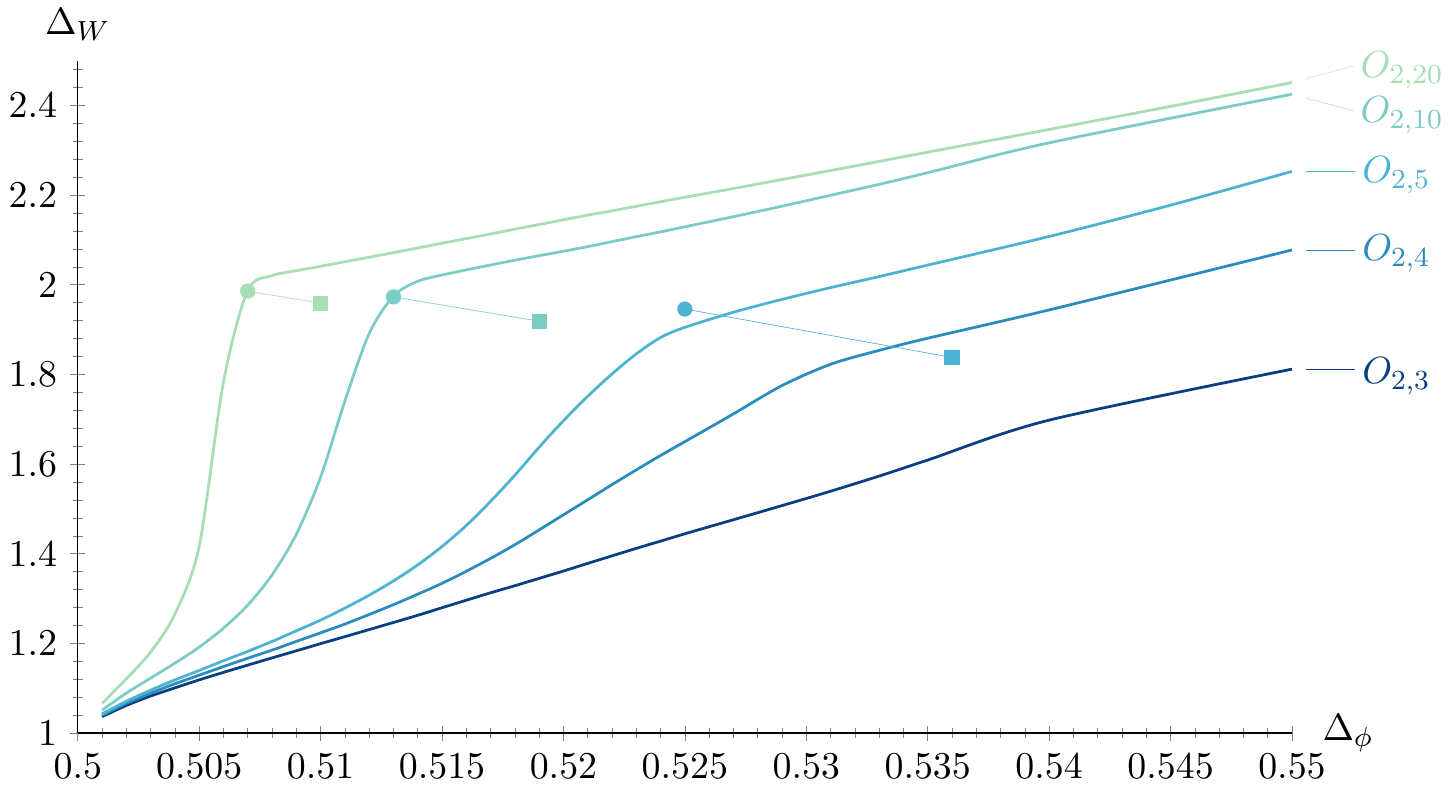}
  \caption{Upper bound on the dimension of the first scalar $W$ operator in
  the $\phi_{ar}\times\phi_{bs}$ OPE as a function of the dimension of $\phi$.
  Areas above the curves are excluded in the corresponding theories.  The
  positions of the fixed points as predicted by the large-$n$ results
  \LargenChiral and \LargenAntichiral for $n=5,10,20$ are also given as
  squares and circles for the chiral and antichiral fixed points,
  respectively. (The lines between squares and circles are added to
  illustrate fixed points with the same symmetry.)} \label{fig:Delta_W_O2n}
\end{figure}

\begin{figure}[ht]
  \centering
  \includegraphics{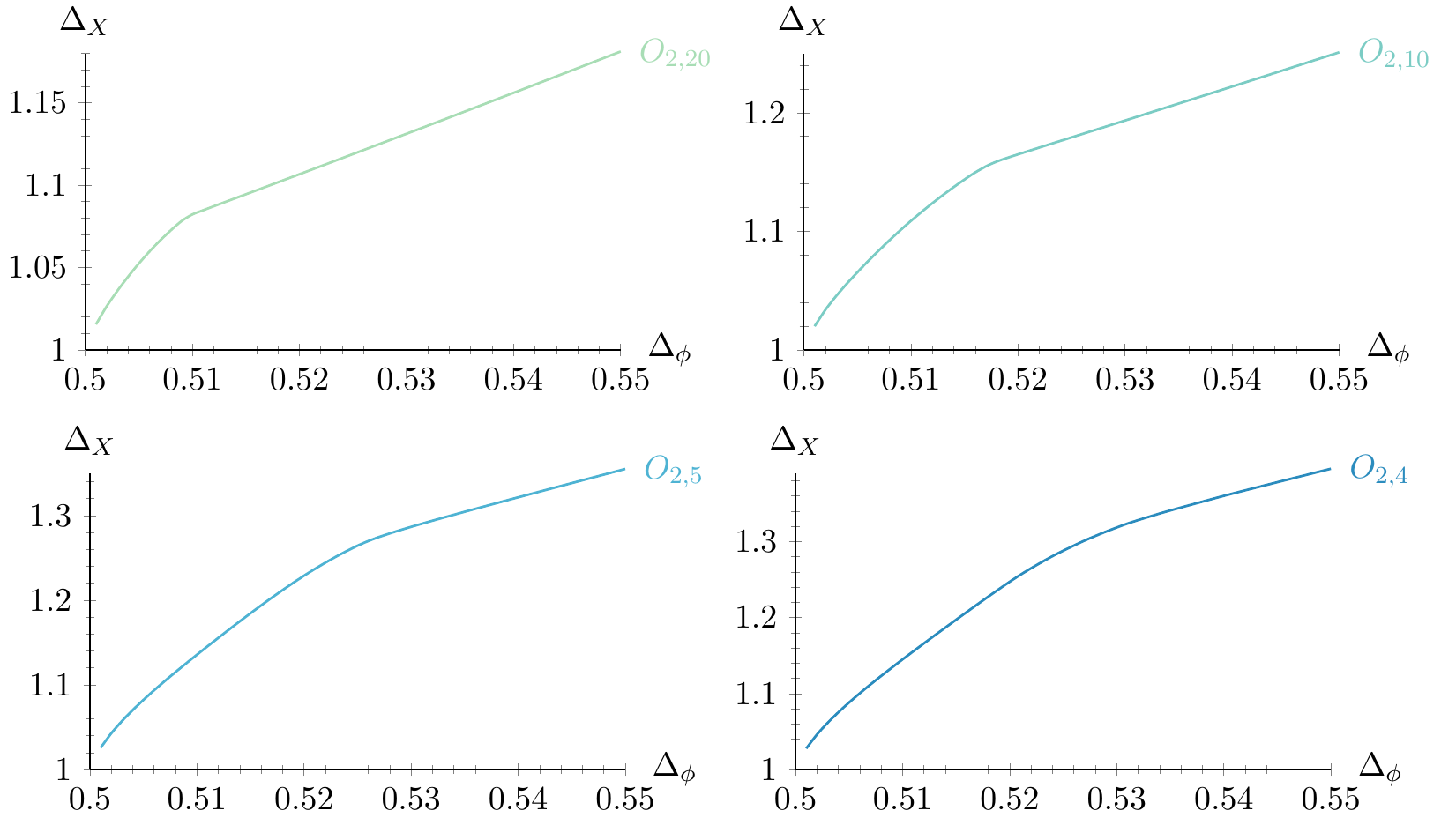}
  \caption{Upper bound on the dimension of the first scalar $X$ operator in
  the $\phi_{ar}\times\phi_{bs}$ OPE as a function of the dimension of $\phi$.
  Areas above the curves are excluded in the corresponding theories.}
  \label{fig:Delta_X_O2n_series}
\end{figure}

\newsec{Numerical bootstrap of \texorpdfstring{$\boldsymbol{O(m)\times
O(n)}$}{O(m)xO(n)} CFTs}[numerics]
\subsec{Single correlator}
In the single-correlator bootstrap, for which the crossing equations are
discussed in Appendix~\ref{appCross}, we have obtained bounds on the
dimensions of the leading scalar operators in the representations
$S,W,X,Y,Z$ as functions of the dimension of $\phi$. The most interesting
results are obtained in the $W$ and $X$ plots. More specifically, using
large-$n$ results we can see that the (mild) kinks that appear in the
$X$-bounds (see Fig.~\ref{fig:Delta_X_O2n}) are due to the chiral fixed
points, while the kinks that appear in the $W$ bounds (see
Fig.~\ref{fig:Delta_W_O2n}) are due to the antichiral fixed points. While
this is clear at large $n$ only, we will assign the same meaning to the
kinks at smaller $n$, but only down to $n=6$ where, as we will see below,
the situation becomes more subtle. As seen in
Fig.~\ref{fig:Delta_W_O2n}, obtained with $m=2$ for different values of
$n$, there are very sharp kinks at large $n$ that get smoothed out as $n$
decreases. In Fig.~\ref{fig:Delta_X_O2n} the kinks are much milder. They
certainly exist at large $n$, but they are not so clear at low $n$---see
Fig.~\ref{fig:Delta_X_O2n_series}---even when at the same $n$, e.g.\ $n=4$,
there is a clear kink in Fig.~\ref{fig:Delta_W_O2n}.

\begin{figure}[ht]
  \centering
  \includegraphics{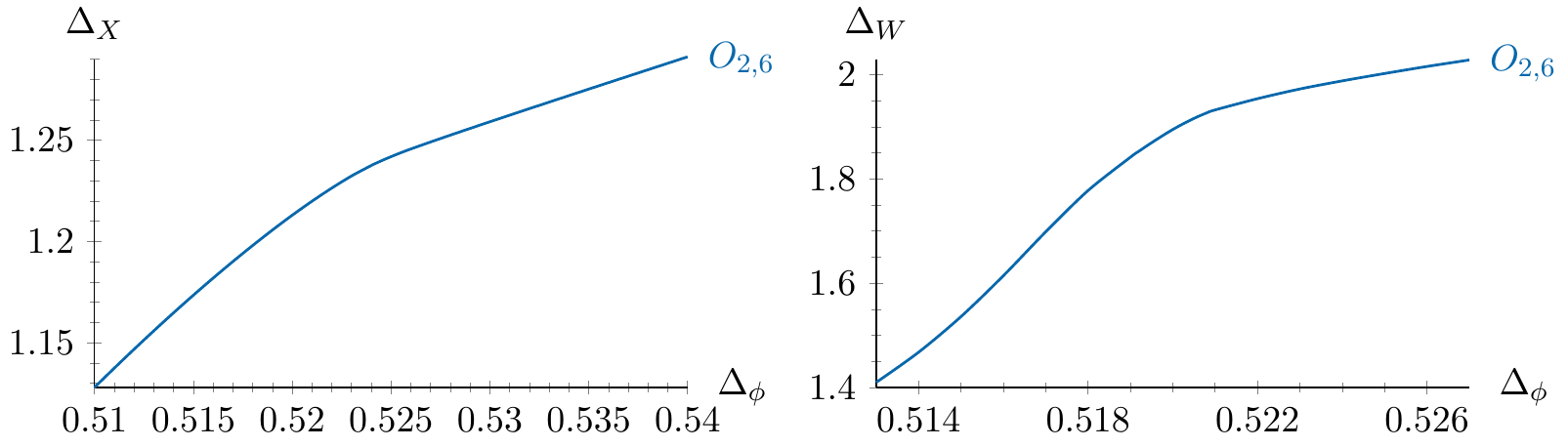}
  \caption{Upper bound on the dimension of the first scalar $X$ and the
  first scalar $W$ operator in the $\phi_{ar}\times\phi_{bs}$ OPE as a function
  of the dimension of $\phi$ in the $O_{2,6}$ theory.}
  \label{fig:Delta_W_Delta_X_O26}
\end{figure}

At $n=6$ we can see from Fig.~\ref{fig:Delta_W_Delta_X_O26} that kinks
exist in both bounds, although the one in the $\Delta_X$ bound is quite
mild.  While a hint of a kink in the $\Delta_X$ bound of the $O_{2,5}$
theory exists, the $\Delta_X$ bounds in the $O_{2,4}$ and $O_{2,3}$
theories are very smooth, although a change in slope can still be seen.
These considerations suggest that the chiral fixed point ceases to exist
for some $n$ between 5 and 6. This is a very rough estimate based on
qualitative features of the $\Delta_X$ bounds. A more accurate estimate
cannot be made based on the presence or absence of kinks as described here.

The persistence of the kinks in the $\Delta_W$ bounds even at small $n$
($n=4,5$ and $n=3$ although the kink is much milder for $n=3$), combined
with their absence in the $\Delta_X$ bounds, is rather puzzling.  After
all, intuition from the $\veps$ expansion dictates that the antichiral and
chiral fixed points, to which we have attributed the kinks in the
$\Delta_W$ and $\Delta_X$ bounds, respectively, annihilate and become
complex fixed points at some $n$ (in this case $n^+(2)$, whose estimated
value in the $\veps$ expansion is 5.96(19)~\cite{Kompaniets:2019xez}).
Since the bootstrap excludes nonunitary theories, both kinks are expected
to disappear at some $n$ around 6, and indeed this is borne out to some
extent for the kinks in the $\Delta_X$ bounds, as we discussed in the
previous paragraph.

One explanation for the persistence of the $\Delta_W$ kinks at small $n$ is
that our numerical bounds are insensitive to the putative nonunitarity of
the antichiral fixed point for small $n$. This scenario could be further
examined by estimating the size of this nonunitarity in perturbation
theory, in a properly quantified sense that we do not discuss here, and
comparing it with that of the chiral fixed point. We do not pursue this
direction here, but it is worth investigating in the future.  Another
possibility is that the kinks in the $\Delta_W$ bounds at small $n$ are due
to another fixed point, which is not the naive continuation of the
antichiral fixed point to small $n$.  Evidence for the existence of such a
fixed point, belonging to the chiral universality class, exists in the
literature; see~\cite[Sec.\ 11.5.3]{Pelissetto:2000ek} and references
therein. A further universality class, typically called collinear, is
supposed to exist for small $n$.\foot{The $O(2)\times O(3)$ and $O(2)\times
O(4)$ theories were also studied with the numerical bootstrap
in~\cite{Nakayama:2014sba}, where a discussion of the chiral and collinear
fixed points of the $O(2)\times O(3)$ and $O(2)\times O(4)$ theories can
also be found. Our numerical bounds are consistent with those
of~\cite{Nakayama:2014sba}.} These chiral and collinear fixed points arise
after resummations of the perturbative beta functions. However, this
approach has been criticized in~\cite{Delamotte_2008, Delamotte:2010ba,
Delamotte_2010}, and the functional RG predicts that they do not
exist~\cite{Tissier:2000tz, Tissier:2001uk, Delamotte:2003dw,
Delamotte:2016acs}.  In section~\ref{secOtwothree} below we will see that,
consistently with the conclusions of~\cite{Nakayama:2014sba}, the chiral
and collinear universality classes appear to exist for $O(2)\times O(3)$
CFTs, although we are not able to conclusively exclude small unitarity
violations.

\begin{figure}[ht]
  \centering
  \includegraphics{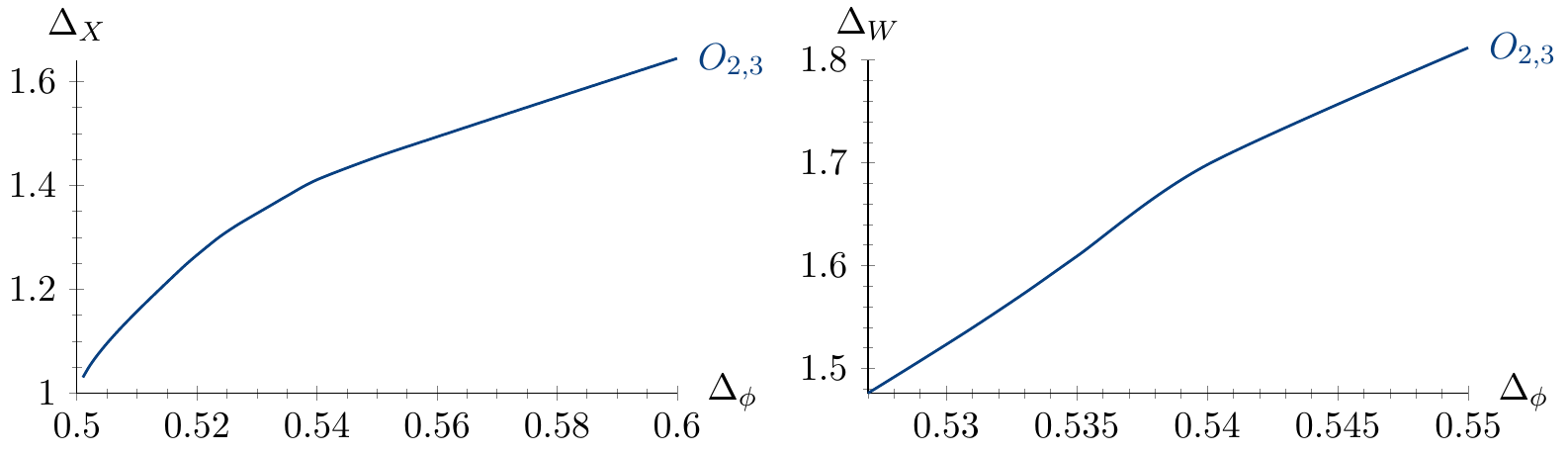}
  \caption{Upper bound on the dimension of the first scalar $X$ and the
  first scalar $W$ operator in the $\phi_{ar}\times\phi_{bs}$ OPE as a function
  of the dimension of $\phi$ in the $O_{2,3}$ theory.}
  \label{fig:Delta_W_Delta_X_O23}
\end{figure}

\subsec{Mixed correlators}
In the mixed correlator system, to find an island around the chiral fixed
point we use the following assumptions (both for the $O_{2,10}$ and the
$O_{2,20}$ theory):
\begin{enumerate}[label=(C-\arabic*), leftmargin=1.7cm]
  \item saturation of $X$ bound of
    Fig.~\ref{fig:Delta_X_O2n},\label{assumCitemI}
  \item existence of conserved current in the $B$ sector, i.e.\
    $\Delta_{J^\mu_B}=2$,\label{assumCitemII}
  \item dimension of next-to-leading vector operator in the $B$ sector,
    $J^{\mu\lsp\prime}_{B}$, above 3, i.e.\
    $\Delta_{J^{\mu\lsp\prime}_{B}}\ge3$,\label{assumCitemIII}
  \item dimension of next-to-leading scalar singlet, $S'$, above
    3, i.e.\ $\Delta_{S'}\ge3$,\label{assumCitemIV}
  \item dimension of next-to-leading bifundamental operator,
    $\phi'$, allowed slightly above $\Delta_\phi$, i.e.\
    $\Delta_{\phi'}\ge\Delta_\phi+0.01$.\label{assumCitemV}
\end{enumerate}
For the antichiral fixed point of the $O_{2,20}$ theory we make the
following assumptions:
\begin{enumerate}[label=(A-$O_{2,20}$-\arabic*), leftmargin=2.7cm]
  \item saturation of $W$ bound of
    Fig.~\ref{fig:Delta_W_O2n},\label{assumA220itemI}
  \item existence of conserved current in the $B$ sector, i.e.\
    $\Delta_{J^\mu_B}=2$,\label{assumA220itemII}
  \item dimension of next-to-leading vector operator in the $B$ sector,
    $J^{\mu\lsp\prime}_{B}$,
    above 3, i.e.\ $\Delta_{J^{\mu\lsp\prime}_{B}}\ge3$,\label{assumA220itemIII}
  \item dimension of next-to-leading scalar singlet, $S'$, above
    1.5, i.e.\ $\Delta_{S'}\ge1.5$.\label{assumA220itemIV}
  \item dimension of next-to-leading bifundamental operator,
    $\phi'$, allowed slightly above $\Delta_\phi$, i.e.\
    $\Delta_{\phi'}\ge\Delta_\phi+0.01$.\label{assumA220itemV}
\end{enumerate}
Finally, for the antichiral fixed point of the $O_{2,10}$ theory we make
the following assumptions:
\begin{enumerate}[label=(A-$O_{2,10}$-\arabic*), leftmargin=2.7cm]
  \item saturation of $W$ bound of
    Fig.~\ref{fig:Delta_W_O2n},\label{assumA210itemI}
  \item existence of conserved current in the $B$ sector, i.e.\
    $\Delta_{J^\mu_B}=2$,\label{assumA210itemII}
  \item dimension of next-to-leading vector operator in the $B$ sector,
    $J^{\mu\lsp\prime}_{B}$, above 3, i.e.\
    $\Delta_{J^{\mu\lsp\prime}_{B}}\ge3$,\label{assumA210itemIII}
  \item dimension of next-to-leading scalar singlet, $S'$, above
    1.6, i.e.\ $\Delta_{S'}\ge1.6$.\label{assumA210itemIV}
  \item dimension of next-to-leading bifundamental operator,
    $\phi'$, above 1.6, i.e.\
    $\Delta_{\phi'}\ge1.6$.\label{assumA210itemV}
\end{enumerate}
Let us note here that even with $\Delta_{S'}\ge3$ we obtain islands around
antichiral fixed points, so long as we keep the gap on $\Delta_{\phi'}$
small. This is inconsistent with the fact that the antichiral fixed point
is unstable, but with our numerical power (see Appendix \ref{appCross}) we
cannot see the inconsistency.  However, when we increase the gap on
$\Delta_{\phi'}$ sufficiently, we do see that the antichiral island
disappears with $\Delta_{S'}\ge3$. This is presumably due to the crossing
equations that arise from the $\langle\phi\phi SS\rangle$ four-point
function, which in the $12\to34$ channel are sensitive to both $\phi'$ and
$S$, while in the $14\to32$ channel they are sensitive to $\phi'$ but not
$S$.

\begin{figure}[H]
  \centering
  \includegraphics{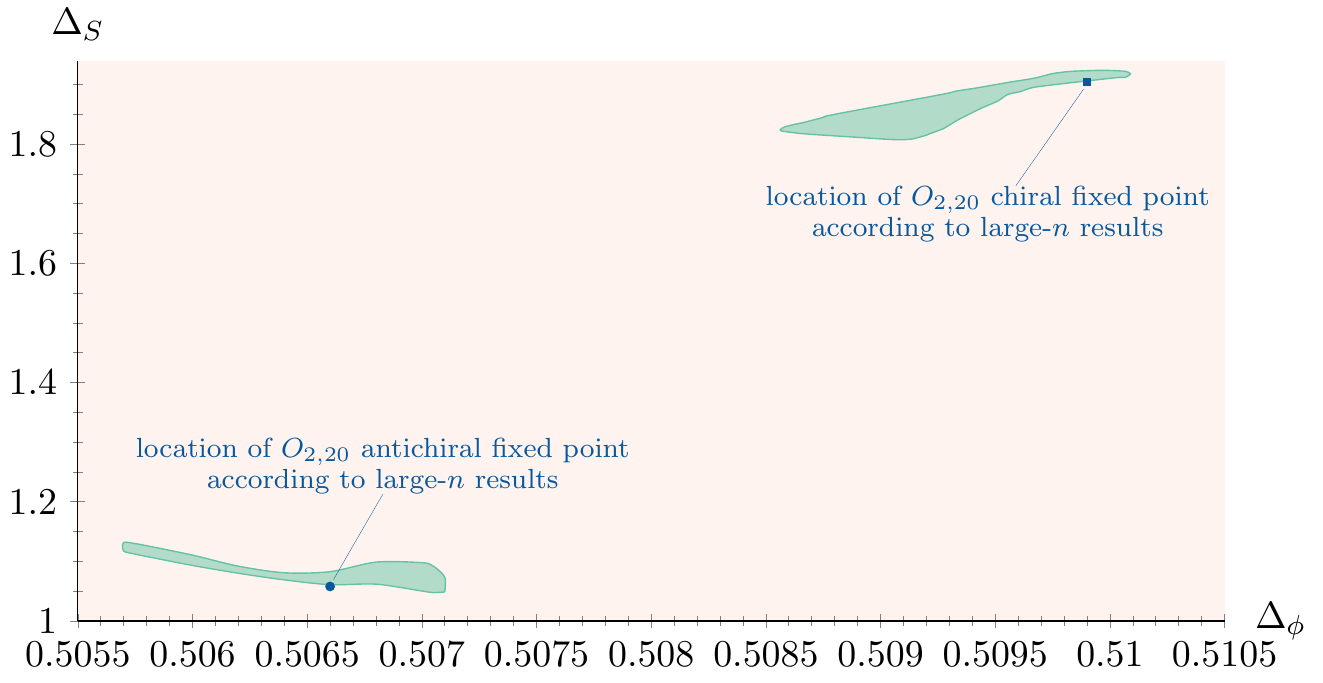}
  \caption{Allowed region (in green) for the $O_{2,20}$ chiral and
    antichiral fixed points and their location according to \LargenChiral.
    The chiral island is obtained with the assumptions
    \ref{assumCitemI}-\ref{assumCitemV}, while the antichiral island is
    obtained with the assumptions \ref{assumA220itemI}-\ref{assumA220itemV}.}
  \label{fig:O2O20_island}
\end{figure}

\begin{figure}[H]
  \centering
  \includegraphics{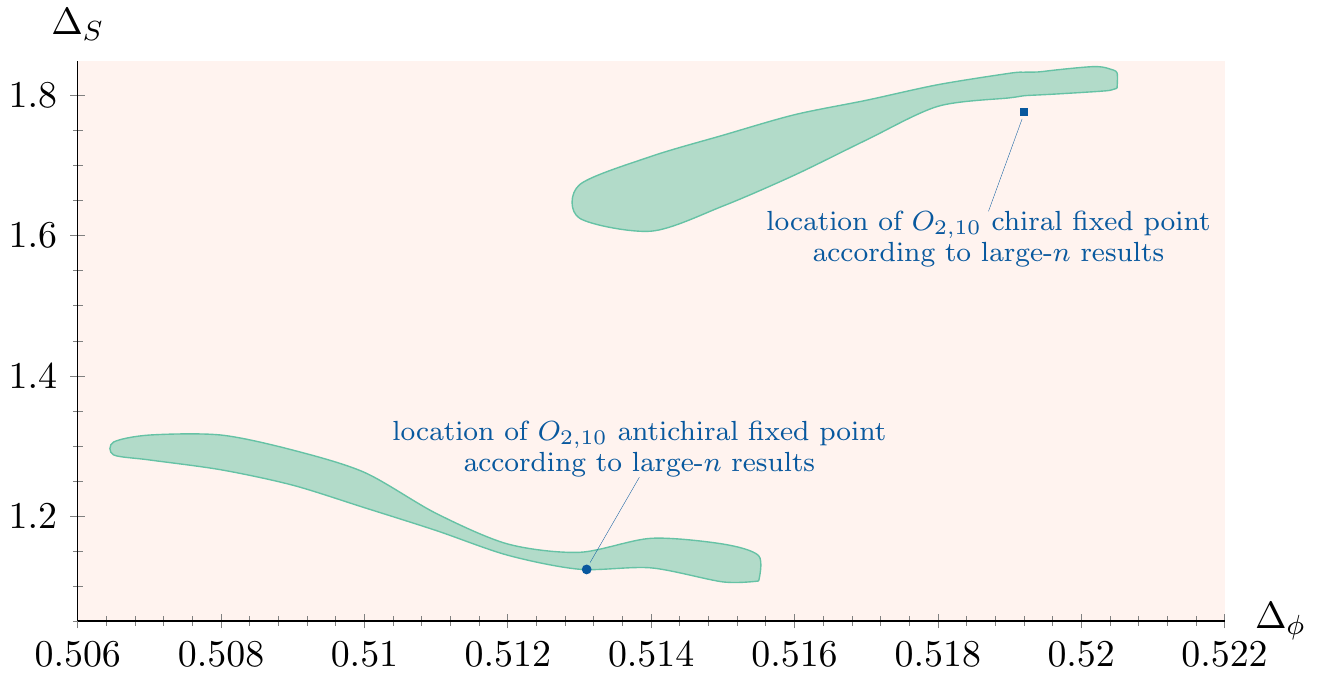}
  \caption{Allowed region (in green) for the $O_{2,10}$ chiral and
    antichiral fixed points and their location according to \LargenAntichiral.
    The chiral island is obtained with the assumptions
    \ref{assumCitemI}-\ref{assumCitemV}, while the antichiral island is
    obtained with the assumptions \ref{assumA210itemI}-\ref{assumA210itemV}.} \label{fig:O2O10_island}
\end{figure}

\begin{figure}[H]
  \centering
  \includegraphics{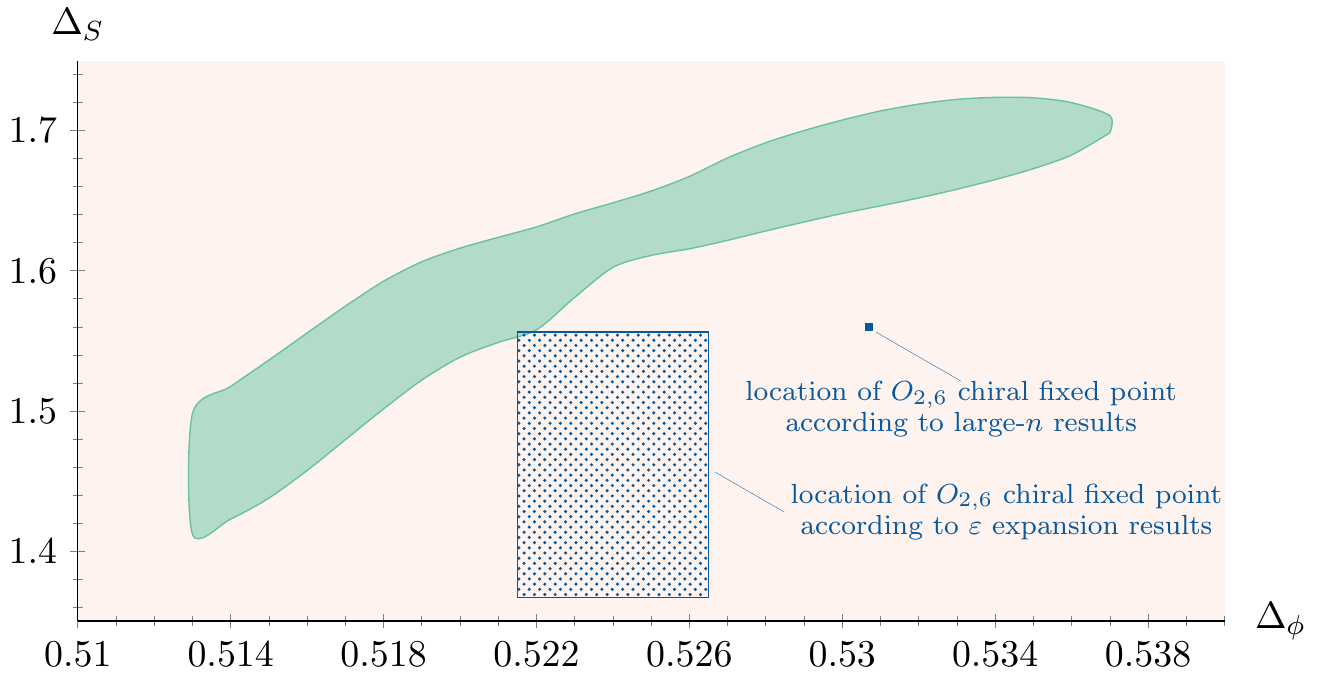}
  \caption{Allowed region (in green) for the $O_{2,6}$ chiral fixed point
    and its location according to \LargenChiral and the $\veps$ expansion
    results of \cite[Table 9]{Kompaniets:2019xez}.  This island is obtained
    with the assumptions \ref{assumCitemI}-\ref{assumCitemV} using
    saturation of the $O_{2,6}$ $X$ bound in
    Fig.~\ref{fig:Delta_W_Delta_X_O26}.}
  \label{fig:O2O6_chiral_island}
\end{figure}

As we have already mentioned, the $\veps$ expansion predicts that
$n^+(2)=5.96(19)$, meaning that a unitary chiral fixed point may exist for
$n=6$. The state-of-the-art analysis of the $O_{2,6}$ theory with the
$\veps$ expansion was performed recently in~\cite{Kompaniets:2019xez}. It
turns out that we can also find an island with our nonperturbative
numerical bootstrap methods, and so we can compare our results with
large-$n$ and $\veps$ expansion results. This is done in
Fig.~\ref{fig:O2O6_chiral_island}.

\subsec{Mixed correlators in the \texorpdfstring{$O(2)\times
O(3)$}{O(2)xO(3)} case}[secOtwothree]
The case $m=2, n=3$ is of particular interest since it is believed to
appear as a symmetry in the continuum limit of frustrated spin models at
criticality. We remind that the fixed points for $m=2, n=3$ arise after
resummations of the perturbative beta functions, i.e.\ they are not found
in the standard perturbative $\veps$ expansion.  There are two fixed
points, called chiral and collinear, with the chiral being of relevance for
the experimentally observed phase transitions in stacked triangular
antiferromagnets~\cite[Sec.\ 11.5]{Pelissetto:2000ek}, and the collinear
potentially relevant for the normal-to-planar superfluid transition in
$^3\text{He}$~\cite{DePrato:2003ra}.  Monte Carlo simulations support the
existence of these fixed points, but functional RG methods come to a
different conclusion, namely that these fixed points do not exist and that
experiments in frustrated magnets are actually seeing weakly first-order
phase transitions~\cite{Tissier:2000tz, Tissier:2001uk, Delamotte:2003dw,
Delamotte:2016acs}.

In what follows we present results regarding putative theories that live on
the $W$ and $Z$ sector single correlator bounds. The $W$ (resp.\ $Z$)
sector bound has a mild kink that appears to correspond to the chiral
(resp.\ collinear) fixed point. This observation was first made
in~\cite{Nakayama:2014sba}. Here we take the next logical step and perform
a mixed correlator bootstrap around these kinks. We also compare to
theoretical predictions and experimental data where applicable. We note
that the general trend of sensitivity to assumptions in the $B$ sector
persists here as well.

\begin{figure}[H]
  \centering
  \includegraphics{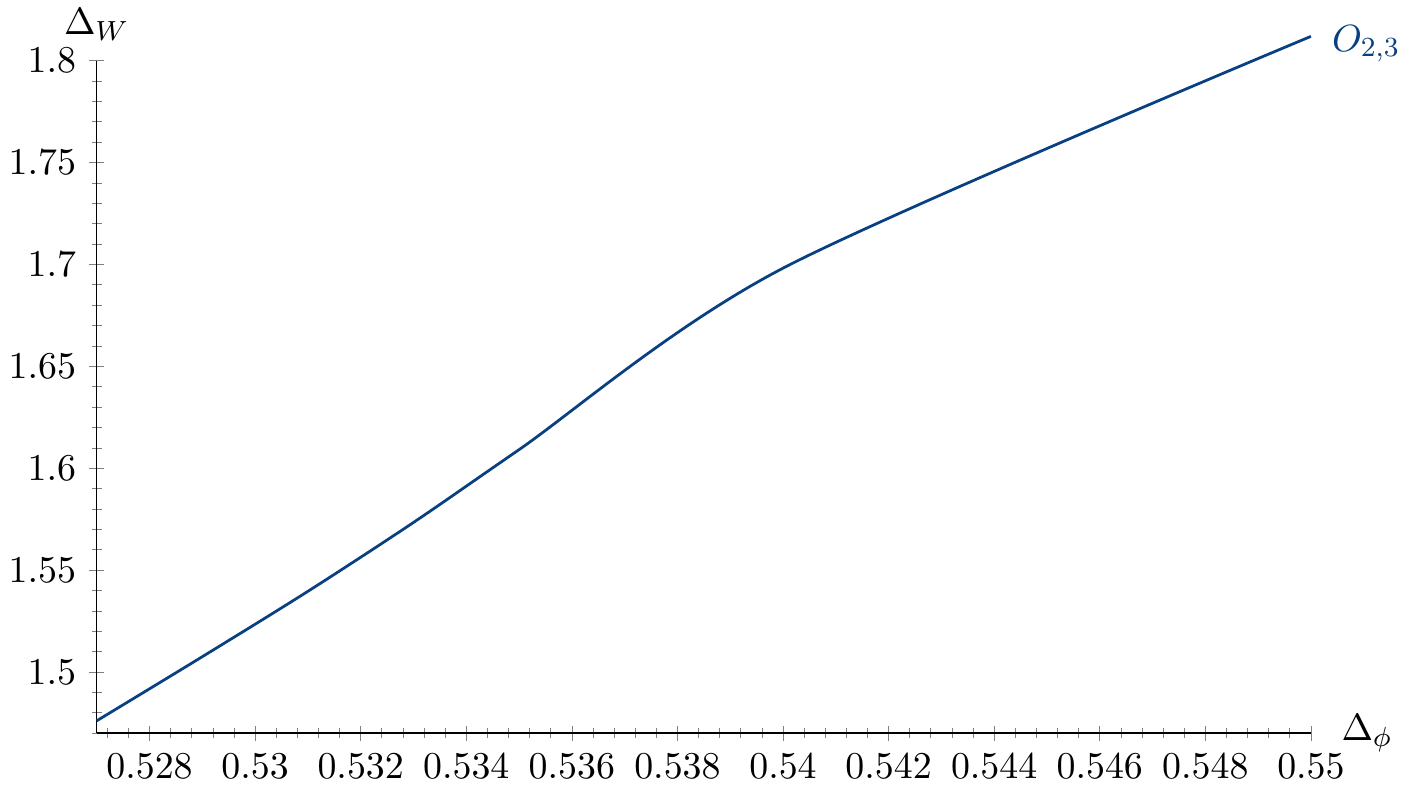}
  \caption{Upper bound on the dimension of the first scalar $W$ operator in
  the $\phi_{ar}\times\phi_{bs}$ OPE as a function of the dimension of $\phi$.
  The area above the curve is excluded.} \label{fig:Delta_W_O23}
\end{figure}

\begin{figure}[H]
  \centering
  \includegraphics{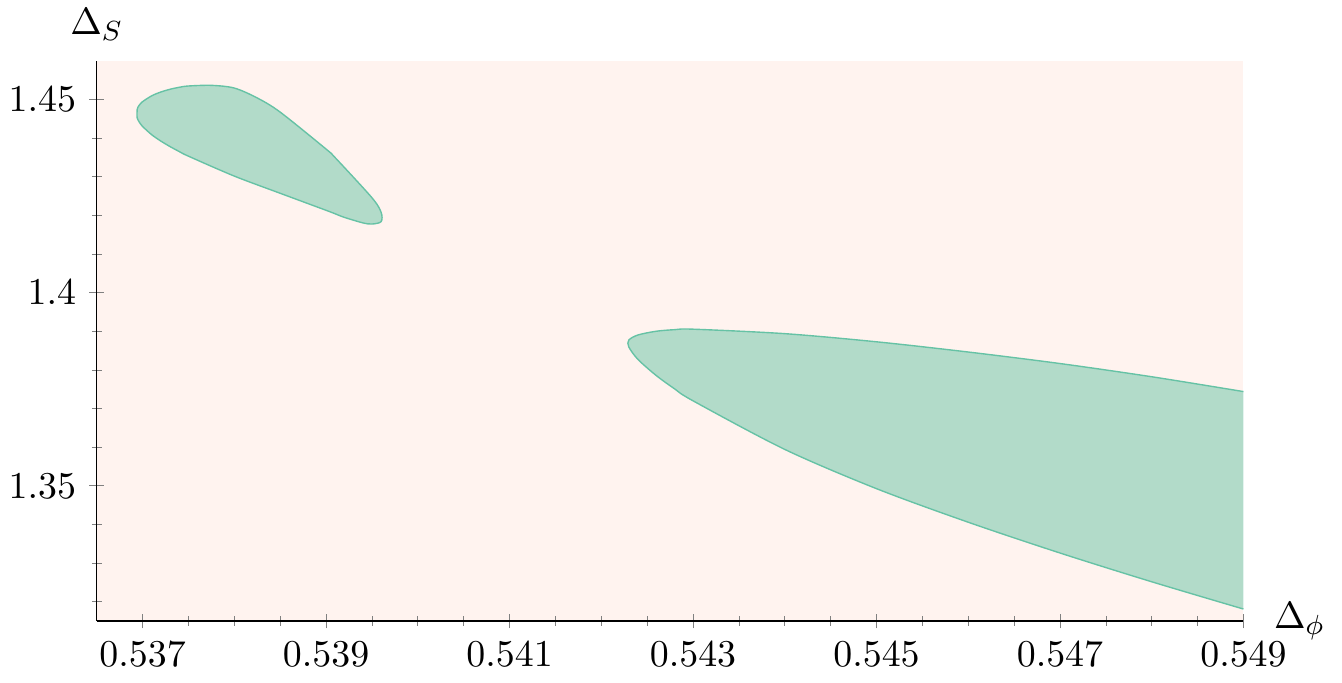}
  \caption{Allowed region (in green) for the $O_{2,3}$ chiral fixed point,
  obtained with the assumptions \ref{assumChiritemI}-\ref{assumChiritemV}.}
  \label{O2O3islandpeninsula}
\end{figure}

Let us start by noticing that, however mild, there seems to be a kink on
the $W$ sector single correlator plot of Fig.~\ref{fig:Delta_W_Delta_X_O23}
around $\Delta_\phi=0.539$. The kink can be seen more clearly in
Fig.~\ref{fig:Delta_W_O23}. As mentioned above, evidence for the existence
of the $O_{2,3}$ chiral fixed point has appeared before in the literature,
based on resummations of perturbative beta functions. Such computations
have been performed at six loops using the massive zero momentum (MZM)
scheme~\cite{Pelissetto:2000ne} and at five loops using the modified
minimal subtraction ($\overline{\text{MS}}$) scheme~\cite{Calabrese:2003ww}
(see also~\cite{Calabrese:2004at}). Monte Carlo simulations have also shown
signs of a critical theory, with the most recent analysis performed
in~\cite{Nagano_2019}. According to results of \cite[Eq.
(2.9)]{Calabrese:2004nt} and \cite[Table III]{Calabrese:2004at}, the theory
at the chiral fixed point has $\Delta_\phi=0.545(20)$ and
$\Delta_W=1.79(9)$ in the $\overline{\text{MS}}$ scheme.\foot{In
\cite[Table III]{Calabrese:2004at}, $y_4=3-\Delta_W$.} In the MZM scheme,
\cite[Table III]{Pelissetto:2000ne} and \cite[Table III]{Calabrese:2004at}
give $\Delta_\phi=0.55(5)$ and $\Delta_W=1.91(5)$. The
$\overline{\text{MS}}$ scheme result is more consistent with the location
of the kink in Fig.~\ref{fig:Delta_W_O23}.
If we assume
\begin{enumerate}[label=(C-$O_{2,3}$-\arabic*), leftmargin=2.5cm]
  \item saturation of $W$ bound of
    Fig.~\ref{fig:Delta_W_O23},\label{assumChiritemI}
  \item existence of a conserved current in the $B$ sector, i.e.\
    $\Delta_{J^\mu_B}=2$,\label{assumChiritemII}
  \item dimension of the next-to-leading vector operator in the $B$ sector,
    $J^{\mu\lsp\prime}_{B}$, above 2.4, i.e.\
    $\Delta_{J^{\mu\lsp\prime}_{B}}\ge2.4$, \label{assumChiritemIII}
  \item dimension of the next-to-leading scalar singlet, $S'$, above
    2, i.e.\ $\Delta_{S'}\ge2$.\label{assumChiritemIV}
  \item dimension of next-to-leading bifundamental operator,
    $\phi'$, above 1.5, i.e.\
    $\Delta_{\phi'}\ge1.5$,\label{assumChiritemV}
\end{enumerate}
then we obtain Fig.~\ref{O2O3islandpeninsula}. Let us note that the island
in Fig.~\ref{O2O3islandpeninsula} remains even if we make the more
constraining assumption $\Delta_{S'}\ge3$, which is compatible with the RG
stability of the $O_{2,3}$ chiral fixed point. If instead of
\ref{assumChiritemIII} we assume that $ J^{\mu\lsp\prime}_{B}$ can appear
with dimension below 2.4, then both the island and the peninsula are part
of a bigger, continuous peninsula that includes both.

The state-of-the-art results in the literature for operator scaling
dimensions of relevance for Fig.~\ref{O2O3islandpeninsula} are as follows:
\eqna{\text{\cite{Pelissetto:2000ne}}&: \qquad \Delta_\phi=0.55(5)\,,\quad
\;\;\;\,\Delta_S=1.18(10)\,,\\
\text{\cite{Calabrese:2004nt}}&:\qquad \Delta_\phi=0.545(20)\,,\quad
\Delta_S=1.41(13)\,,\\
\text{\cite{Nagano_2019}}&:\qquad \Delta_\phi=0.50(4)\,,\quad
\;\;\;\,\Delta_S=1.08(4)\,.}[critsChiral]
Even with the large error bars in \critsChiral, we see that agreement is
best with the results of \cite{Calabrese:2004nt}, mainly due to
$\Delta_S$.\foot{Note that the six-loop $\overline{\text{MS}}$ beta
functions for $O(m)\times O(n)$ theories have recently been obtained
in~\cite{Kompaniets:2019xez}, so the five-loop analysis
of~\cite{Calabrese:2004nt} can perhaps be extended to six loops.} In
conjunction with the $\Delta_W$ results mentioned earlier, it is clear that
the $\overline{\text{MS}}$ results of \cite{Calabrese:2004nt} and
\cite{Calabrese:2004at} agree best with our bootstrap results for the chiral
fixed point.

Experimental results for transitions described by the $O_{2,3}$ chiral
fixed point can be found in~\cite[Table 37]{Pelissetto:2000ek}. The
agreement of our results for the critical exponent $\nu=1/(3-\Delta_S)$ is
very good, although the same cannot be said for the critical exponent
$\beta=\Delta_\phi/(3-\Delta_S)$.

Refs.\ \cite{Pelissetto:2000ne, Calabrese:2004nt, Nagano_2019} indicate
that their fixed points are of the focus type, which means that they are
nonunitary, while our study pertains to unitary theories. It is unclear to
us how sizable nonunitarities of the type discussed in
\cite{Pelissetto:2000ne, Calabrese:2004nt, Nagano_2019} could have been
missed by our bootstrap results. We note that the numerical bootstrap has
previously found islands for theories that are believed to be nonunitary,
namely the five-dimensional $O(N)$ models~\cite{Li:2016wdp} and the Ising
model in $d=4-\veps$~\cite{Behan:2016dtz}. In the former case, increasing
the constraining power of the numerics led to the disappearance of the
allowed region. It is possible that also our island in
Fig.~\ref{O2O3islandpeninsula} will disappear with stronger numerics.

\begin{figure}[H]
  \centering
  \includegraphics{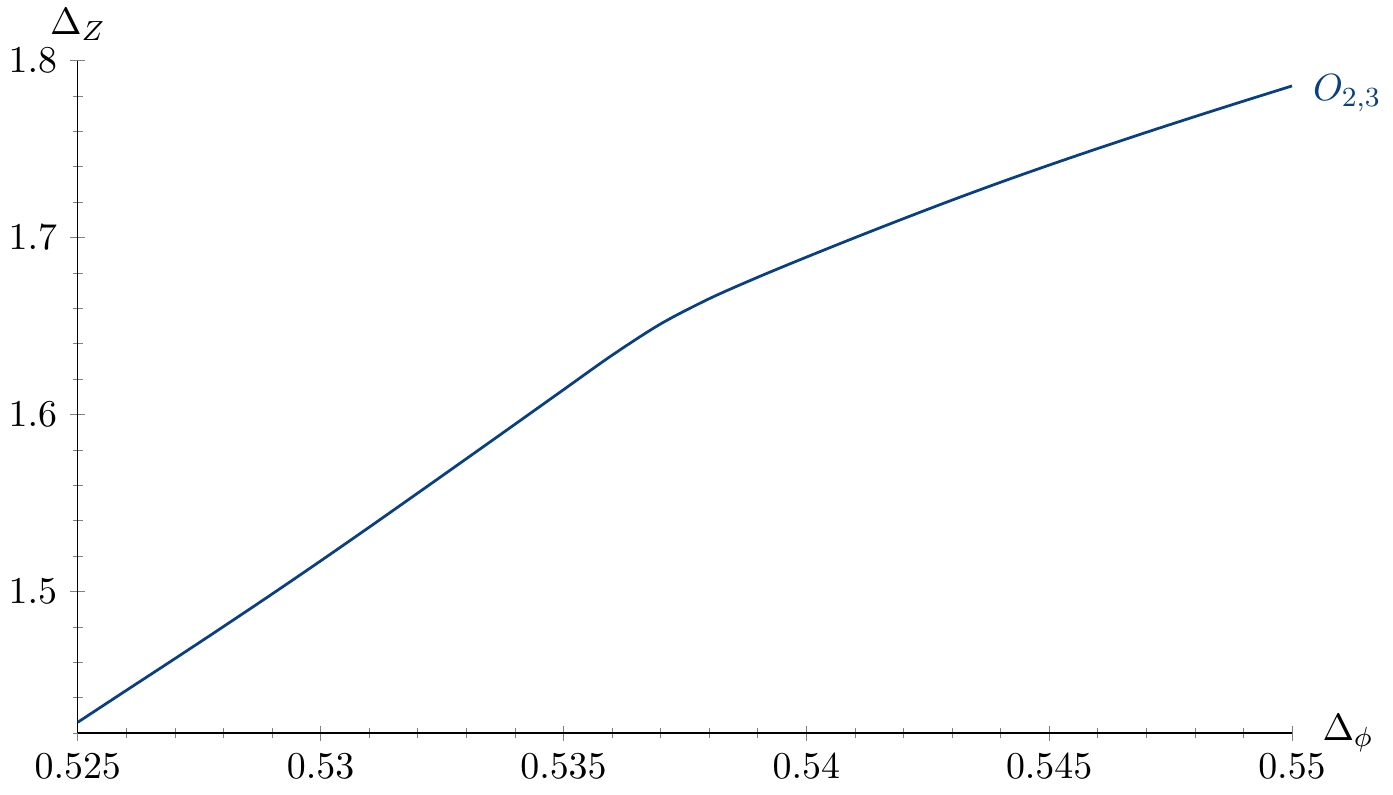}
  \caption{Upper bound on the dimension of the first scalar $Z$ operator in
  the $\phi_{ar}\times\phi_{bs}$ OPE as a function of the dimension of $\phi$.
  The area above the curve is excluded.} \label{fig:Delta_Z_O23}
\end{figure}

\begin{figure}[H]
  \centering
  \includegraphics{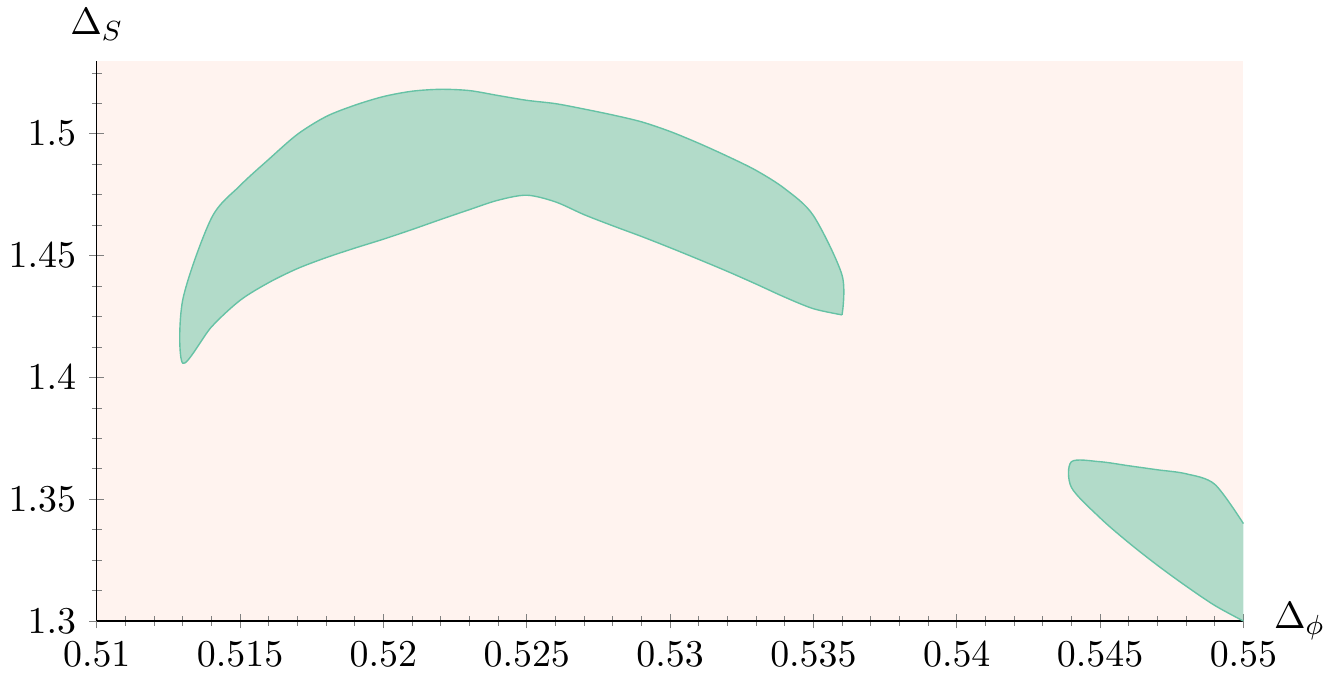}
  \caption{Allowed region (in green) for the $O_{2,3}$ collinear
    fixed point with the assumptions
    \ref{assumCollitemI}-\ref{assumCollitemV}.}
  \label{fig:O2O3_collinear_island_peninsula}
\end{figure}

The $O_{2,3}$ collinear fixed point corresponds to a kink in the bound of
the first scalar operators in the $Z$ irrep; see
Fig.~\ref{fig:Delta_Z_O23}. According to results of \cite{DePrato:2003ra}
and \cite[Table III]{Calabrese:2004at}, the theory at the collinear fixed
point has $\Delta_\phi=0.543(12)$ and $\Delta_Z=1.8(1)$ in the
$\overline{\text{MS}}$ scheme.\foot{In \cite[Table III]{Calabrese:2004at},
$y_1=3-\Delta_Z$.} In the MZM scheme, \cite{DePrato:2003ra} and \cite[Table
III]{Calabrese:2004at} give $\Delta_\phi=0.5395(35)$ and
$\Delta_Z=1.75(10)$. The consistency of the MZM scheme result with the
location of the kink in Fig.~\ref{fig:Delta_Z_O23} appears to be slightly
better than that of the $\overline{\text{MS}}$ scheme result.
With the assumptions
\begin{enumerate}[label=(Coll-\arabic*), leftmargin=2.1cm]
  \item saturation of $Z$ bound of
    Fig.~\ref{fig:Delta_Z_O23},\label{assumCollitemI}
  \item existence of conserved current in the $B$ sector, i.e.\
    $\Delta_{J^\mu_B}=2$,\label{assumCollitemII}
  \item dimension of next-to-leading vector operator in the $B$ sector,
    $J^{\mu\lsp\prime}_{B}$, above 2.5, i.e.\
    $\Delta_{J^{\mu\lsp\prime}_{B}}\ge2.5$, \label{assumCollitemIII}
  \item dimension of next-to-leading scalar singlet, $S'$, above
    3, i.e.\ $\Delta_{S'}\ge3$.\label{assumCollitemIV}
  \item dimension of next-to-leading bifundamental operator,
    $\phi'$, slightly above $\Delta_\phi$, i.e.\ $\Delta_{\phi'}\ge\Delta_\phi+0.01$,\label{assumCollitemV}
\end{enumerate}
we find a rather large island that appears to end slightly to the left of
the position of the kink in Fig.~\ref{fig:Delta_Z_O23}; see
Fig.~\ref{fig:O2O3_collinear_island_peninsula}. If we weaken the assumption
\ref{assumCollitemIII} to $\Delta_{J^{\mu\lsp\prime}_{B}}\ge2.4$, then
there is no separate island and peninsula, but rather a continuous
peninsula that gets narrow in the region between the island and peninsula
of Fig.~\ref{fig:O2O3_collinear_island_peninsula}. According
to~\cite{DePrato:2003ra}, in the $\overline{\text{MS}}$ scheme the
$O_{2,3}$ collinear fixed point has $\Delta_\phi=0.543(12)$ and
$\Delta_S=1.41(20)$, while in the MZM scheme it has
$\Delta_\phi=0.5395(35)$ and $\Delta_S=1.31(11)$. Here the MZM scheme
result for $\Delta_\phi$ and the $\overline{\text{MS}}$ scheme result for
$\Delta_S$ appear to agree better with our island in
Fig.~\ref{fig:O2O3_collinear_island_peninsula}. Just like the $O_{2,3}$
chiral fixed point analyzed above, it would be very interesting to study
the effect of stronger numerics in this case as well.

\subsec{Single correlator in the \texorpdfstring{$O(2)\times
O(2)$}{O(2)xO(2)} case}[secOtwotwo]
In the $O_{2,2}$ case, the collinear fixed point is equivalent to the
$O(2)$ fixed point~\cite{Calabrese:2004at}, and so we will only be
examining the proposed chiral fixed point~\cite{Pelissetto:2000ne,
Calabrese:2004at}. As discussed in Appendix~\ref{appCross}, the
single-correlator crossing equations of the $O_{2,2}=O(2)^2/\mathbb{Z}_2$
theory are equivalent to those of the $M\lnsp N_{2,2}=O(2)^2\rtimes S_2$
theory discussed in~\cite{Stergiou:2019dcv}. A strong kink was obtained in
the $X$ sector of~\cite{Stergiou:2019dcv} (see Fig.~1 there), which
corresponds to the $Z$ sector in \eqref{crEqOnOn} below. This bound is
shown in Fig.~\ref{fig:Delta_Z_O22}. According to \cite[Table
III]{Pelissetto:2000ne} and \cite[Table III]{Calabrese:2004at}, in the MZM
scheme the $O_{2,2}$ chiral fixed point has $\Delta_\phi=0.54(5)$,
$\Delta_S=1.25(9)$, $\Delta_{W\lnsp X}=1.75(4)$, $\Delta_Y=0.97(7)$ and
$\Delta_Z=0.46(12)$, while, in the $\overline{\text{MS}}$ scheme,
\cite[Eq.\ (2.8)]{Calabrese:2004nt} and \cite[Table III]{Calabrese:2004at}
give $\Delta_\phi=0.545(20)$, $\Delta_S=1.46(14)$, $\Delta_{W\lnsp
X}=1.66(15)$, $\Delta_Y=1.00(15)$ and $\Delta_Z=0.63(15)$.\foot{The
notation $S, W\lnsp X, Y, Z$ is explained in Appendix~\ref{appCross}.}
Therefore, we conclude that the kink in Fig.~\ref{fig:Delta_Z_O22} does not
correspond to the $O_{2,2}$ chiral fixed point. In~\cite{Stergiou:2019dcv}
it was suggested that this kink may correspond to the fully-interacting
theory of the $\veps$ expansion analyzed in~\cite{Shpot, Shpot2, Mudrov,
Mudrov2}.
\begin{figure}[H]
  \centering
  \includegraphics[scale=0.9]{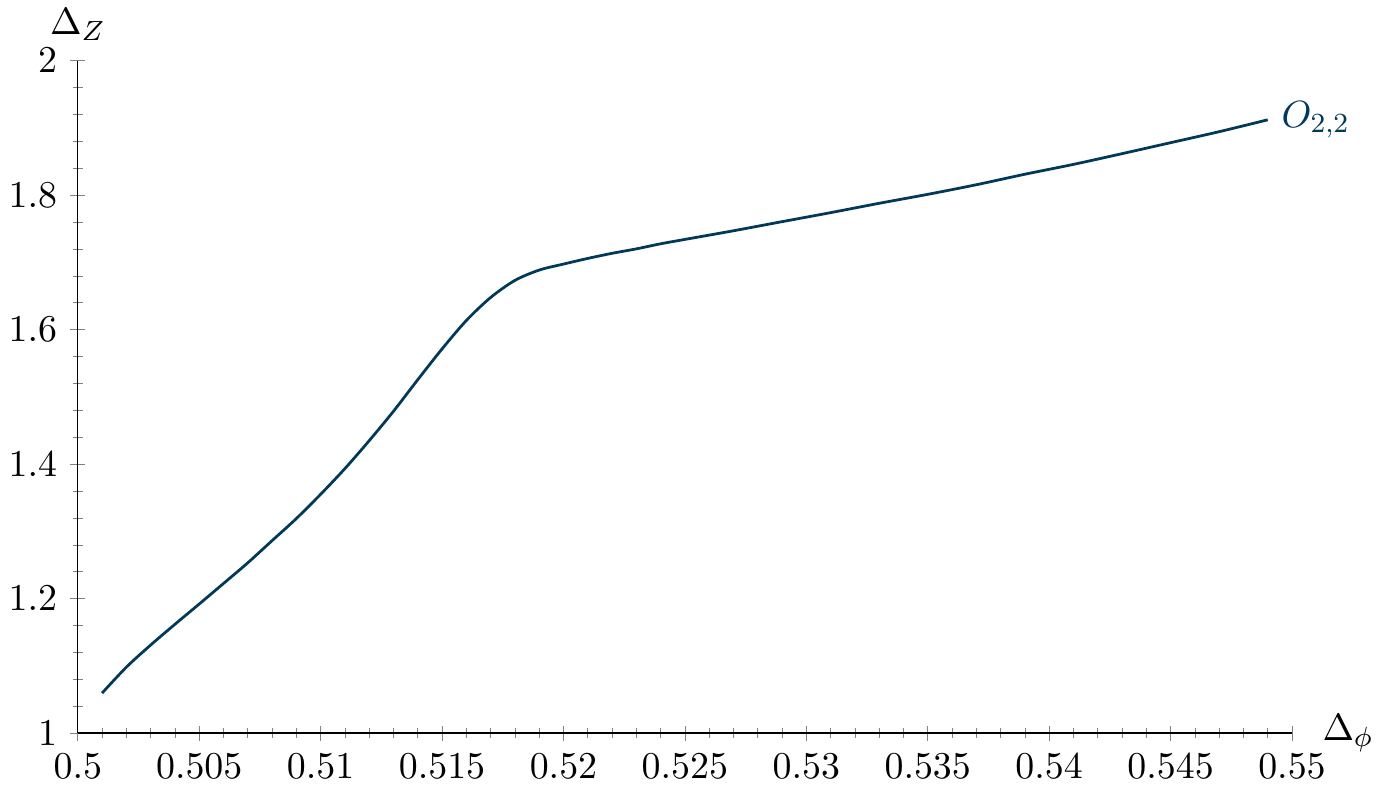}
  \caption{Upper bound on the dimension of the first scalar $Z$ operator in
  the $\phi_{ar}\times\phi_{bs}$ OPE as a function of the dimension of $\phi$.
  The area above the curve is excluded.} \label{fig:Delta_Z_O22}
\end{figure}

\begin{figure}[H]
  \centering
  \includegraphics[scale=0.9]{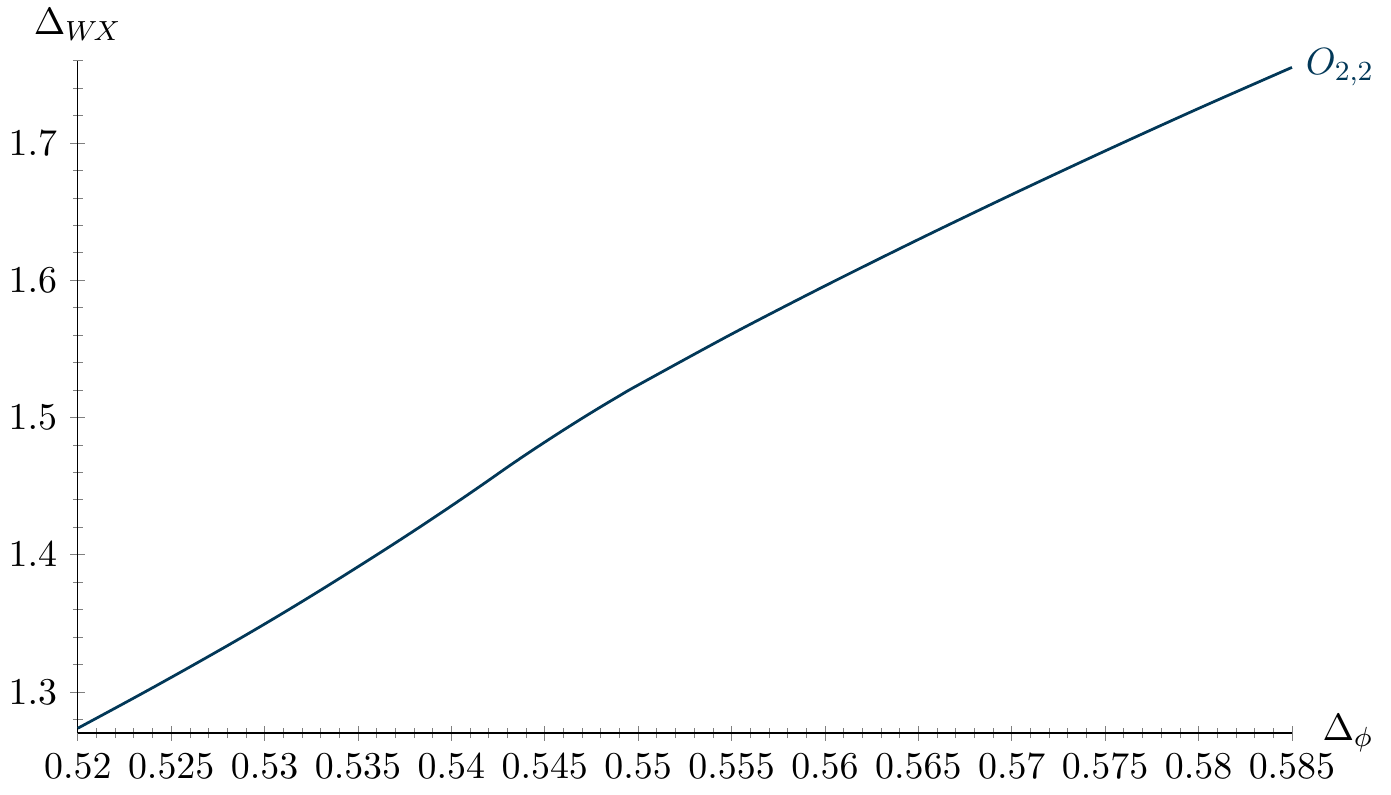}
  \caption{Upper bound on the dimension of the first scalar $W\lnsp X$
  operator in the $\phi_{ar}\times\phi_{bs}$ OPE as a function of the dimension
  of $\phi$.  The area above the curve is excluded.}
  \label{fig:Delta_WX_O22}
\end{figure}
To continue our search for the $O_{2,2}$ chiral fixed point, we obtain a
bound on the dimension of the first scalar operator in the $W\lnsp X$
representation. The bound is shown in Fig.~\ref{fig:Delta_WX_O22}. An
extremely mild change in the slope of the bound is observed at
$\Delta_\phi=0.547(2)$, at which point $\Delta_{W\lnsp X}=1.507(10)$. These
numbers are in relatively good agreement with the $\overline{\text{MS}}$
scheme results mentioned in the previous paragraph. The mildness of the
kink in Fig.~\ref{fig:Delta_WX_O22}, however, indicates that we need
stronger numerics in order to reach solid conclusions. We leave this for
future work. Finally, we note that there exists a second kink in the
$\Delta_Z$-$\Delta_\phi$ bound, with $\Delta_\phi\approx0.615$ and
$\Delta_Z\approx2.7$ \cite[Fig.\ 3]{Stergiou:2019dcv}, which was suggested
in \cite{Stergiou:2019dcv} to be relevant for phase transitions in the
helimagnets Ho, Dy and the structural phase transition of NbO$_2$.

\newsec{Summary and conclusion}[conc]
Conformal field theories with $O(m)\times O(n)$ global symmetry in $d=3$
spacetime dimensions have generated intense interest and deep questions for
many years.  Standard methods like the $\veps$ and large-$n$ expansions
have been employed for their study, and have revealed a rather intricate
structure of fixed points depending on the values of $m$ and $n$ (see
Fig.~\ref{fig:Omn_fixed_points}). For example, it is widely believed that
for $m=2$ the large $n$ fixed points (chiral and antichiral) survive down
to $n=6$ as unitary fixed points, while for some $n$ between 5 and 6 they
collide and become nonunitary for $n=2,3,4,5$. Similar results hold for
$m>2$ and corresponding values of $n$.

In this work we applied analytic and numerical bootstrap techniques to the
study of $O(m)\times O(n)$ CFTs. Our analytic bootstrap results both
corroborated and extended earlier results in the $\veps$ and large-$n$
expansions. Indeed, we found the large-$n$ chiral and antichiral fixed
points purely from analytic bootstrap considerations, and verified some of
their properties as established with older methods.  Our new analytic
results consisted of $\veps$ and large-$n$ expansion expressions for
scaling dimensions of $\phi$-bilinears in all representations of operators
that appear in the $\phi\times\phi$ OPE for spins zero and one, where
$\phi$ transforms in the bifundamental representation of $O(m)\times O(n)$.
These results allowed us to unequivocally identify kinks in our numerical
bootstrap bounds with the known fixed points at large $n$.  We also
obtained analytic results for OPE coefficients related to the central
charge $C_T$ and the coefficients $C_J$ of current-current two-point
functions.

Our numerical bootstrap bounds were focused on the case $m=2$. For $n=10,
20$ we were able to find islands in the $(\Delta_\phi,\Delta_S)$ parameter
space, where $\Delta_S$ is the dimension of the leading scalar singlet in
the $\phi\times\phi$ OPE, corresponding to the chiral and antichiral fixed
points and located in the region predicted by large-$n$ results (see
Figs.~\ref{fig:O2O20_island} and \ref{fig:O2O10_island}).  For the edge
case of $m=2, n=6$ we also obtained an island (see
Fig.~\ref{fig:O2O6_chiral_island}). In all these cases we used a mixed
correlator bootstrap. In the $m=2,n=6$ case we were able to compare our
nonperturbative bootstrap results with the state-of-the-art resummed
$\veps$ expansion results of~\cite{Kompaniets:2019xez}. The agreement is
reasonably good, but our island is rather large. We need stronger numerics
and more refined bootstrap techniques in order to make our island smaller
and obtain more accurate determinations of critical exponents.

CFTs with $O(2)\times O(3)$ and $O(2)\times O(2)$ global symmetry have been
argued to have various applications to observed critical phenomena.
However, when analyzed with different RG methods, notably resummations of
perturbative beta functions, Monte Carlo and functional RG, the obtained
results are not in mutual agreement. More specifically, beta function
resummations~\cite{Pelissetto:2000ne, DePrato:2003ra, Calabrese:2004nt} and
Monte Carlo computations~\cite{Nagano_2019} indicate the presence of fixed
points beyond the ones found with perturbative methods, while the
functional RG concludes that such fixed points do not
exist\cite{Tissier:2000tz, Tissier:2001uk, Delamotte:2003dw,
Delamotte:2016acs}.  Correspondingly, there are two conflicting
suggestions, namely that experiments are seeing critical (second-order) or
weakly first-order behavior. It is important to note here that the beta
function resummation and Monte Carlo fixed points are suggested to be
nonunitary (focus type), with next-to-leading scalar singlets of complex
scaling dimensions. We attempted to address these issues in this work. Our
results provided support for the existence of these fixed points, but we
saw no signs of nonunitarity.\footnote{Ideally, assuming there is
nonunitarity, we would expect the islands to disappear at some higher, a
priori unknown, number of derivatives included in the bootstrap functional.
Also, we emphasize that it is natural we do not see the nonunitarity in the
analytic bootstrap since it does not capture the new fixed points found in
resummations.}  Overall, we were unable to provide conclusive answers, but
we believe that more dedicated bootstrap work with stronger numerics will
be able to reach definitive conclusions in the near future.

Another area of interest concerns the chiral phase transition of two-flavor
massless quantum chromodynamics~\cite{Pisarski:1983ms} (with global
symmetry $SU(2)\times SU(2)$) which has not been conclusively demonstrated
to be first- or second-order. A conformal bootstrap study of this question,
as related to the three-dimensional $O(2)\times O(4)$ theory after assuming
that the axial $U(1)$ symmetry is restored at the transition point so that
the full global symmetry group is $U(1)\times SU(2)\times SU(2)$, has
appeared in~\cite{Nakayama:2014sba} (for a review
see~\cite{Nakayama:2015ikw}). Based on the existence of kinks in bootstrap
bounds, e.g.\ the one in Fig.~\ref{fig:Delta_W_O2n}, it was suggested in
\cite{Nakayama:2014sba} that the phase transition is second order. The
$O(2)\times O(4)$ case is similar to the $O(2)\times O(3)$ and $O(2)\times
O(2)$ CFTs studied in this paper, in that the existence of the suggested
fixed points has been questioned by other methods. We believe that with
stronger numerics and perhaps larger mixed-correlator systems than those
considered here it will be possible to settle this question with the
conformal bootstrap.

We conclude with the Tables~\ref{tab:tableChiral}--\ref{tab:tableO2O2},
which summarize our results of critical exponents for $O(2)\times O(3)$ and
$O(2)\times O(2)$ theories, as well as corresponding results from the
literature.\footnote{The error bars in the first lines of Tables
\ref{tab:tableChiral} and \ref{tab:tableAntiChiral} correspond to the size
of the islands, but are not rigorous due to the assumption that the
theories must saturate the $\Delta_W$ and $\Delta_Z$ exclusion bounds,
respectively. For the crossover exponents we use the values of $\Delta_W$
and $\Delta_Z$ that correspond to the positions where the slope changes in
the corresponding exclusion bounds. Note also that Table
\ref{tab:tableO2O2} refers to the theory saturating the bound in
Fig.~\ref{fig:Delta_Z_O22}; this theory is expected to be the one found in
the $\varepsilon$ expansion. Lastly, the error bars for the numerical
bootstrap in Table \ref{fig:Delta_Z_O22} are due to the use of the extremal
functional method and are also not rigorous.}

\begin{table}[H]
  \small
  \begin{center}
    \caption{$O(2)\times O(3)$ ``chiral'' critical exponents.}
    \label{tab:tableChiral}
    \begin{tabular}{l|c|c|c|c|c|c|c}
      \textbf{Method} & $\alpha$ & $\beta$ & $\nu$ & $\gamma$ & $\delta$ & $\eta$ & $\phi_W$\\
      \hline
      This work (Figs.\ \ref{fig:Delta_W_O23}, \ref{O2O3islandpeninsula})& 0.082(22)  & 0.344(5) &0.639(7) &1.23(3) &4.573(14) &0.077(3) &0.818(16)\\
      $\overline{\text{MS}}$ scheme \cite{Calabrese:2004nt,
      Calabrese:2004at} & 0.11(15)&0.34(4) &0.63(5) &1.20(24) &4.5(2) &0.09(4) & 0.76(12)\\
      MZM scheme \cite{Pelissetto:2000ne, Calabrese:2004at} &0.35(9)  & 0.30(4)&0.55(3) &1.04(18) &4.5(5) &0.1(1) &0.58(6)\\
      Monte Carlo \cite{Nagano_2019} & 0.44(3) &0.26(3) &0.52(1) &1.04(9) &5.0(5) &0.00(8) &--\\
    \end{tabular}
  \end{center}
\end{table}
\begin{table}[h!]
  \small
  \begin{center}
    \caption{$O(2)\times O(3)$ ``antichiral/collinear'' critical exponents.}
    \label{tab:tableAntiChiral}
    \begin{tabular}{l|c|c|c|c|c|c|c}
      \textbf{Method} & $\alpha$ & $\beta$ & $\nu$ & $\gamma$ & $\delta$ & $\eta$ & $\phi_Z$\\
      \hline
      This work (Figs.\ \ref{fig:Delta_Z_O23}, \ref{fig:O2O3_collinear_island_peninsula}) &0.05(7)   &0.341(19)  &0.650(23) &1.27(11) &4.72(13) &0.049(23) &0.89(4)\\
      $\overline{\text{MS}}$ scheme \cite{Calabrese:2004nt,
      Calabrese:2004at} &0.11(24) &0.34(5) &0.63(8) &1.2(3) &4.52(12) &0.086(24) &0.75(16)\\
      MZM scheme \cite{Pelissetto:2000ne, Calabrese:2004at} &0.22(12)  &0.319(23) &0.59(4) &1.14(16) &4.56(4) &0.079(7) &0.74(11)\\
    \end{tabular}
  \end{center}
\end{table}
\begin{table}[h!]
  \small
  \begin{center}
    \caption{$O(2)\times O(2)$ critical exponents.}
    \label{tab:tableO2O2}
    \begin{tabular}{l|c|c|c|c|c|c}
      \textbf{Method} & $\alpha$ & $\beta$ & $\nu$ & $\gamma$ & $\delta$ & $\eta$  \\
      \hline
      Bootstrap (Fig.\ \ref{fig:Delta_Z_O22} and \cite{Stergiou:2019dcv}) &0.302(18)   & 0.293(3) &0.566(6) &1.112(24) &4.7952(21) &0.0353(4) \\
      $\varepsilon$ expansion \cite{Mudrov2} &$-0.14(3)$ &0.370(6) &0.715(10) &1.404(25) &4.801(11) &0.0343(20) \\
    \end{tabular}
  \end{center}
\end{table}
\noindent Hints for the controversial chiral $O(2)\times O(2)$ model where
found in Fig.~\ref{fig:Delta_WX_O22}, but we consider the strength of our
numerics insufficient to provide estimates for critical exponents in this
case.

\ack{We are grateful to B.\ Delamotte, J.\ Gracey and D.\ Mouhanna for
comments on the manuscript and for pointing out some relevant literature.
We would like to thank the organizers of ``Bootstrap 2019'' at Perimeter
Institute where this work was initiated. Research at Perimeter Institute is
supported in part by the Government of Canada through the Department of
Innovation, Science and Economic Development Canada and by the Province of
Ontario through the Ministry of Economic Development, Job Creation and
Trade. This research used resources provided by the Los Alamos National
Laboratory Institutional Computing Program, which is supported by the U.S.\
Department of Energy National Nuclear Security Administration under
Contract No.\ 89233218CNA000001. This research used resources of the
National Energy Research Scientific Computing Center (NERSC), a U.S.\
Department of Energy Office of Science User Facility operated under
Contract No.\ DE-AC02-05CH11231. Some numerical computations in this paper
were run on the LXPLUS cluster at CERN and the Metropolis cluster at the
Crete Center for Quantum Complexity and Nanotechnology. AS is grateful for
an allocation of computing time on the Caltech High Performance Cluster,
provided by the Simons  Foundation grant 48865
(\href{https://bootstrapcollaboration.com/}{Simons Collaboration on the
Nonperturbative Bootstrap}). The research work of SRK was supported by the
Hellenic Foundation for Research and Innovation (HFRI) under the HFRI PhD
Fellowship grant (Fellowship Number: 1026).  Research presented in this
article was supported by the Laboratory Directed Research and Development
program of Los Alamos National Laboratory under project number
20180709PRD1.

\vspace{-18pt}
\begin{figure}[H]
  \flushright
  \includegraphics[scale=0.45]{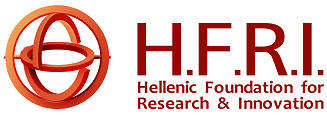}
\end{figure}
}
\vspace{-1.5cm}

\begin{appendices}

\section{Description of ancillary data file}
\label{sec:ancillarydata}
In the ancillary data file of the arXiv submission, we give full results for a number of quantities, presented on the form $\langle$\emph{quantity}$\rangle\langle$\emph{fx-pt}$\rangle\langle$\emph{expansion}$\rangle$, where $\langle$\emph{fx-pt}$\rangle$ ranges over the fixed points \texttt{ON}, \texttt{Chiral} and \texttt{Anti}; and $\langle$\emph{expansion}$\rangle$ over the expansions \texttt{Eps} and \texttt{N}. In Table~\ref{table:auxdata} we list the options for $\langle$\emph{quantity}$\rangle$.

The expressions make use of the constants
\begin{align}
\texttt{sqRmn}&=\sqrt{R_{mn}}=\sqrt{m^2-10 m n+n^2-4 m-4 n+52}\,,
\\
\texttt{eta1}&=\eta_1^{O(N)}=\frac{(\mu -2)\Gamma (2 \mu -1)}{   \Gamma (\mu )\Gamma (\mu +1)\pi \csc(\pi\mu)}\,,
\end{align}
implemented as \texttt{sqRmnval} and \texttt{eta1val} respectively (see \eqref{eq:eta1def}).
The values that we present for \texttt{deltaScalars}$\langle$\emph{fx-pt}$\rangle$\texttt{Eps} contain results from the literature at order $\veps^2$, extracted from \cite{Osborn:2017ucf}.

\begin{table}[H]
\centering
\caption{Quantities given in the ancillary data file.}\label{table:auxdata}
\begin{tabular}{|llll|}\hline
Expression & Quantity & Order & Depends on
\\\hline
\texttt{deltaPhi}$\langle$\emph{fx-pt}$\rangle$\texttt{Eps} & $\Delta_\phi$
& $\veps^3$ & $m,n,\veps,\texttt{sqRmn}$
\\
\texttt{deltaScalars}$\langle$\emph{fx-pt}$\rangle$\texttt{Eps} & $\{\Delta_{\phi^2_R}\}_{R=1,\ldots,5}$ & $\veps^2$ & $m,n,\veps,\texttt{sqRmn}$
\\
\texttt{deltaSpinning}$\langle$\emph{fx-pt}$\rangle$\texttt{Eps} & $\{\Delta_{R,\ell}\}_{R=1,\ldots,9}$ & $\veps^3$ & $\ell,m,n,\veps,\texttt{sqRmn}$
\\
\texttt{CT}$\langle$\emph{fx-pt}$\rangle$\texttt{Eps} & $C_T/C_{T,\text{free}}$ & $\veps^3$ & $m,n,\veps,\texttt{sqRmn}$
\\
\texttt{CJA}$\langle$\emph{fx-pt}$\rangle$\texttt{Eps} & $C_{J_A}/C_{J_A,\text{free}}$ & $\veps^3$ & $m,n,\veps,\texttt{sqRmn}$
\\
\texttt{CJB}$\langle$\emph{fx-pt}$\rangle$\texttt{Eps} & $C_{J_B}/C_{J_B,\text{free}}$ & $\veps^3$ & $m,n,\veps,\texttt{sqRmn}$
\\\hline
\texttt{deltaPhi}$\langle$\emph{fx-pt}$\rangle$\texttt{N} & $\Delta_\phi$ & $1/n$ & $m,n,\mu,\texttt{eta1}$
\\
\texttt{deltaScalars}$\langle$\emph{fx-pt}$\rangle$\texttt{N} & $\{\Delta_R\}_{R=1,\ldots,5}$ & $1/n$ & $m,n,\mu,\texttt{eta1}$
\\
\texttt{deltaSpinning}$\langle$\emph{fx-pt}$\rangle$\texttt{N} & $\{\Delta_{R,\ell}\}_{R=1,\ldots,9}$ & $1/n$ & $\ell,m,n,\mu,\texttt{eta1}$
\\
\texttt{CT}$\langle$\emph{fx-pt}$\rangle$\texttt{N} & $C_T/C_{T,\text{free}}$ & $1/n$ & $m,n,\mu$
\\
\texttt{CJA}$\langle$\emph{fx-pt}$\rangle$\texttt{N} & $C_{J_A}/C_{J_A,\text{free}}$ & $1/n$ & $m,n,\mu$
\\
\texttt{CJB}$\langle$\emph{fx-pt}$\rangle$\texttt{N} & $C_{J_B}/C_{J_B,\text{free}}$ & $1/n$ & $m,n,\mu$
\\
\texttt{ope}$\langle$\emph{fx-pt}$\rangle$\texttt{N}  & $\{a_{R,\ell}\}_{R=1,\ldots,9} $ & $1/n$&  $\ell,m,n,\mu,\texttt{eta1}$
\\\hline
\end{tabular}
\end{table}
Data for \texttt{ope}$\langle fx\text{-}pt\rangle$\texttt{Eps} include long
expressions that we do not include in the ancillary data file but can be
made available upon request.

\section{Explicit formulas used in the main text}
\label{app:explicitresults}
Here we give a few explicit formulas used in the main text. In \eqref{eq:OPEcoefseps2} we have
\begin{align}\nonumber
a^{\mathrm{GFF}}_{0,\ell}=\frac{2\Gamma(\ell+1)^2}{\Gamma(2\ell+1)}\Big[&
1+\left(S_1(2\ell)-2S_1(\ell)\right)\left(\veps-2\gamma_\phi^{(2)}\veps^2\right)\\
&+\frac{\veps^2}4\left(
8S_1(\ell)(S_1(\ell)-S_1(2\ell))+2S_1(2\ell)^2-3S_2(\ell)+2S_2(2\ell)
\right)
\Big]\,,
\label{eq:aGFF2}
\end{align}
which is an expansion of the general formula for the OPE coefficients of generalized free fields \cite{Fitzpatrick:2011dm},
\begin{equation}
\label{eq:GFFOPE}
\hspace{-2pt}a^{\mathrm{GFF}}_{n,\ell}|_{\Delta_\phi}=\frac{2(\Delta_\phi+1-\mu)_n^2(\Delta_\phi)_{n+\ell}^2}{\ell !\, n!\, (\ell+\mu)_n(2\Delta_\phi+n+1-2\mu)_n(2\Delta_\phi+\ell+n-\mu)_n(2\Delta_\phi+2n+\ell-1)_\ell},
\end{equation}
where $(a)_n=\frac{\Gamma(a+n)}{\Gamma(a)}$ is the Pochhammer symbol.

In Inversion~\ref{inv:fromcurrents}, $E_\pm$ is given by
\begin{align}\nonumber
E_\pm=\alpha \kappa^2\frac{2\hb-1}{4}\Bigg[&\frac{2\pi \csc(\pi\mu)\left((\mu-2)(\SS[\mu-1](\hb)+\pi\cot(\pi\mu))-1\right)\pm 2\left(\frac{3-\mu}{\mu-2}-S_1(\mu-2)\right)}{J^2(\mu-2)^2}\\
&\pm\frac{\mu(\mu-1)B(\mu,\hb)+2(S_2(\mu-2)-\zeta_2)+\frac{2(\mu-3)}{(\mu-2)^2}}{(J^2-(\mu-1)(\mu-2))(\mu-2)}-\frac{2\pi\csc(\pi\mu)}{J^4(\mu-2)}
\Bigg]\,,
\label{eq:Eplusminus}
\end{align}
where
\begin{equation}
B(\mu,\hb)=\frac{{\displaystyle {_4F_3}\!\left(\!\left. {1,1,2,\mu+1}~\atop~{\!\!3,3-\hb,\hb+2} \right|1\right)}}{J^2(J^2-2)}
-\frac{2 \pi  \Gamma (\hb) \Gamma (\mu +\hb-1)}{J^2 \Gamma (\mu +1) \sin (\pi  \hb)  \Gamma (2 \hb)} {_3F_2}\!\left(\!\left. {\hb-1,\hb,\hb+\mu -1}~\atop~{\!\!2 \hb,\hb+1} \right|1\right)\,,
\end{equation}
and in expressions \eqref{eq:chargesChiral} and \eqref{eq:chargesAnti} we have
\begin{align}
c_1&=\left(
\frac{4+2\mu-\mu^2}{2\mu(2-\mu)}
+\pi\cot(\pi\mu)+S_1(2\mu-2)
\right)\frac{\eta_1^{O(N)}}{\mu+1}\,,
\nonumber\\
c_2&=\frac{2\mu-1}{\mu(\mu-1)}\eta_1^{O(N)}\,,
\nonumber\\
c_3&=\left(
\frac{\mu^3-6\mu^2+11\mu-4}{\mu(\mu-1)(2-\mu)}
+\pi\cot(\pi\mu)+S_1(2\mu-3)
\right)\eta_1^{O(N)}\,.
\label{eq:cconstants}
\end{align}

\newsec{Crossing equations in \texorpdfstring{$\boldsymbol{O(m)\times O(n)}$}{O(m)xO(n)} theories}[appCross]
For the single-correlator numerical bootstrap, i.e.\ that of the four-point
function $\langle\phi_{ar}\phi_{bs}\phi_{ct}\phi_{du}\rangle$, we can
easily work out the crossing equations by noticing that, when thinking of
$\phi$ as a matrix, $O(m)$ acts on the row index and $O(n)$ on the column
index. Consequently, the projectors that allow for the decomposition of the
four-point function into invariant subspaces can be formed as products of
the well-known $O(N)$ projectors,
\eqn{\hat{P}^{S}_{N;\lsp ijkl}=\tfrac{1}{N}\delta_{ij}\delta_{kl}\,,\quad
\hat{P}^{T}_{N;\lsp ijkl}=\tfrac12(\delta_{ik}\delta_{jl}+\delta_{il}\delta_{jk}
-\tfrac{2}{N}\delta_{ij}\delta_{kl})\,,\quad
\hat{P}^{A}_{N;\lsp ijkl}=\tfrac12(\delta_{ik}\delta_{jl}
-\delta_{il}\delta_{jk})\,,}[]
where the subscript $N$ is introduced to capture the relation
$\delta_{ij}\delta^{ij}=N$. The four-point function
$\langle\phi_{ar}\phi_{bs}\phi_{ct}\phi_{du}\rangle$ can be decomposed as
\eqn{\langle\phi_{ar}(x_1)\phi_{bs}(x_2)\phi_{ct}(x_3)\phi_{du}(x_4)
\rangle=\frac{1}{(x_{12}^{2}x_{34}^{2})^{\Delta_\phi}}
\sum_R\sum_{\mathcal{O}_R}c_{\mathcal{O}_R}^2
P^R_{arbsctdu}\lsp G_{\Delta_R,\ell_R}(u,v)\,,}[decomp]
where the sum over $R$ runs over the representations $S,W,X,Y,Z,A,B,C,D$,
$x_{ij}=x_i-x_j$, $c_{\O_R}^2$ are squared OPE coefficients and
$G_{\Delta_R,\ell_R}(u,v)$ are conformal blocks\foot{In our numerical
bootstrap considerations we define conformal blocks using the conventions
of~\cite{Behan:2016dtz}.} that are functions of the conformally-invariant
cross-ratios defined in \crossratios.
The projectors in \decomp are
\begin{align}
P^S_{arbsctdu}&=\hat{P}^S_{m;\lsp abcd}\hat{P}^S_{n;\lsp rstu}\,,\quad
P^W_{arbsctdu}=\hat{P}^T_{m;\lsp abcd}\hat{P}^S_{n;\lsp rstu}\,,\quad
P^X_{arbsctdu}=\hat{P}^S_{m;\lsp abcd}\hat{P}^T_{n;\lsp rstu}\,,\nonumber\\
P^Y_{arbsctdu}&=\hat{P}^T_{m;\lsp abcd}\hat{P}^T_{n;\lsp rstu}\,,\quad
P^Z_{arbsctdu}=\hat{P}^A_{m;\lsp abcd}\hat{P}^A_{n;\lsp rstu}\,,\quad
P^A_{arbsctdu}=\hat{P}^A_{m;\lsp abcd}\hat{P}^S_{n;\lsp rstu}\,,\nonumber\\
P^B_{arbsctdu}&=\hat{P}^S_{m;\lsp abcd}\hat{P}^A_{n;\lsp rstu}\,,\quad
P^C_{arbsctdu}=\hat{P}^A_{m;\lsp abcd}\hat{P}^T_{n;\lsp rstu}\,,\quad
P^D_{arbsctdu}=\hat{P}^T_{m;\lsp abcd}\hat{P}^A_{n;\lsp rstu}\,.
\end{align}
For the crossing equation we define
\eqn{F_{\Delta,\lsp\ell}^{\pm}(u,v)=v^{\Delta_\phi}g_{\Delta,\lsp\ell}(u,v)
\pm u^{\Delta_\phi}g_{\Delta,\lsp\ell}(v,u)\,,}[Fpmdef]
and we find\foot{In \eqref{crEqOmOn} we suppress the labeling of the
$F_{\Delta,\ell}$'s and $c_\O^2$'s with the appropriate index $I$.
The appropriate labeling, however, is obvious from the overall sum in each
term.}
\begin{align}
  &\sum_{S^+}c_\O^2\begin{pmatrix}
  F^-_{\Delta,\lsp\ell}\\
  0\\
  0\\
  0\\
  0\\
  F^+_{\Delta,\lsp\ell}\\
  0\\
  0\\
  0
\end{pmatrix}+
\sum_{W^+}c_\O^2\begin{pmatrix}
  -\frac{2}{m}F^-_{\Delta,\lsp\ell}\\
  0\\
  F^-_{\Delta,\lsp\ell}\\
  F^-_{\Delta,\lsp\ell}\\
  0\\
  -\frac{2}{m}F^+_{\Delta,\lsp\ell}\\
  0\\
  -F^+_{\Delta,\lsp\ell}\\
  F^+_{\Delta,\lsp\ell}
\end{pmatrix}+
\sum_{X^+}c_\O^2\begin{pmatrix}
  -\frac{2}{n}F^-_{\Delta,\lsp\ell}\\
  F^-_{\Delta,\lsp\ell}\\
  F^-_{\Delta,\lsp\ell}\\
  0\\
  0\\
  -\frac{2}{n}F^+_{\Delta,\lsp\ell}\\
  F^+_{\Delta,\lsp\ell}\\
  F^+_{\Delta,\lsp\ell}\\
  0
\end{pmatrix}+
\sum_{Y^+}c_\O^2\begin{pmatrix}
  (1+\frac{4}{mn})F^-_{\Delta,\lsp\ell}\\
  (1-\frac{2}{m})F^-_{\Delta,\lsp\ell}\\
  -2\lsp(\frac{1}{m}+\frac{1}{n})F^-_{\Delta,\lsp\ell}\\
  (1-\frac{2}{n})F^-_{\Delta,\lsp\ell}\\
  2\llsp F^-_{\Delta,\lsp\ell}\\
  -(1-\frac{4}{mn})F^+_{\Delta,\lsp\ell}\\
  -(1+\frac{2}{m})F^+_{\Delta,\lsp\ell}\\
  -2\lsp(\frac{1}{m}-\frac{1}{n})F^+_{\Delta,\lsp\ell}\\
  -(1+\frac{2}{n})F^+_{\Delta,\lsp\ell}
\end{pmatrix}\nonumber\\
&\hspace{4.5cm}+
\sum_{Z^+}c_\O^2\begin{pmatrix}
  F^-_{\Delta,\lsp\ell}\\
  -F^-_{\Delta,\lsp\ell}\\
  0\\
  -F^-_{\Delta,\lsp\ell}\\
  2\llsp F^-_{\Delta,\lsp\ell}\\
  -F^+_{\Delta,\lsp\ell}\\
  F^+_{\Delta,\lsp\ell}\\
  0\\
  F^+_{\Delta,\lsp\ell}
\end{pmatrix}+
\sum_{A^-}c_\O^2\begin{pmatrix}
  0\\
  0\\
  -F^-_{\Delta,\lsp\ell}\\
  F^-_{\Delta,\lsp\ell}\\
  0\\
  0\\
  0\\
  F^+_{\Delta,\lsp\ell}\\
  F^+_{\Delta,\lsp\ell}
\end{pmatrix}+
\sum_{B^-}c_\O^2\begin{pmatrix}
  0\\
  F^-_{\Delta,\lsp\ell}\\
  -F^-_{\Delta,\lsp\ell}\\
  0\\
  0\\
  0\\
  F^+_{\Delta,\lsp\ell}\\
  -F^+_{\Delta,\lsp\ell}\\
  0\\
\end{pmatrix}\nonumber\\
&\hspace{4cm}+\sum_{C^-}c_\O^2\begin{pmatrix}
  -F^-_{\Delta,\lsp\ell}\\
  -F^-_{\Delta,\lsp\ell}\\
  \frac{2}{n}F^-_{\Delta,\lsp\ell}\\
  (1-\frac{2}{n})F^-_{\Delta,\lsp\ell}\\
  2\llsp F^-_{\Delta,\lsp\ell}\\
  F^+_{\Delta,\lsp\ell}\\
  F^+_{\Delta,\lsp\ell}\\
  -\frac{2}{n}F^+_{\Delta,\lsp\ell}\\
  -(1+\frac{2}{n})F^+_{\Delta,\lsp\ell}
\end{pmatrix}+
\sum_{D^-}c_\O^2\begin{pmatrix}
  -F^-_{\Delta,\lsp\ell}\\
  (1-\frac{2}{m})F^-_{\Delta,\lsp\ell}\\
  \frac{2}{m}F^-_{\Delta,\lsp\ell}\\
  -F^-_{\Delta,\lsp\ell}\\
  2\llsp F^-_{\Delta,\lsp\ell}\\
  F^+_{\Delta,\lsp\ell}\\
  -(1+\frac{2}{m})F^+_{\Delta,\lsp\ell}\\
  \frac{2}{m}F^+_{\Delta,\lsp\ell}\\
  F^+_{\Delta,\lsp\ell}
\end{pmatrix}=
\begin{pmatrix}
  0\\
  0\\
  0\\
  0\\
  0\\
  0\\
  0\\
  0\\
  0
\end{pmatrix}.
\label{crEqOmOn}
\end{align}
The signs that appear as superscripts in the various irrep symbols indicate
the spins of the operators we sum over in the corresponding term: even when
positive and odd when negative.

When $m=n$ there is a reduction in the number of crossing equations.  In
this case, instead of separate projectors $P^W_{arbsctdu}$ and
$P^X_{arbsctdu}$, we only have the projector $P^{W\lnsp
X}_{arbsctdu}=P^W_{arbsctdu}+P^X_{arbsctdu}$, and similarly
$P^{AB}_{arbsctdu}=P^A_{arbsctdu}+P^B_{arbsctdu}$ and $P^{C\lnsp
D}_{arbsctdu}=P^C_{arbsctdu}+P^D_{arbsctdu}$, always with $m=n$.  The
crossing equation in the $m=n$ case is
\begin{align}
&\sum_{S^+}c_\O^2\begin{pmatrix}
  F^-_{\Delta,\lsp\ell}\\
  0\\
  0\\
  0\\
  F^+_{\Delta,\lsp\ell}\\
  0
\end{pmatrix}+
\sum_{W\lnsp X^+}c_\O^2\begin{pmatrix}
  -\tfrac{4}{n\lsp} F^-_{\Delta,\lsp\ell}\\
  F^-_{\Delta,\lsp\ell}\\
  F^-_{\Delta,\lsp\ell}\\
  0\\
  -\tfrac{4}{n}\lsp F^+_{\Delta,\lsp\ell}\\
  F^+_{\Delta,\lsp\ell}
\end{pmatrix}+
\sum_{Y^+}c_\O^2\begin{pmatrix}
  (1+\tfrac{4}{n^2})\lsp F^-_{\Delta,\lsp\ell}\\
  (1-\tfrac{2}{n})\lsp F^-_{\Delta,\lsp\ell}\\
  -\frac{2}{n}\lsp F^-_{\Delta,\lsp\ell}\\
  F^-_{\Delta,\lsp\ell}\\
  -(1-\tfrac{4}{n^2})\lsp F^+_{\Delta,\lsp\ell}\\
  -(1+\tfrac{2}{n})\lsp F^+_{\Delta,\lsp\ell}
\end{pmatrix}
+\sum_{Z^+}c_\O^2\begin{pmatrix}
  F^-_{\Delta,\lsp\ell}\\
  -F^-_{\Delta,\lsp\ell}\\
  0\\
  F^-_{\Delta,\lsp\ell}\\
  -F^+_{\Delta,\lsp\ell}\\
  F^+_{\Delta,\lsp\ell}
\end{pmatrix}\nonumber\\
&\hspace{6.5cm}+
\sum_{AB^-}c_\O^2\begin{pmatrix}
  0\\
  F^-_{\Delta,\lsp\ell}\\
  -F^-_{\Delta,\lsp\ell}\\
  0\\
  0\\
  F^+_{\Delta,\lsp\ell}
\end{pmatrix}+
\sum_{C\lnsp D^-}c_\O^2\begin{pmatrix}
  -F^-_{\Delta,\lsp\ell}\\
  -\tfrac{1}{n}\lsp F^-_{\Delta,\lsp\ell}\\
  \frac{1}{n}\lsp F^-_{\Delta,\lsp\ell}\\
  F^-_{\Delta,\lsp\ell}\\
  F^+_{\Delta,\lsp\ell}\\
  -\tfrac{1}{n}\lsp F^+_{\Delta,\lsp\ell}
\end{pmatrix}=
\begin{pmatrix}
  0\\
  0\\
  0\\
  0\\
  0\\
  0
\end{pmatrix}.
\label{crEqOnOn}
\end{align}

When $n=2$, \eqref{crEqOnOn} is equivalent to the crossing equation derived
for the global symmetry group $O(2)^2\rtimes S_2$
in~\cite[Eq.~(2.15)]{Stergiou:2019dcv}.  The representations $S,W\lnsp
X,Y,Z,AB,C\lnsp D$ in \eqref{crEqOnOn} correspond to the representations
$S,Z,Y,X,A,B$, respectively, in \cite[Eq.~(2.15)]{Stergiou:2019dcv}.
Theories with $O(2)^2/\mathbb{Z}_2$ and $O(2)^2\rtimes S_2$ global symmetry
are equivalent at the Lagrangian level in $d=4-\veps$
dimensions~\cite{Pelissetto:2000ek, Osborn:2017ucf}. The groups
$O(2)^2/\mathbb{Z}_2$ and $O(2)^2\rtimes S_2$ each have a fully-symmetric
four-index invariant tensor, as well as an additional four-index invariant
tensor whose symmetry properties are such that it does not generate quartic
invariants in the corresponding Lagrangians. For the $O(2)^2/\mathbb{Z}_2$
case the relevant invariant tensor was called $w_2$
in~\cite[Eq.~(5.81)]{Osborn:2017ucf}, and for $O(2)^2\rtimes S_2$ it was
called $w$ in~\cite[Eq.~(5.96)]{Osborn:2017ucf}. The relevance of the
$O(2)^2/\mathbb{Z}_2$ theory for experiments in stacked triangular
antiferromagnets, helimagnets and structural phase transitions has been
discussed extensively in~\cite{Stergiou:2019dcv} and references therein.

In this paper we also consider a mixed correlator bootstrap involving the
four-point functions $\langle\phi\phi\phi\phi\rangle$, $\langle\phi\phi S
S\rangle$ and $\langle S S S S\rangle$, where $S$ is the first scalar
singlet in the $\phi\times\phi$ OPE. The crossing equations for this system
are straightforward to derive due to the fact that $S$ transforms in the
singlet representation, and so in the OPE $\phi\times S$ we find only
operators that transform in $\phi$'s representation under the global
symmetry.

As usual, our numerical treatment involves two steps, namely generation of
an $\texttt{xml}$ file that encodes the problem and subsequently its
solution with a numerical algorithm. For the first step we use
\texttt{PyCFTBoot}~\cite{Behan:2016dtz}, and for the second
\texttt{SDPB}~\cite{Landry:2019qug}. For the single correlator bootstrap we
use the numerical parameters $\texttt{m\_max}=8, \texttt{n\_max}=11,
\texttt{k\_max}=42$ in \texttt{PyCFTBoot}, and we include spins up to
$\texttt{l\_max}=36$. For the mixed correlator bootstrap we use the
numerical parameters $\texttt{m\_max}=7, \texttt{n\_max}=9,
\texttt{k\_max}=40$ in \texttt{PyCFTBoot}, and we include spins up to
$\texttt{l\_max}=30$. The binary precision for the \texttt{xml} files in
both cases is 660 digits.  \texttt{SDPB} is run with the options
\texttt{--precision=860}, \texttt{--findPrimalFeasible},
\texttt{--findDualFeasible}, \texttt{--primalErrorThreshold=1e-30} and
finally\linebreak \texttt{--dualErrorThreshold=1e-15}. Default values are
used for other parameters of the solver.

\newsec{Mellin bootstrap for any global symmetry}
\label{sec:Mellinbootstrap}
In this appendix we will revisit some equations from the analytic boostrap in Mellin space
and apply them to the $\veps $ expansion for $\phi^4$ theories with arbitrary global symmetry.
The framework is described in detail in \cite{Gopakumar:2016cpb} and was generalized
to global symmetry in \cite{Dey:2016mcs}, focusing on $O(N)$ and hypercubic symmetry.
We will show how the Mellin space bootstrap reproduces the system of equations
\eqref{eq:eqsystScalars}, which we used to find the perturbative fixed points for a given global
symmetry. In addition, we will show that
\begin{equation}
c_{\phi\phi\phi^2_R}^2=a_{R,0}(1-g_R\veps)+\text O(\veps^2)\,,
\label{eq:alpharel}
\end{equation}
i.e.\ that $\alpha_R=-g_R$ in \eqref{eq:c2phi2def}, a result needed for our
computations in section~\ref{sec:generalsoleps} at order $\veps^3$.

The Mellin space bootstrap considers manifestly crossing symmetric expressions
in the Mellin space variables $s,t$, equivalent to sums over Witten diagrams for exchanges
of operators parametrized by $\Delta,\ell$. Consistency with the OPE implies
equations derived from the cancellation of spurious poles. For poles generated by the Mellin variable $s$ one
gets the equation
\begin{equation}
\sum_{\Delta,\ell}\left.\left(
c^{ijkl(s)}_{\Delta,\ell}M^{(s)}_{\Delta,\ell}(s,t)+
c^{ijkl(t)}_{\Delta,\ell}M^{(t)}_{\Delta,\ell}(s,t)+
c^{ijkl(u)}_{\Delta,\ell}M^{(u)}_{\Delta,\ell}(s,t)
\right)\right|_{s=\Delta_\phi}=0\,,
\label{eq:Mellinstart}
\end{equation}
valid for all $t$, where we have generalized the notation of \cite{Gopakumar:2016cpb}
by adding global symmetry indices. From \eqref{eq:Mellinstart}, one derives the
\emph{constraint equations}, by projecting to a given spin $\ell$ in the $s$-channel of the
$R$ representation, using the fact that the $M_{\Delta,\ell}(s,t)$ can be expressed in terms of continuous Hahn polynomials. The first term becomes simply $\sum_\Delta c^R_{\Delta,\ell}$, but for the other
two terms, operators of any spin $\ell'$ contribute. We arrive at constraint
equations of the form
\begin{align}
\sum_\Delta
c^R_{\Delta,\ell}\lsp
q_{\Delta,\ell}^{R(2,s)}+2\sum_{\rp}M_{R\rp}\sum_{\Delta',\ell'}c_{\Delta',\ell'}^{\rp}\lsp q_{\Delta',\ell|\ell'}^{\rp(2,t)}
&=0\,,
\label{eq:Mellindirect}
\\
\sum_\Delta
c^R_{\Delta,\ell}\lsp q_{\Delta,\ell}^{R(1,s)}
+
2M_{RS}\,q_{\1}^{(1,t)}
+2\sum_{\rp}M_{R\rp}\sum_{\Delta',\ell'}c_{\Delta',\ell'}^{\rp}\lsp q_{\Delta',\ell|\ell'}^{\rp(1,t)}
&=0\,,
\label{eq:Mellinderiv}
\end{align}
where, again, the exact form of the involved quantities can be found in
\cite{Gopakumar:2016cpb}. We have collected the $t$ and $u$ channel
contributions under the label $t$, and the appearance of the matrix
$M_{R\rp}$ follows from using the projectors of \eqref{eq:correlatordef}.
Equations \eqref{eq:Mellindirect} and \eqref{eq:Mellinderiv} are
generalizations of equations (2.39)--(2.44) of \cite{Dey:2016mcs}. We will
evaluate them for $\ell=0$, assuming that
$\Delta_{\phi^2_R}=2-\veps+g_R\lsp\veps+\text O(\veps^2)$ and
$c_{\phi\phi\phi^2_R}^2=a_{R,0}(1+\alpha_R\lsp\veps)+\text O(\veps^2)$, and
expanding to order $\veps$. Only $\ell'=0$ contributes to the sum, and following \cite{Gopakumar:2016cpb} we
substitute
\begin{align}
c^R_{\Delta,0}q_{\Delta,0}^{R(2,s)}&=-a_{R,0}\, g_R(g_R-1)\frac\veps2+\text O(\veps^2)\,,
\\
c^R_{\Delta,0}q_{\Delta,0}^{R(1,s)}&=a_{R,0}\left(1+(\alpha_R+g_R-1-\gamma_E)\veps\right)+\text O(\veps^2)\,,
\\
c_{\Delta',0'}^{\rp}q_{\Delta',0|0'}^{\rp(2,t)}&=-a_{\rp,0}\,g_{\rp}^2\frac{\veps}2+\text O(\veps^2)\,,
\\
c_{\Delta',0'}^{\rp}q_{\Delta',0|0'}^{\rp(1,t)}&=0+\text O(\veps^2)\,,
\\
q_{\1}^{(1,t)}&=-1+(1+\gamma_E)\veps+\text O(\veps^2)\,,
\end{align}
where $\gamma_E$ is the Euler--Mascheroni constant.

With these substitutions we can solve the constraint equations.
From \eqref{eq:Mellinderiv} at order $\veps^0$ we get $a_{R,0}=2M_{RS}$, in agreement
with \eqref{eq:treeOPE} above. Feeding this into \eqref{eq:Mellindirect} we get
\begin{equation}
-M_{RS}\lsp g_R(g_R-1)\lsp\veps-2\sum_{\rp}M_{R\rp}M_{\rp S}\lsp
g_{\rp}^2\lsp\veps=0+\text O(\veps^2)\,,
\end{equation}
which exactly agrees with \eqref{eq:eqsystScalars}. Finally,
by looking at \eqref{eq:Mellinderiv} at order $\veps$ we get
$\alpha_R+g_R=0$, proving \eqref{eq:alpharel}.

In \cite{Dey:2016mcs}, the CFT-data was computed to order
$\veps^3$ for $O(N)$ and hypercubic symmetry, and we believe that this will generalize to
arbitrary global symmetry. From such implementation, one can derive the order $\veps^3$ correction to
 the OPE coefficient $c_{\phi\phi\phi^2_R}^2$ (taking $\gamma_{\phi^2_R}^{(3)}$ as input),
 an observable that is inaccessible from large spin perturbation theory in its present formulation.

\end{appendices}

\bibliography{analytic_numerical_bootstrap_OmOn}

\end{document}